\documentclass[usenatbib]{mn2e}

\usepackage{graphicx}
\usepackage{amsmath}
\usepackage{amssymb}
\usepackage{url}
\usepackage{multirow}
\usepackage{color}

\usepackage[normalem]{ulem}

\usepackage[norounding]{rccol}
\rcDecimalSign{.}

\newcommand{\cpar}[1]{\left( #1 \right)}
\newcommand{\spar}[1]{\left[ #1 \right]}
\newcommand{\bpar}[1]{\left\{ #1 \right\}}

\newcommand{\gaa}{\text{\AA}}

\newcommand{\ww}{\textwidth}

\newcommand{\lmin}{\mbox{$\lambda_{\rm min}$}}
\newcommand{\lmax}{\mbox{$\lambda_{\rm max}$}}
\newcommand{\siiv}{\mbox{Si\,{\sc iv}}}
\newcommand{\civ}{\mbox{C\,{\sc iv}}}
\newcommand{\ciii}{\mbox{C\,{\sc iii}]}}
\newcommand{\mgii}{\mbox{Mg\,{\sc ii}}}
\newcommand{\nevi}{\mbox{[Ne\,{\sc vi}]}}
\newcommand{\oii}{\mbox{[O\,{\sc ii}]}}
\newcommand{\neiii}{\mbox{[Ne\,{\sc iii}]}}
\newcommand{\ha}{\mbox{H$\alpha$}}
\newcommand{\hb}{\mbox{H$\beta$}}
\newcommand{\hg}{\mbox{H$\gamma$}}
\newcommand{\hd}{\mbox{H$\delta$}}
\newcommand{\oiii}{\mbox{[O\,{\sc iii}]}}
\newcommand{\hei}{\mbox{He\,{\sc i}}}
\newcommand{\nii}{\mbox{[N\,{\sc ii}]}}
\newcommand{\sii}{\mbox{[Si\,{\sc ii}]}}
\newcommand{\feii}{\mbox[Fe\,{\sc ii]}}
\newcommand{\qsfit}{{\sc QSFit}}
\newcommand{\gfit}{{\sc GFIT}}
\newcommand{\qsfitcat}{{\tt QSFIT}}
\newcommand{\version}{{1.2}}

\newcommand{\nn}{71,251}

\def\aj{AJ}  
\def\araa{ARA\&A}
\def\apj{ApJ}
\def\apjl{ApJ}
\def\apjs{ApJS}
\def\aap{A\&A}
\def\mnras{MNRAS}
\def\pasp{PASP}

\title[QSFit: automatic analysis of optical AGN spectra] {QSFit:
  Automatic analysis of optical AGN spectra.}

\author[G. Calderone et al.]
       {G. Calderone,$^{1}$\thanks{E-mail: {\tt calderone@oats.inaf.it}}
         L. Nicastro,$^{2}$
         G. Ghisellini,$^{3}$
         M. Dotti,$^{4}$
         T. Sbarrato,$^{4}$ \newauthor
         F. Shankar$^{5}$
         and M. Colpi$^{4}$\\
         $^{1}$INAF--Osservatorio Astronomico di Trieste, Via Tiepolo 11, I-34131, Trieste (Italy)\\
         $^{2}$INAF--Istituto di Astrofisica Spaziale e Fisica Cosmica, Via Piero Gobetti 101, I-40129 Bologna (Italy)\\
         $^{3}$INAF--Osservatorio Astronomico di Brera, Via E. Bianchi 46, I-23807 Merate (Italy)\\ 
         $^{4}$Dip. di Fisica ``G. Occhialini'', Universit\`a di Milano -- Bicocca, Piazza della Scienza 3, I-20126 Milano (Italy)\\
         $^{5}$Department of Physics and Astronomy, University of Southampton, Highfield SO17 1BJ (UK)
}

\begin{document}

\pagerange{\pageref{firstpage}--\pageref{lastpage}} \pubyear{2011}

\maketitle

\label{firstpage}

\begin{abstract}
We present \qsfit{}, a new software package to automatically perform
the analysis of Active Galactic Nuclei (AGN) optical spectra.  The
software provides luminosity estimates for the AGN continuum, the
Balmer continuum, both optical and UV iron blended complex, host
galaxy and emission lines, as well as width, velocity offset and
equivalent width of 20 emission lines.

Improving on a number of previous studies on AGN spectral analysis,
QSFit fits all the components simultaneously, using an AGN continuum
model which extends over the entire available spectrum, and is thus a
probe of the actual AGN continuum whose estimates are scarcely
influenced by localized features (e.g. emission lines) in the
spectrum.

We used \qsfit{} to analyze \nn{} optical spectra of Type 1 AGN at
$z<2$ (obtained by the Sloan Digital Sky Survey, SDSS) and to produce
a publicly available catalog of AGN spectral properties.  Such catalog
allowed us (for the first time) to estimate the AGN continuum slope
and the Balmer continuum luminosity on a very large sample, and to
show that there is no evident correlation between these quantities the
redshift.

All data in the catalog, the plots with best fitting model and
residuals, and the IDL code we used to perform the analysis, are
available on a dedicated website.  The whole fitting process is
customizable for specific needs, and can be extended to analyze
spectra from other data sources.  The ultimate purpose of \qsfit{} is
to allow astronomers to run standardized recipes to analyze the AGN
data, in a simple, replicable and shareable way.
\end{abstract}

\begin{keywords}
  methods: data analysis, catalogues, galaxies: active, quasars: emission lines
\end{keywords}

\section{Introduction}
Statistical studies on large samples of Active Galactic Nuclei (AGN)
have proven to be valuable tools to understand the AGN phenomenon
\citep[e.g.][]{1974-Khachikian-type1vs2,
  1989-Kellermann-def_radio_loudness, 1989-sanders-torusReproBBB,
  1995-Urry-unifiedscheme}.  

The ongoing and future large--scale extragalactic surveys require more
appropriate, rapid, and flexible tools to properly reduce and
statistically analyze the vast amounts of already available and
incoming data.  For AGN, some examples of key quantities to extract
from the data are the composite broad--band SEDs
\citep{1994-Elvis-atlasQuasar, 2006-richards-meanSED}, composite
optical/UV spectra
\citep{1991-francis-composite,2001-vanden-composite,2002-telfer-UVPropOfQSO},
or the compilation of catalogs of spectral properties
\citep[e.g.][]{2011-shen-catdr7}.  These works allow to highlight
specific features of large samples, which are typically hidden when
considering individual sources.

The data analysis procedures employed so far, even if meticulously
described, have been very often based on ``private'' algorithms, which
do not allow other research groups to replicate the analysis, check
for inconsistencies, customize the procedure for specific needs, or
apply the analysis recipes to new data
\citep[e.g.][]{2008-Vestergaard-BHMassFunction, 2011-shen-catdr7}.
Nevertheless, the employed software is (at least) good enough to
produce results worth a publication, therefore it is likely good
enough to be publicly
released.\footnote{\url{http://www.nature.com/news/2010/101013/full/467753a.html}}
Moreover, the availability of a well tested software to perform data
analysis may help to standardize the analysis recipes, which would
otherwise be heterogeneously developed, and the results possibly hard
to compare directly.  In the field of X--ray data analysis, a very
important example of such a standardization is the {\sc XSPEC}
package,\footnote{\url{https://heasarc.gsfc.nasa.gov/xanadu/xspec/}}
which allowed researchers to focus on the interpretation of results,
instead of dealing with the many technical details of the analysis
procedures.

In the field of optical/UV AGN spectral analysis there is not yet a
commonly accepted and widely used software tool to estimate the
relevant spectral quantities, namely the AGN continuum luminosities,
the slopes, the emission line widths, luminosities, equivalent widths
etc.  although several works have tackled the problem of estimating
these quantities.  For instance the software by
\citet{2016-CalistroRivera-MCMC-SED} requires data from IR to X--rays
and does not provide emission lines estimates.  The works by
\citet{1992-boroson-emlineprop-irontempl} and \citet{2011-shen-catdr7}
provide the same estimates provided by \qsfit{}, but they did not
release the software code.  Finally, general purpose packages such as
SPECFIT \citep{1994-Kriss-specfit} provide just the basic fitting
functionalities, but do they lack a complete analysis recipe to go
from the spectrum in the FITS file to the final estimate of a spectral
quantities.

In this work we present, and make publicly available to the community,
our new and extensively tested software named \qsfit{} ({\it Quasar
  Spectral Fitting} package).  In its first release (version
\version{}) the software provides estimates of the spectral quantities
by simultaneously fitting all components (AGN continuum, Balmer
continuum, emission lines, host galaxy, iron blended lines) of the
observed optical spectrum.  The software is written in {\sc IDL}, and
it is released under the GPL license (\S\ref{sec:qsfit}).  It
currently only operates on spectra observed by the Sloan Digital Sky
Survey, Data Release 10
(SDSS--DR10\footnote{\url{https://www.sdss3.org/dr10/}}) and works
non--interactively, in order to provide the simplest possible
interface to the user.  However, \qsfit{} can be customized for
specific needs, and can be extended to analyze spectra from other data
sources or other detectors, e.g. to perform a spectral analysis using
both optical and UV data.  The ultimate purpose of \qsfit{} is to
allow astronomers to run standardized recipes to analyze the AGN data
in a simple, replicable and shareable way.

In this work we also present a catalog of spectral quantities obtained
by running \qsfit{} on a sample of \nn{} Type 1 AGN at $z<2$ observed
by SDSS--DR10.  All data in the catalog, as well as the plots with
best fitting model and residuals, the IDL code required to replicate
all our data analysis, are available
online.\footnote{\url{http://qsfit.inaf.it/}} Our catalog is similar
to the one from \citet[][hereafter S11]{2011-shen-catdr7}, although
the data analysis follows a different approach (\S\ref{sec:cmpS11}).
Moreover, we perform a comparison of our results with those from S11
in order to assess the reproducibility of the results in both
catalogs, and discuss the differences.  Finally, we check whether the
AGN continuum slopes with \qsfit{} show any evolution with the
redshift.

This paper is organized as follows: in \S\ref{sec:qsfit} we present
the \qsfit{} software in detail; in \S\ref{sec:catalog} we discuss the
Type 1 AGN sample and the data analysis employed to build the
\qsfitcat{} catalog; in \S\ref{sec:cmpS11} we compare our results with
those of S11; in \S\ref{sec:discussion} we discuss our results and in
\S\ref{sec:conclusion} we draw our conclusions.  Finally, in
\S\ref{sec:qsfitManual} we describe the procedures for \qsfit{}
installation, usage and customization.

In this work we adopt a standard $\Lambda$CDM cosmology with $H_0=70$
km s$^{-1}$ Mpc$^{-1}$, $\Omega_{\rm M}=0.3$ and
$\Omega_{\Lambda}=0.7$ (exactly the same as S11).

\section{QSFit: Quasar Spectral fitting procedure}
\label{sec:qsfit}

\qsfit{} (version \version{}) is an IDL\footnote{The \qsfit{} project
  started a few years ago, when the IDL language seemed the most
  viable option because of its widespread usage in the astronomy
  community.  Today, Python appears more appropriate since it would
  allow \qsfit{} to be a complete open source solution.  Hence, we
  started the Python porting, which will however maintain most of
  \qsfit{} current architecture in order to facilitate the transition
  to Python even to those who will start using \qsfit{} in its IDL
  version.}  package able to perform AGN spectral analysis at
optical/UV wavelengths.  It is ``free
software'',\footnote{\url{https://www.gnu.org/philosophy/free-sw.html}}
released under the GPL license, and available for download on
Github.\footnote{\url{https://github.com/gcalderone/qsfit}}

\qsfit{} operates on the optical spectrum of a single source and
provides estimates of: AGN continuum luminosities and slopes at
several rest frame wavelengths; Balmer continuum luminosity;
luminosities, widths and velocity offsets of 20 emission lines (\ha{},
\hb{}m \mgii{}, \oiii{}, \civ{}, etc.); luminosities of iron blended
lines at optical and UV wavelengths; host galaxy luminosities at
5500\AA{} (for sources with $z \le 0.8$).  Beyond the spectral
quantities, the software also provides several ``quality flags'' to
assess the reliability of the results.

The purpose is similar to that of previous works on AGN spectral
analysis, but we followed a different approach:\footnote{ see also
  \citet{2008-Vestergaard-BHMassFunction,
    2012-Shen-ComparingSEVEstimators}.} instead of focusing on a
single emission line and estimate the continuum in the surrounding
region, we fit all the components simultaneously using an AGN
continuum component which extends over the entire available spectrum.
and is thus a probe of the actual AGN continuum whose estimates are
scarcely influenced by localized features in the spectrum.\footnote{In
  the few cases where this approach is not appropriate (because of
  peculiar continuum shapes, strong absorptions or atypical iron
  emission patterns), the user can customize the source code to find a
  sensible fit.  In our catalog (\S\ref{sec:catalog}) we excluded the
  sources with $z > 2$ (since their continuum cannot be fit with a
  single power law) and the BAL sources (to avoid strong absorptions).
  The goodness of the fit for the remaining cases can be inspected by
  checking the $\chi^2$ of the fit and the plots.}

The model used to fit the data is a collection of several
``components'', each representing a specific contribution to the
observed optical/UV spectrum.  The list of components currently
employed is:
\begin{enumerate}
\item AGN continuum (\S\ref{sec:comp-continuum}): we use a single
  power--law to describe the AGN continuum over the entire
  (rest--frame) wavelength coverage
  \citep{1990-Cristiani-CompositeQuasar};

\item Balmer continuum (\S\ref{sec:balmer}): we follow the recipe by
  \citet{1982-Grandi-3000Bump} and \citet{2002-Dietrich} to model the
  Balmer continuum.  We also account for the high order Balmer lines
  ($7 \le n \le 50$, i.e. H$\epsilon$ and higher) which are not fitted
  individually, using the line ratios given in
  \citet{1995-Storey-BalmerLines};

\item host galaxy (\S\ref{sec:comp-galaxy}): we use the template of an
  elliptical galaxy component to account for the host galaxy
  contribution (this component is relevant only for sources with
  $z \le 0.8$);

\item iron blended emission lines (\S\ref{sec:comp-iron}): we used the
  iron template from \citet{2004-veron-spectra-izw1} and
  \citet{2001-vestergaard-UV-iron} at optical and UV wavelengths
  respectively;

\item emission lines (\S\ref{sec:comp-lines}): we used a Gaussian
  profile to represent all the emission lines (both ``narrow'' and
  ``broad'') in the spectrum.  Beyond the ``known'' emission lines
  expected to be relevant in the considered wavelength range (see
  Tab.~\ref{tab:knownline}), we also considered a list of ``unknown''
  emission lines, i.e. priori not associated to any known line
  (\S\ref{sec:comp-linesunk}).  By considering these additional
  components we may account for the lack of an iron template in the
  wavelength range 3100--3500\AA, or for asymmetric profiles in known
  emission lines.
\end{enumerate}
The actual model is the sum of all the above mentioned components.  No
absorption line is considered, since its parameters would be highly
degenerate with either the host galaxy or the emission lines.

\qsfit{} relies on the {\sc
  MPFIT}\footnote{\url{http://www.physics.wisc.edu/~craigm/idl/fitting.html}}
\citep{2009-Markwardt-MPFIT} procedure as minimization routine: during
the fitting process the component's parameters are varied until the
differences between the data and the model are minimized, following a
Levenberg--Marquardt least--squares minimization algorithm.

Currently, \qsfit{} operates only on SDSS--DR10 spectra of sources
whose redshift is $z<2$ (to avoid the absorptions troughs in the
neighborhood of the hydrogen Lyman--$\alpha$ line).  The whole
analysis process is tuned to operate on the SDSS spectra and to
generate the \qsfitcat{} catalog (\S\ref{sec:catalog}), but it can be
modified to analyze spectra from other sources or to customize the
fitting recipe (\S\ref{sec:qsfitManual}).  The details of the fit
procedure are discussed in \S\ref{sec:fit-proc}.

\subsection{Units}
The units and symbols used throughout the paper are as follows:
\begin{itemize}
\item $\lambda$ [\AA]: wavelength;
\item $\nu$ [Hz]: frequency;
\item $L_{\lambda}$ [$10^{42}$~erg~s$^{-1}$ {\AA}$^{-1}$]: luminosity
  density;
\item $L$ [$10^{42}$~erg~s$^{-1}$]: emission line integrated
  luminosity, continuum $\nu L_{\nu}$ or $\lambda L_{\lambda}$
  luminosities, etc.;
\item $\alpha_{\lambda}$ AGN spectral slope, defined as $L_{\lambda}
  \propto \lambda^{\alpha_{\lambda}}$;
\item \lmin{} and \lmax{} [\AA]: minimum and maximum rest frame
  wavelength for a specific spectrum;
\item FWHM and V$_{\rm off}$ [km s$^{-1}$]: full width at half maximum
  and velocity offset of emission lines.  Both values are calculated
  as the width/displacement of the emission line, normalized by the
  reference wavelength, and multiplied by the speed of light.
  Positive values of V$_{\rm off}$ means the line is blue--shifted.
\end{itemize}
All quantities are given in the source rest frame.

\subsection{AGN continuum}
\label{sec:comp-continuum}
The AGN broad--band continuum is modeled as a power law in the form:
\begin{equation}
  \label{eq-pl}
  L_{\lambda} = A
  \cpar{\frac{\lambda}{\lambda_{\rm b}}}^{\alpha_{\lambda}}
\end{equation}
where $\lambda_{\rm b}$ is a reference break wavelength, $A$ is the
luminosity density at $\lambda=\lambda_{\rm b}$ and $\alpha_{\lambda}$
is the spectral slope at $\lambda \ll \lambda_{\rm b}$.  In the
current \qsfit{} implementation the $A$ parameter is constrained to be
positive; $\lambda_{\rm b}$ is fixed to be in the middle of the
available spectral wavelength range; $\alpha_{\lambda}$ is constrained
in the range [$-3$, 1].

The AGN continuum component extends through the whole observed
wavelength, hence the luminosity and slopes estimates are scarcely
influenced by localized features in the spectrum, and are assumed to
be reliable probes of the actual shape of the ``real'' AGN continuum.

If the spectral coverage is sufficiently large the single power law
model may not be suitable to fit the data.  In these cases the
continuum component can be customized to behave as a smoothly broken
power law model (see \S\ref{sec:sbpl}).  For the \qsfitcat{} catalog
we chose to use a simple power law since the SDSS spectral coverage
provides only weak constraints to a second power law component.  Note
also that when the SDSS data are used and the host galaxy is
significantly more luminous than the AGN continuum the slope parameter
becomes highly degenerate with the galaxy normalization, and the two
parameters can not be reliably constrained.  This occur typically for
sources with $z \le 0.6$ (Fig.~\ref{fig:galaxyContRatio}), hence for
these sources we fix the continuum slope to $\alpha_{\lambda} = -1.7$,
i.e. the average value for the sources with $z\sim 0.7$ (see
Fig.~\ref{fig:slopeavg_z} and discussion in \S\ref{sec:mainResults}).

\subsection{Balmer continuum and pseudo--continuum}
\label{sec:balmer}

The Balmer continuum \citep{1985-Wills-SBB} is modeled according to
\citet{1982-Grandi-3000Bump, 2002-Dietrich}:
\begin{equation}
  L_{\lambda} = A \times B_\lambda(T_{\rm e}) 
  \bpar{1 - \exp{\spar{\tau_{\rm BE} \cpar{\frac{\lambda}{\lambda_{\rm BE}}}^3}}}
\end{equation}
where $A$ is the luminosity density at 3000\AA{}, in units of the
continuum (\S\ref{sec:comp-continuum}) luminosity density at the same
wavelength, $B_\lambda(T_{\rm e}) $ is the black body function at the
electron temperature $T_{\rm e}$, $\tau_{\rm BE}$ is the optical depth
at the Balmer edge and $\lambda_{\rm BE}$ is the edge wavelength
(3645\AA{}).  To avoid degeneracies among parameters we fixed the
electron temperature at $T_{\rm e} = 15,000$ K and $\tau_{\rm BE} = 1$
\citep{2002-Dietrich}.  Also, we defined the Balmer continuum
normalization in terms of the continuum luminosity at 3000\AA{} since
we expect the former to be a fraction of the latter, and this
parametrization allows to always start with a reasonable guess
parameter ($A = 0.1$).

The \qsfit{} Balmer component also accounts for high order Balmer
lines ($7 \le n \le 50$, i.e. H$\epsilon$ and higher) which are not
fitted individually (\S\ref{sec:comp-lines}) and blends into a Balmer
``pseudo--continuum'' at wavelengths longer than 3645\AA{}.  The line
ratios are taken from \citet{1995-Storey-BalmerLines}, for a fixed
electron density of 10$^9$ cm$^{-3}$ and a temperature of 15,000 K.
The luminosity ratio $R$ of the high order blended line complex to the
Balmer continuum (at the Balmer edge) is a free parameter in the fit.

\begin{figure}
  \includegraphics[width=9cm]{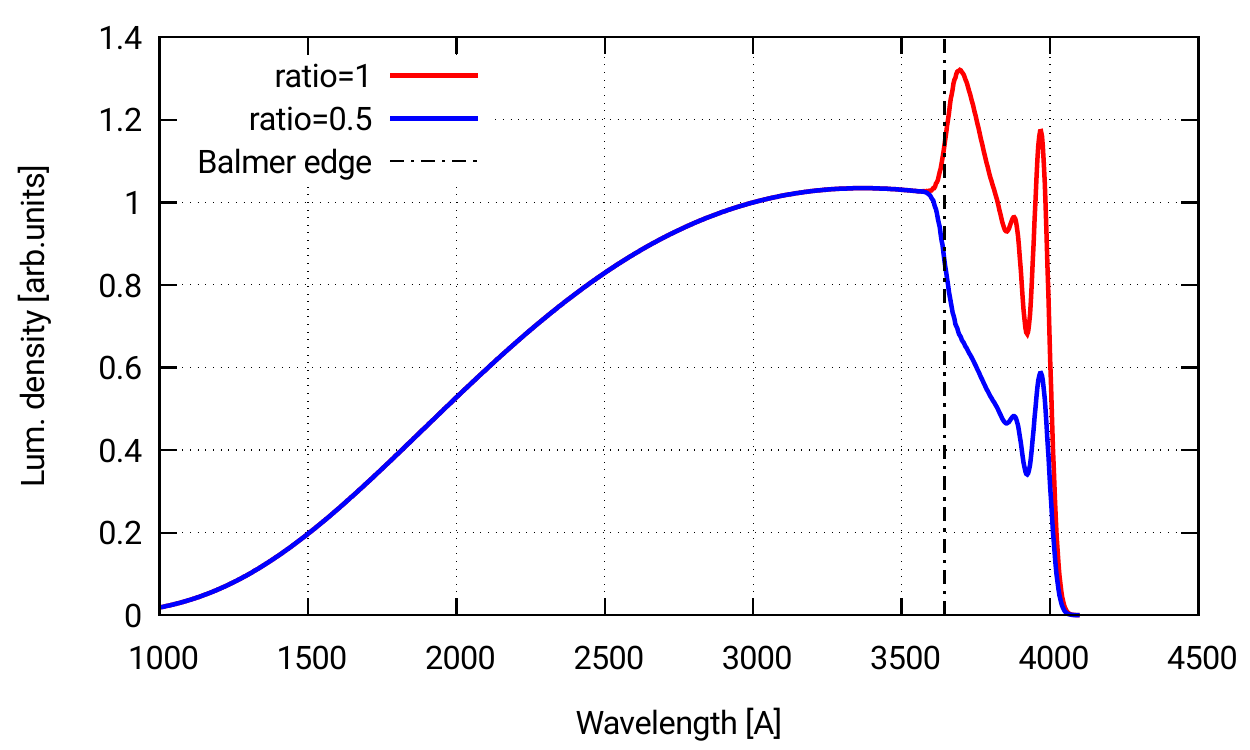}
  \caption{Plot of the Balmer continuum (at wavelengths shorter than
    the edge, 3645\AA{}) and pseudo--continuum template (at
    wavelengths longer than the edge).  The latter is the
    super--position of high order Balmer lines ($7 \le n \le 50$,
    i.e. H$\epsilon$ and higher).  The ratio $R$ of the
    pseudo--continuum to continuum luminosities at the Balmer edge is
    a free parameter in the fit.  In the above plot two ratios are
    shown: 1 (red) and 0.5 (blue).  The two templates overlaps at
    wavelengths shorter than the edge, hence only the blue is
    visible.}
  \label{fig:balmer}
\end{figure}

The whole component (sum of Balmer continuum and high order lines) is
finally broadened by a Gaussian profile of 5000 km s$^{-1}$.  A plot
of the Balmer template, for two values of the ratio of the high order
blended line complex to the Balmer continuum (1 and 0.5 respectively)
is shown in Fig.~\ref{fig:balmer}.  The electron temperature and
density, optical depth at Balmer edge and width of broadening profiles
are fixed by default, but can be left free to vary for specific
purposes.  Also, for sources with $z \ge 1.1$ we observe the spectrum
at wavelengths shorter than $\sim$4000\AA{}, i.e. we miss the most
characterizing part of the Balmer continum.  At $\lambda
\le$~3500\AA{} the Balmer continuum resembles a simple power law, and
it would become degenerate with the continuum component.  Therefore we
fixed $A = 0.1$ and $R$ = 0.3 for sources with $z \ge 1.1$ (see
Fig.~\ref{fig:BalmerContRatio}).

\subsection{Host galaxy}
\label{sec:comp-galaxy}
\begin{figure}
  \includegraphics[width=9cm]{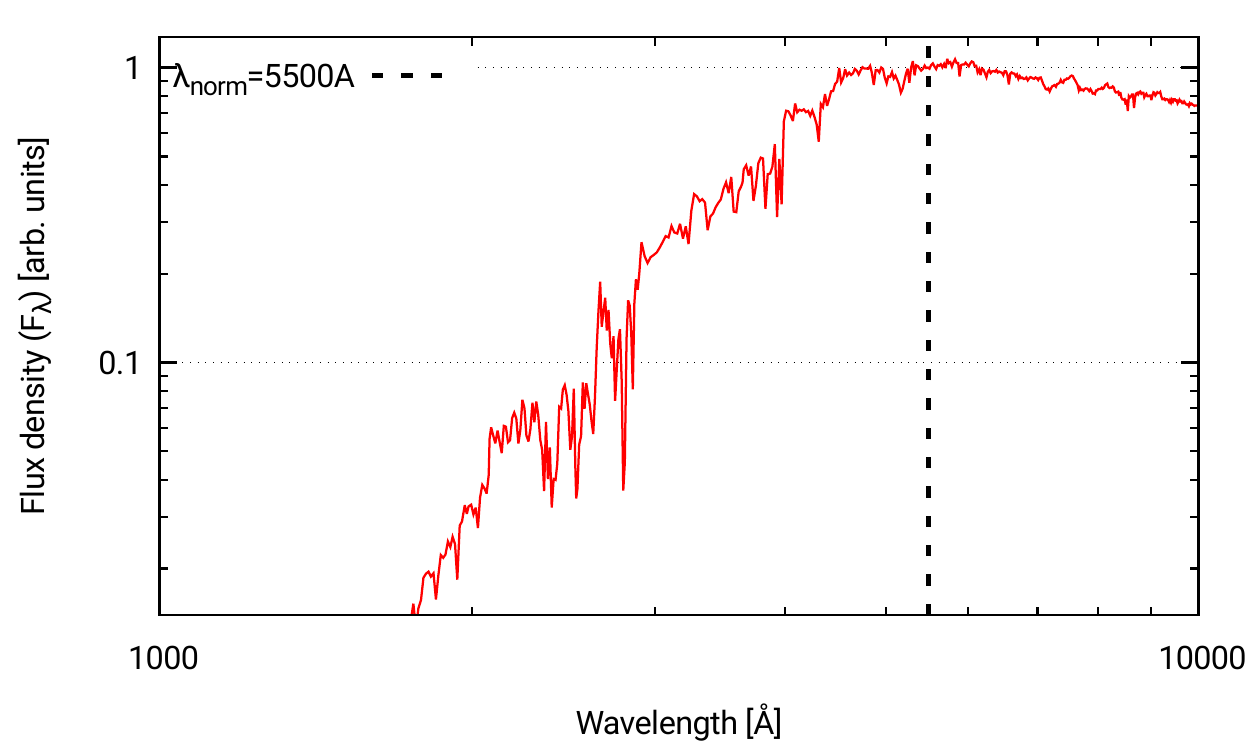}
  \caption{The normalized elliptical galaxy template from
    \citet{2007-Polletta-SWIRETemplates} used to model the host galaxy
    emission.}
  \label{fig:galtempl}
\end{figure}

The host galaxy may contribute to the observed luminosity of an AGN,
especially at optical/IR wavelengths.  It is therefore necessary to
disentangle its contribution from the AGN continuum to provide
reliable estimates of the latter
\citep[e.g.][]{2008-Vestergaard-BHMassFunction}.  The galaxy spectra
typically show a change of slope at 4000\AA{}, corresponding to the
Wien's tail of black body spectra from stars, as shown in
Fig.~\ref{fig:galtempl}.  This change of slope can be detected in SDSS
spectra of low--luminosity, low redshift ($z \le 0.8$) AGN
(e.g. Fig.~3 in \citealt{2012-calderone-torus}) and can be used in
automatic fitting procedures to estimate the host galaxy contribution.

In the current \qsfit{} implementation we used a simulated 5 Gyr old
elliptical galaxy template
\citep{1998-Silva-GRASIL,2007-Polletta-SWIRETemplates} with a
normalization factor as the only parameter.
The template is normalized to have luminosity density of 1 at
5500\AA{}, hence the value of the normalization parameter is the
luminosity density of the host galaxy at 5500\AA.  We chose the
elliptical template since luminous AGN (as those considered in the
\qsfitcat{} catalog, \S\ref{sec:catalog}) are likely hosted in
elliptical galaxies \citep[e.g.][]{2004-Floyd-HostOfLumQSO}.  For
sources with $z>0.8$ we disabled the host galaxy component since the
SDSS rest frame spectral range covers wavelengths shorter than
5000\AA{}, providing weak constraints on the host galaxy component.

Our approach in fitting the host galaxy contribution with a single
template is clearly simplistic, but it allowed us to homogeneously
analyze a very large sample.  More sophisticated AGN/host galaxy
decomposition methods have been proposed by other authors
\citep[e.g.][]{2015-Barth-LickMonitoring, 2015-Matsuoka-SDSS-RM}, and
could in principle be used also with \qsfit{} by customizing it for
specific needs (\S\ref{sec:qsfitManual}).

\subsection{Blended iron lines}
\label{sec:comp-iron}
Modeling of both broad and narrow (permitted and forbidden) iron lines
is essential for a proper estimate of broad--band continuum and
contaminated lines (such as \mgii{} and \hb{}), especially if the
source shows small widths of broad lines
\citep{1992-boroson-emlineprop-irontempl}.  In the current \qsfit{}
implementation we used the iron template of
\citet{2001-vestergaard-UV-iron} at UV wavelengths, and the iron
template of \citet{2004-veron-spectra-izw1} at optical wavelengths.

\subsubsection{Iron emission lines at optical wavelengths}
\label{sec:comp-ironopt}
At optical wavelengths we used the spectral decomposition of I Zw 1
made by \citet{2004-veron-spectra-izw1}, spanning the wavelength range
from 3500 to 7200\AA{} (rest frame).

\begin{figure}
  \includegraphics[width=9cm]{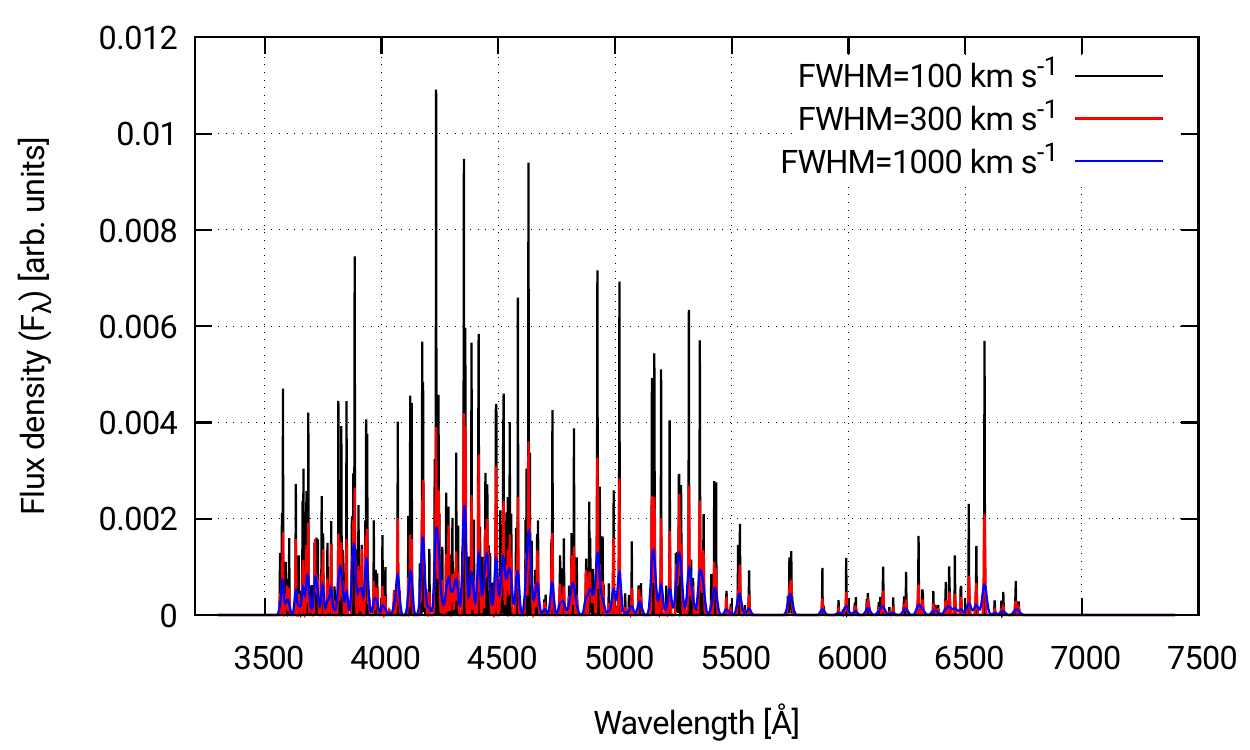}\\
  \includegraphics[width=9cm]{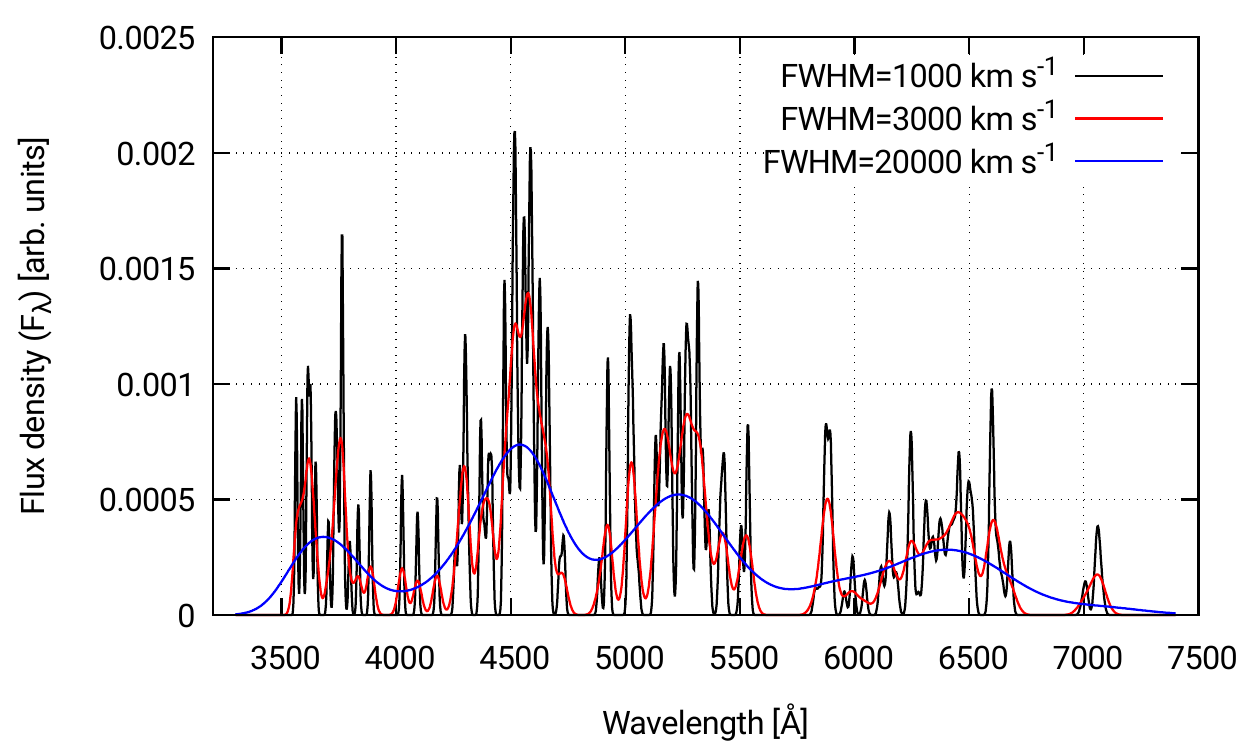}
  \caption{Three examples of narrow (upper panel) and broad (lower
    panel) optical iron templates used in the fit. These templates
    were obtained by summing a Gaussian line profile for each
    entry in tables A.1 and A.2 of \citet{2004-veron-spectra-izw1}
    (excluding the hydrogen Balmer lines).  The FWHM of the template
    is the FWHM of the Gaussian profile used to generate the
    template. All these templates are normalized to 1.}
  \label{fig:optiron}
\end{figure}

To compare the observed data with the optical iron template we
prepared a grid of templates, each with its own value of FWHM.  To
generate each template we sum a Gaussian profile (of the given FWHM)
for each line listed in tables A.1 and A.2 of
\citet{2004-veron-spectra-izw1} (excluding the hydrogen Balmer lines).
The relative line intensities are fixed to those reported in the
tables,\footnote{We considered only the William Herschel Telescope
  (WHT) intensities} and the final template is normalized to 1.
Hence, the model parameters are the width of the lines (which
identifies a template in our grid) and its overall normalization.  The
``broad'' and ``narrow'' templates (tables A.1 and A.2 respectively)
are kept separated, each with its own parameters.  The FWHM in our
grid vary from 10$^3$ to 10$^4$ km s$^{-1}$ for the broad templates,
and from 10$^2$ to 10$^3$ km s$^{-1}$ for the narrow templates.  A
plot of the iron template, for three values of FWHM (10$^3$, 3$\times
10^3$ and 10$^4$ km s$^{-1}$) is shown in Fig.~\ref{fig:optiron}.
The iron emission at optical wavelengths is typically weak, and the
SDSS spectral resolution is not sufficient to provide robust
constraints to the FWHM values, hence we keep it fixed at 3000 km
s$^{-1}$ for the broad component and to 500 km s$^{-1}$ for the narrow
component.  These constraints can be relaxed when using data with
higher spectral resolution.

\subsubsection{Iron emission lines at UV wavelengths}
\label{sec:comp-ironuv}
\begin{figure}
  \includegraphics[width=.48\ww]{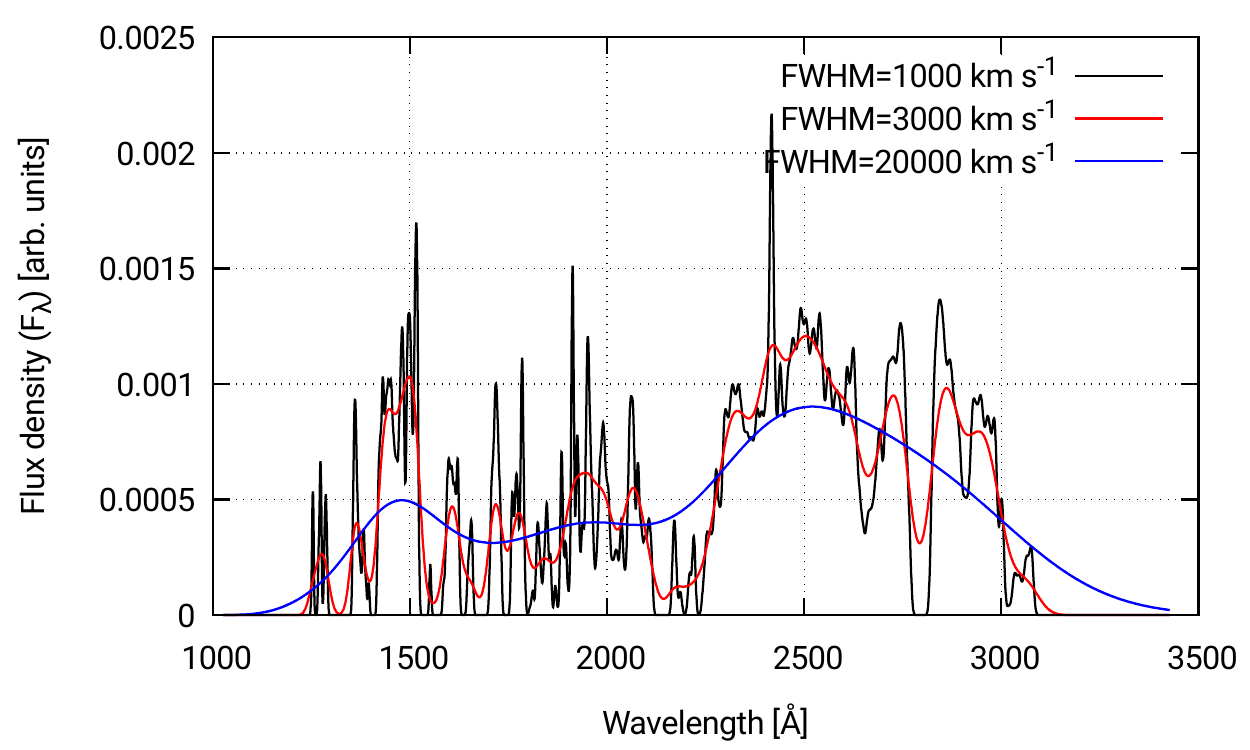}
  \caption{Three examples of the broadened and normalized UV iron
    templates.  These templates were obtained by convolving the
    original iron template in \citet{2001-vestergaard-UV-iron} with a
    Gaussian of FWHM equal to $10^3$, $3\times 10^3$ and $2 \times
    10^4$ km s$^{-1}$ respectively.}
  \label{fig:uviron}
\end{figure}

We considered the ``B'' template and the FeII UV191 and FeIII UV47
multiplets of the iron template described in
\citet{2001-vestergaard-UV-iron}, spanning the wavelength range from
1250\AA\ to 3090\AA.  The original template is based on the
observations of I Zw 1, whose FWHM of broad lines is 900 km s$^{-1}$.
To adapt the template to sources with broader emission lines we
prepared a grid of broadened templates by convolving the original one
with a set of Gaussian profiles, whose FWHM range from 10$^3$ to
2$\times 10^4$ km s$^{-1}$.  In this case there is no separation
between ``broad'' and ``narrow'' contributions since the original
template in \citet{2001-vestergaard-UV-iron} is the sum of both.  A
plot of the template, for three values of FWHM (10$^3$, 3$\times 10^3$
and $2 \times 10^4$ km s$^{-1}$) is shown in Fig.~\ref{fig:uviron}.

The model parameters are the equivalent width (integrated over the
whole wavelength range 1250\AA\ to 3090\AA) and the FWHM of the
profile used to broaden the template.  The FWHM values are constrained
in the range 10$^3$--10$^4$ km s$^{-1}$.  The SDSS spectral resolution
is not sufficient to provide robust constraints to the FWHM value,
hence we keep it fixed at 3000 km s$^{-1}$.  This constraint can be
relaxed when using data with higher spectral resolution.  Also, for
sources with $z \le 0.4$ we fix the iron UV equivalent width to
138\AA{} (i.e. the median of the values measured at larger redshifts)
since only a small part of the whole template is visible, and the iron
normalization would become degenerate with the continuum slope.
Moreover fixing the UV iron helps in constraining the continuum slope
to have similar distributions at redshifts below and above 0.4.

\subsection{Main emission lines}
\label{sec:comp-lines}
The broad and narrow emission lines considered in \qsfit{} are shown in
Tab.~\ref{tab:knownline}.  Each line is modeled with a Gaussian
profile, whose fitting parameters are the total (integrated) line
luminosity, the FWHM and velocity offset (with respect to the
reference wavelength) of the line profile.
\begin{table}
  \begin{center}
    \caption{The list of all the considered emission lines.  The third
      column ({\it type}) tells whether the line is modeled with a
      broad (B), narrow (N) component. The fourth column ({\it Field
        name}) shows the prefix of the FITS column name for each
      considered emission line. The H$\alpha$ line is modeled with
      three components to account for the very broad feature below
      that line.}
    \label{tab:knownline}
    \begin{tabular}{lccl}
      \hline\hline
      {\bf Line}             &
      {\bf Wavelength [\AA]} &
      {\bf Type}             &
      {\bf Field name}       \\
      \hline
      \siiv{}   &  1399.8   & B   & \verb|br_siiv_1400|  \\
      \civ{}    &  1549.48  & B   & \verb|br_civ_1549|   \\
      \ciii{}   &  1908.734 & B   & \verb|br_ciii_1909|  \\
      \mgii{}   &  2799.117 & B   & \verb|br_mgii_2798|  \\
      \nevi{}   &  3426.85  & N   & \verb|na_nevi_3426|  \\
      \oii{}    &  3729.875 & N   & \verb|na_oii_3727|   \\
      \neiii{}  &  3869.81  & N   & \verb|na_neiii_3869| \\
      \hd{}     &  4102.89  & B   & \verb|br_hd|         \\
      \hg{}     &  4341.68  & B   & \verb|br_hg|         \\
      \hb{}     &  4862.68  & B   & \verb|br_hb|         \\
                &           & N   & \verb|na_hb|         \\
      \oiii{}   &  4960.295 & N   & \verb|na_oiii_4959|  \\
      \oiii{}   &  5008.240 & N   & \verb|na_oiii_5007|  \\
      \hei{}    &  5877.30  & B   & \verb|br_hei_5876|   \\
      \nii{}    &  6549.86  & N   & \verb|na_nii_6549|   \\
      \ha{}     &  6564.61  & B   & \verb|br_ha|         \\
                &           & N   & \verb|na_ha|         \\
                &           & (see text)   & \verb|line_ha_base|  \\
      \nii{}    &  6585.27  & N   & \verb|na_nii_6583|   \\
      \sii{}    &  6718.29  & N   & \verb|na_sii_6716|   \\
      \sii{}    &  6732.67  & N   & \verb|na_sii_6731|   \\
      \hline\hline
    \end{tabular}
  \end{center}
  \end{table}
The FWHM of narrow and broad lines is constrained in the range
[10$^2$, 2$\times 10^3$] km s$^{-1}$ and [9$\times 10^2$, 1.5$\times
  10^4$] km s$^{-1}$ respectively. The velocity offset of narrow and
broad lines is constrained in the range $\pm 10^3$ km s$^{-1}$ and
$\pm 3\times10^3$ km s$^{-1}$ respectively.  For the \ha{} and \hb{}
emission lines we used both a broad and narrow components, but we had
to limit the FWHM of the narrow component to [10$^2$, $10^3$] km
s$^{-1}$ to allow a good decomposition of the line profile.  Also, we
had to limit the \mgii{} broad component velocity offset to $\pm 10^3$
km s$^{-1}$ to avoid degeneracies with the UV iron template.  Finally,
we used three components to fit the H$\alpha$ to account for the very
broad feature below that line.  The FWHM of the H$\alpha$ ``base''
line is constrained in the range [10$^4$, $3\times 10^4$] km s$^{-1}$.

A line component is ignored if the resulting line profile would fall
outside (even partially) the available wavelength range.  Moreover, a
line component is ignored if the amount of dropped spectral channels
(whose quality mask is not 0, \S\ref{sec:prep}), in the wavelengths
relevant to the emission line, exceeds 40\%.  These precautions were
taken to ensure the emission line parameters can actually be
constrained by the available data.

\subsection{``Unknown'' emission lines}
\label{sec:comp-linesunk}
Beyond the ``known'' emission lines (\S\ref{sec:comp-lines}) expected
to be relevant in any AGN, we also considered a list of 10 ``unknown''
emission lines, i.e. not a priori associated to any known line.  These
emission lines are added after all the other components (see
\S\ref{sec:fit-proc}), and their initial wavelength is set at the
position where there is the maximum positive residual.  By considering
these further components we may account for the lack of an iron
template in the wavelength range 3100--3500\AA, or for asymmetric
profiles in known emission lines.

\subsection{Fitting procedure and data reduction}
\label{sec:fit-proc}

In this section we describe the spectral analysis procedure, which
runs through four distinct steps: (i) preparation of the spectrum,
(ii) model fitting, and (iii) data reduction.

\subsubsection{Spectrum preparation}
\label{sec:prep}
The current \qsfit{} implementation operates only on spectra observed
by the the Sloan Digital Sky Survey, data release 10
\citep[SDSS--DR10,][]{2000-york-sdss-summary, 2014-Ahn-SDSS-DR10}.

To prepare the spectrum we drop the 100 spectral channels at the
beginning and end of each spectrum to avoid artifacts from instrument
or pipeline.  Then we drop the spectral channels whose quality mask is
not equal to zero.  If the remaining ``good'' channels accounts for
less than 75\% of the original channels, corresponding to $\sim$~2700
channels, \qsfit{} issues an error and skips the analysis.  In the
vast majority of cases the final wavelength range in the observer
frame is $\sim$~4150--8740\AA, and the number of channels is
$\sim$~3450.  The spectral resolution is $\sim$~150 km s$^{-1}$.

The spectrum is corrected for Galactic extinction using the CCM
parametrization by \citet{1989-cardelli-extinction} and
\citet{1994-odonnell-updateCCM}, assuming a total selective extinction
A(V)~/~E(B-V)~=~3.1.  We neglected any intrinsic (i.e. rest frame)
reddening of the spectrum since most of sources show small or no
reddening at all \citep[e.g.][]{2010-Grupe-SEDOfXRaySelectedAGN}.  To
build the \qsfitcat{} catalog (\S\ref{sec:catalog}) we used the same
color excess and redshift estimates as S11.  Finally, the spectrum is
transformed to the rest frame using the redshift estimate.  The
fitting process is performed by comparing the data and the model, in
units of $10^{42}$~erg~s$^{-1}$ {\AA}$^{-1}$.

\subsubsection{Model fitting}
\label{sec:modelFitting}

The fitting process occurs by varying the component parameters until
the $\chi^2$, as obtained by comparing the data and the model, is
minimized.  The minimization procedure follows a Levenberg--Marquardt
algorithm, hence the result may depend significantly from the initial
(guess) parameter values.  In order to develop an ``automatic''
spectral analysis (without human intervention) we should find a proper
way to get rid of such dependency on initial parameters.  We found
that a promising approach is to build the model step by step, i.e. by
iteratively adding a component and re--running the minimization
procedure.

\begin{figure*}
  \includegraphics[width=.43\ww]{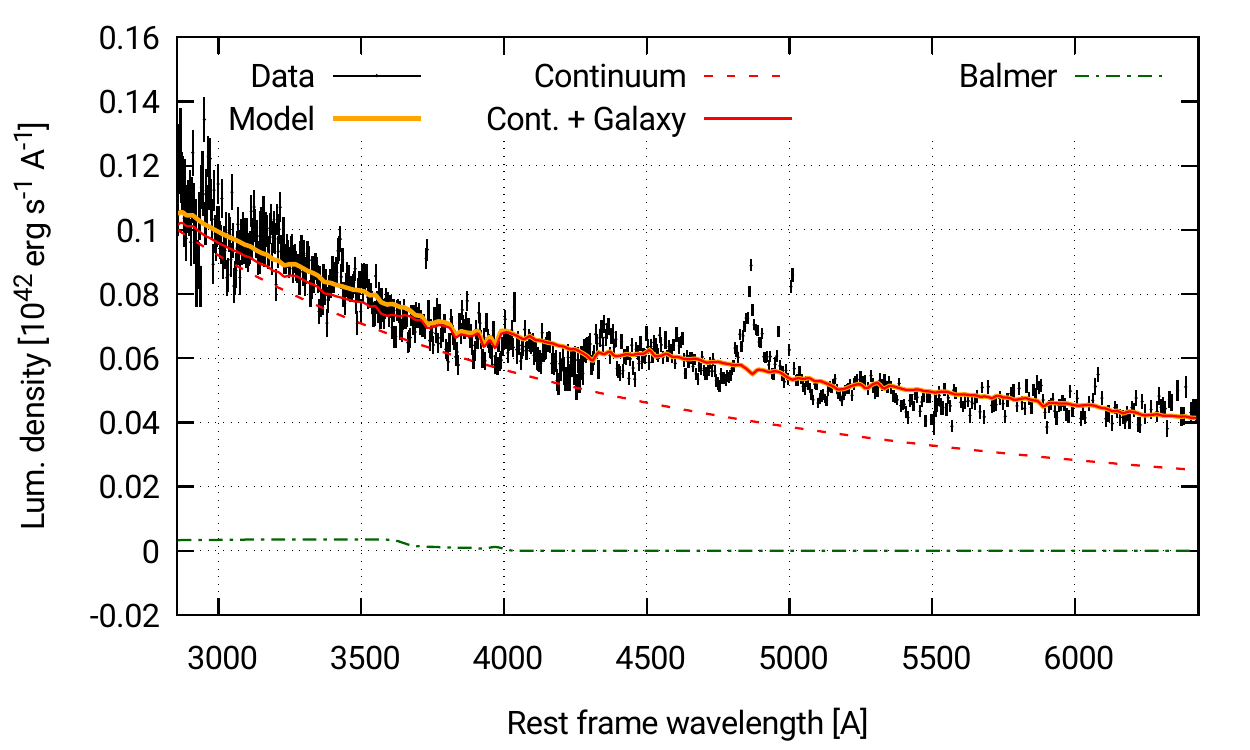}
  \hspace{-0.3cm}
  \includegraphics[width=.43\ww]{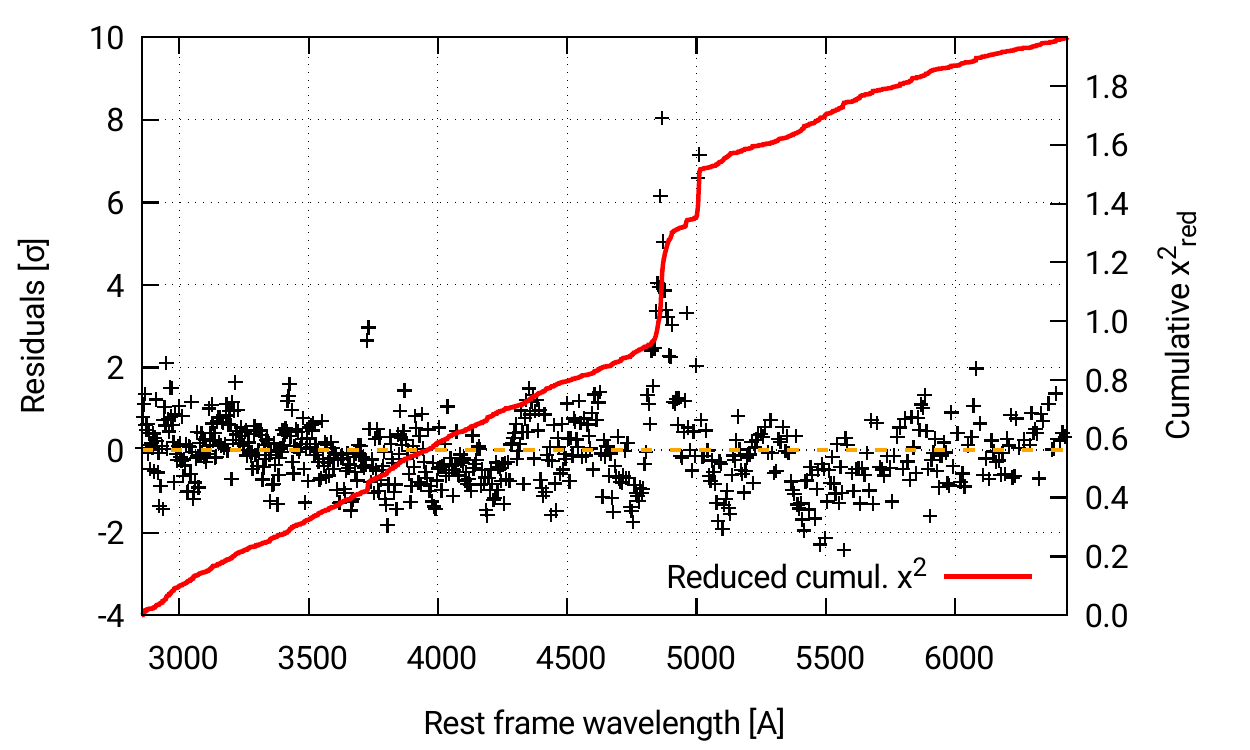}\\
  \includegraphics[width=.43\ww]{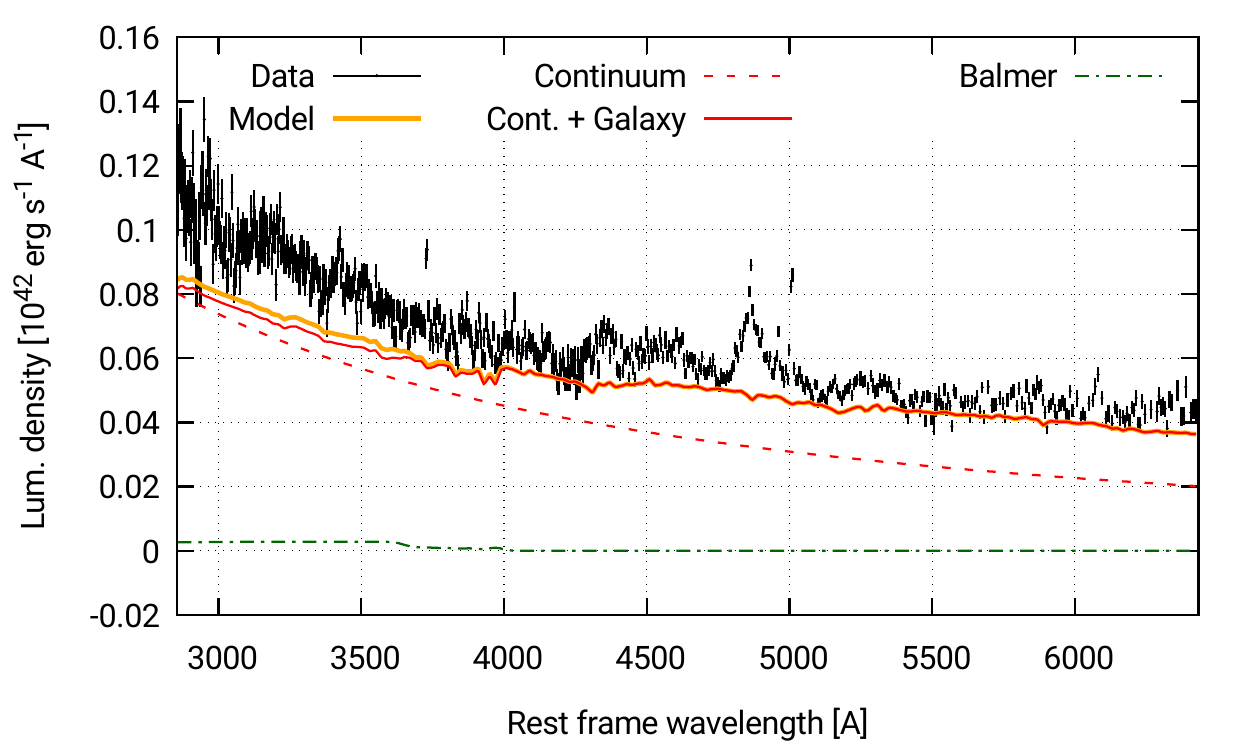}
  \hspace{-0.3cm}
  \includegraphics[width=.43\ww]{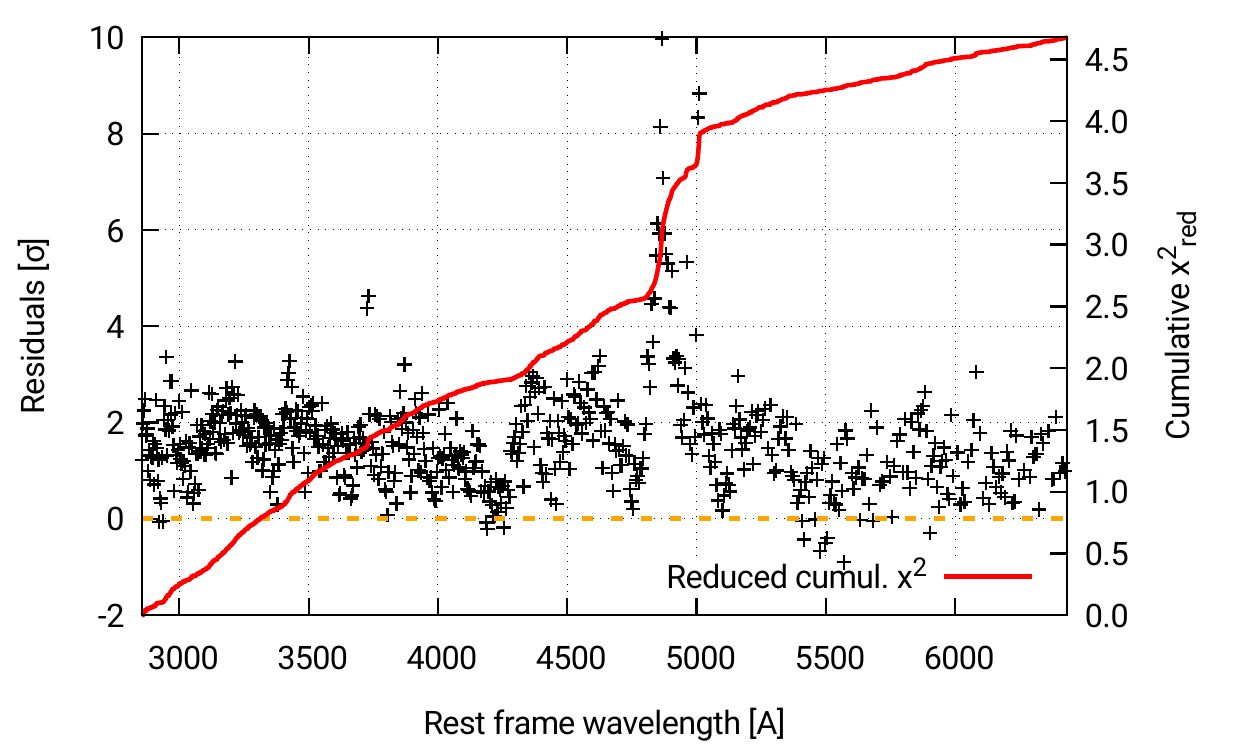}\\
  \includegraphics[width=.43\ww]{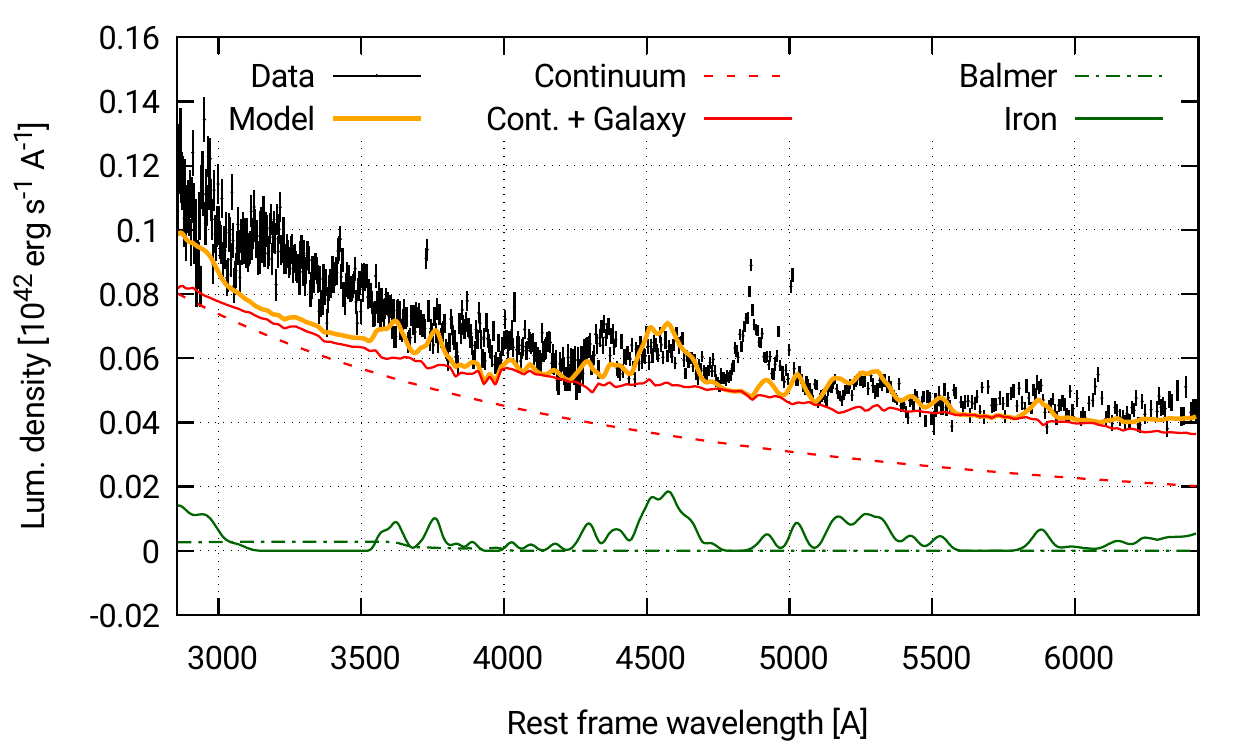}
  \hspace{-0.3cm}
  \includegraphics[width=.43\ww]{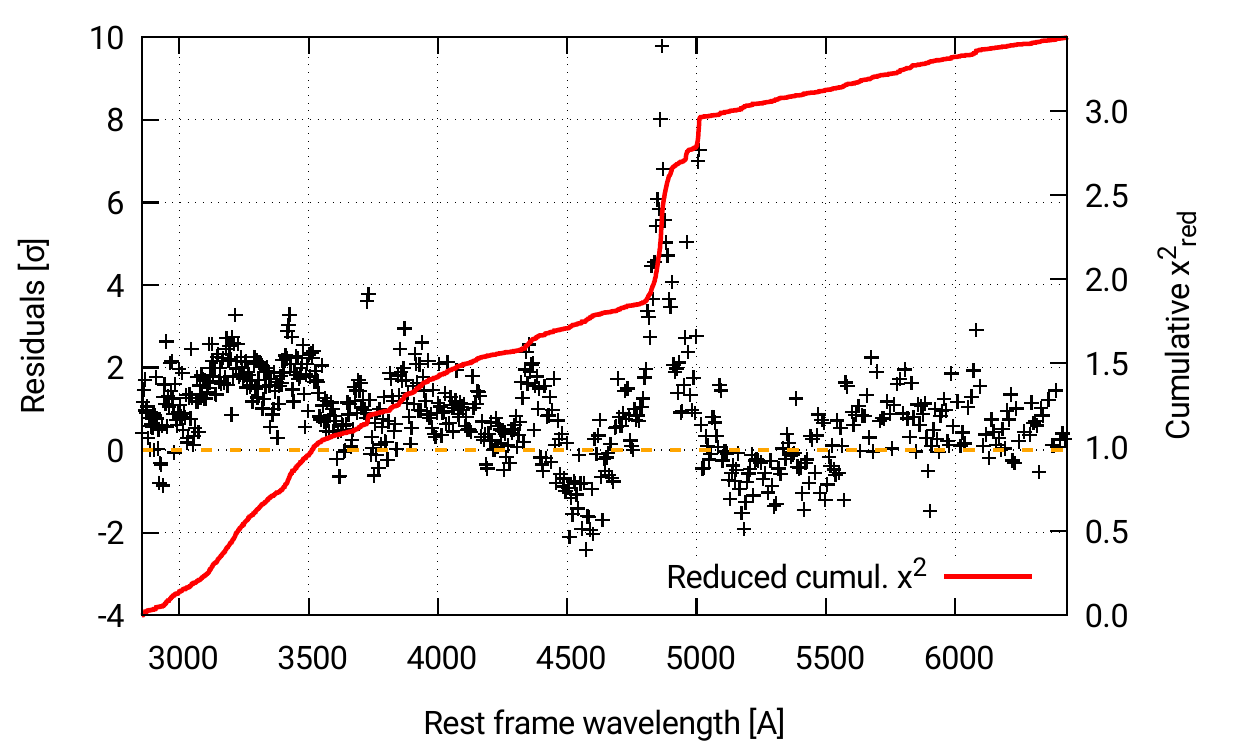}\\
  \includegraphics[width=.43\ww]{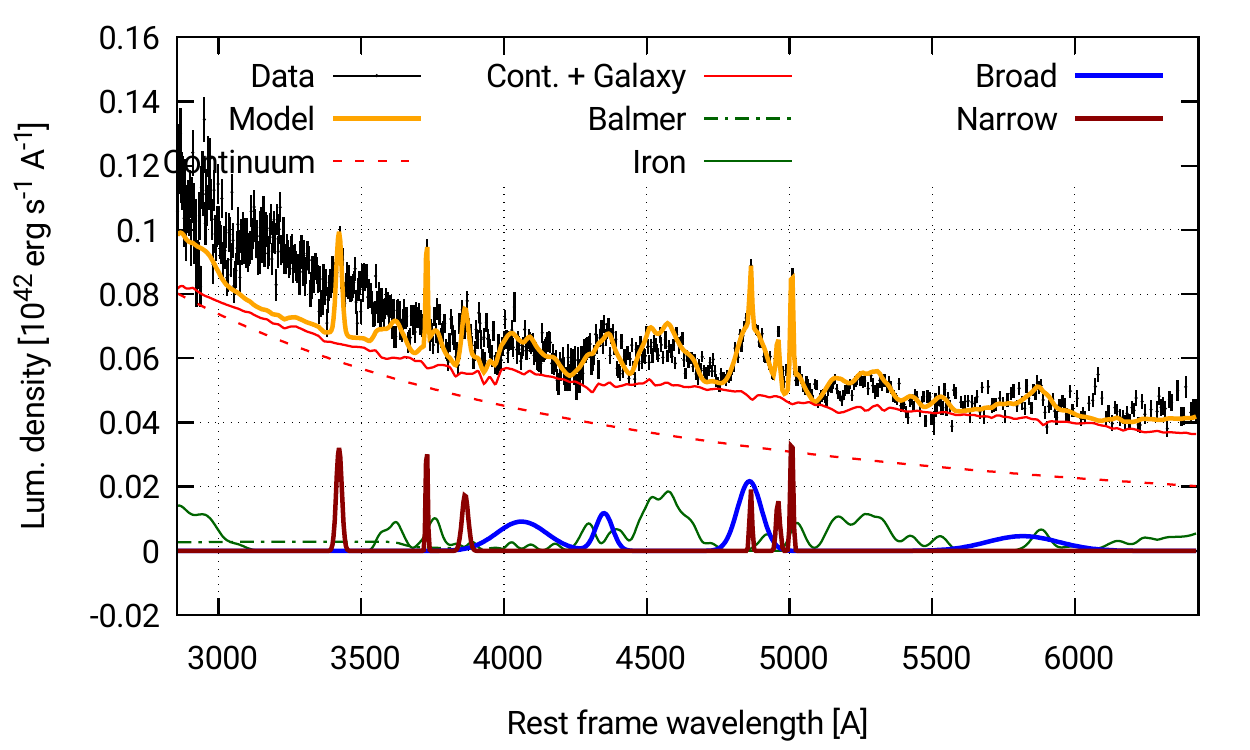}
  \hspace{-0.3cm}
  \includegraphics[width=.43\ww]{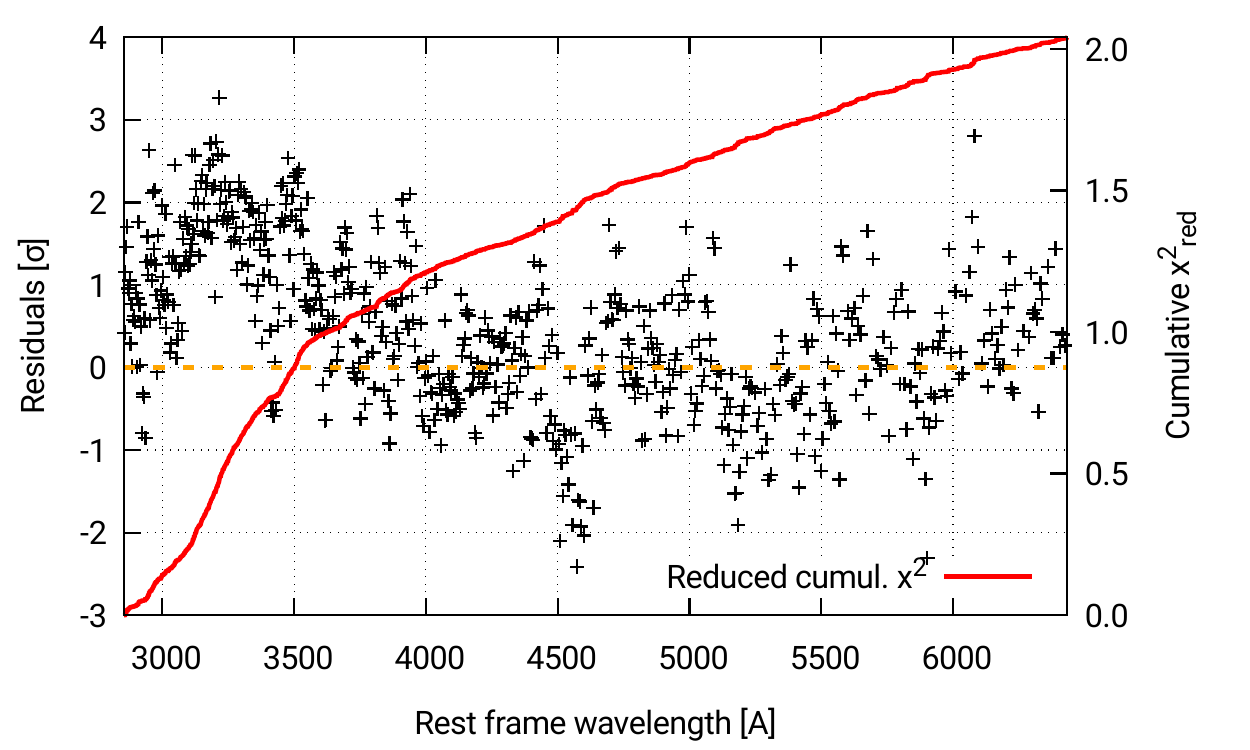}\\
  \includegraphics[width=.43\ww]{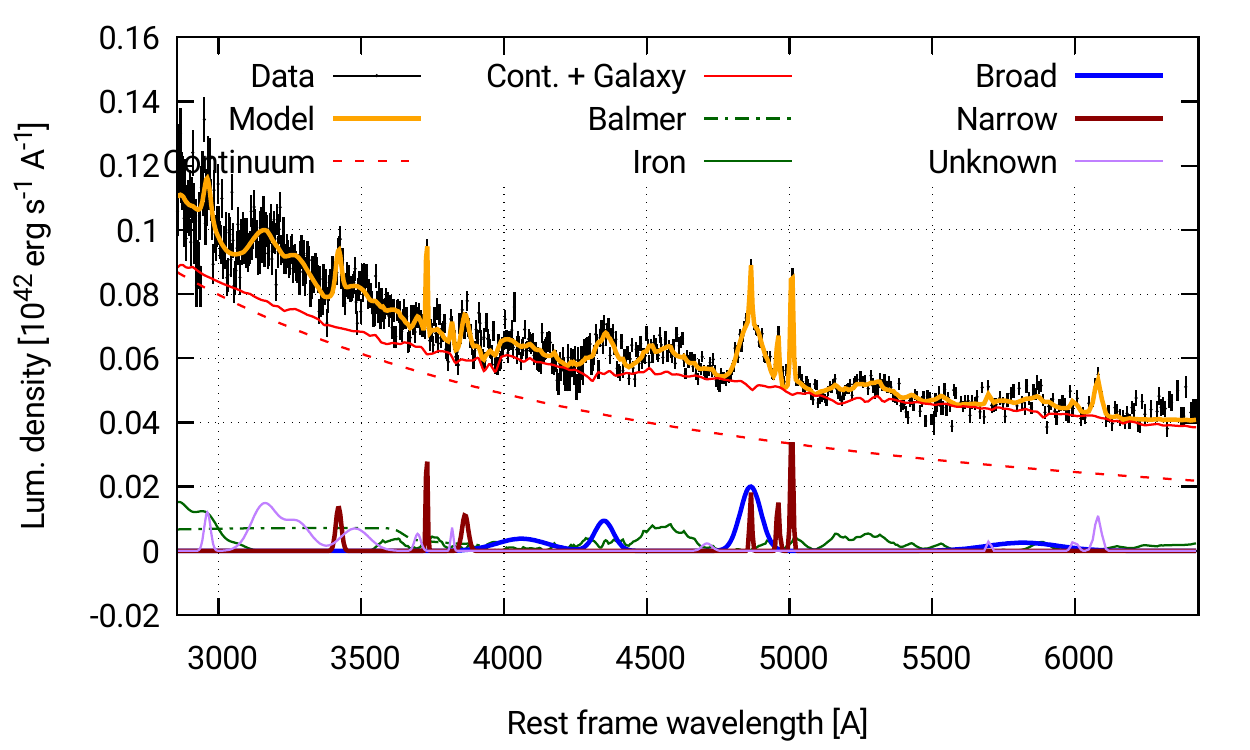}
  \hspace{-0.3cm}
  \includegraphics[width=.43\ww]{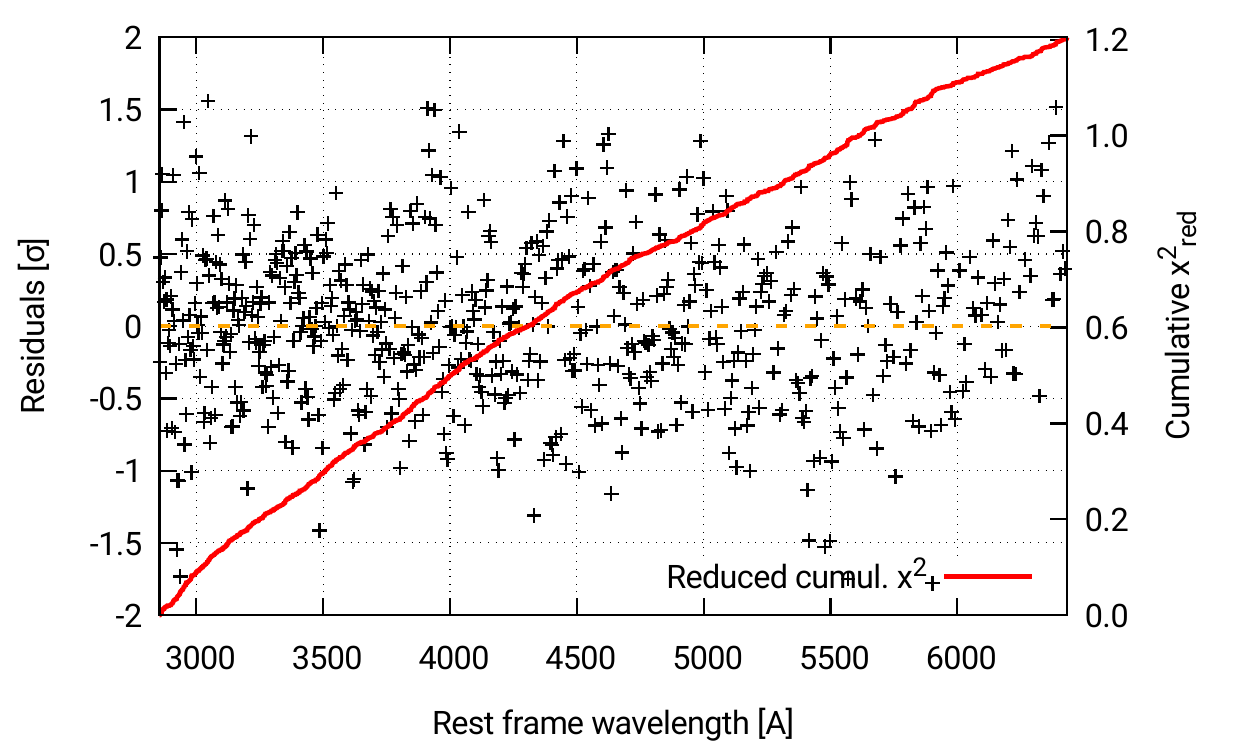}\\
  \caption{The plots in the left column show the comparison between
    the data and the overall model, as the individual components are
    being added.  The five steps of the fitting procedure
    (corresponding to the five rows in the plots above) are described
    in \S\ref{sec:modelFitting}.  In order to produce clearer plots,
    all data, as well as the model, were rebinned by a factor of
    5.  The final plot at full resolution is shown in
    Fig.~\ref{fig:ex1}.}
  \label{fig:stepbystep}
\end{figure*}

In the following we will discuss the five steps involved in the
fitting process. The plots of data, model and residuals at each step
of the fitting procedure are shown in Fig.~\ref{fig:stepbystep}. In
order to produce clearer plots all data, as well as the model, were
rebinned by a factor of 5.  The analysis, however, has been
carried out at full resolution. Additionally, the catalog browser
web tools allows the user to view the data using any rebinning factor.

\begin{enumerate}
\item The first components being added are the AGN continuum (actually
  a power law, \S\ref{sec:comp-continuum}) and the host galaxy
  template (\S\ref{sec:comp-galaxy}).  A first minimization is
  performed, which will bring the ``best fit'' model to pass through
  the data, without accounting for any emission line.  The amount of
  positive residuals, calculated as {\it (data $-$ model) /
    uncertainty}, is approximately equal to the amount of negative
  residual.

\item In order to provide room for further components (namely the
  emission lines) we lower the continuum normalization until the
  positive residuals reach $\sim 90$\%,\footnote{Although there appear
    to be some arbitrariness in performing this step, we found that
    the final results (after all the remaining steps have been
    performed) depends weakly on the fraction of positive residuals,
    as long as the latter is above $\sim 70$\%.}  and fix all
  parameters of both AGN continuum and host galaxy components for the
  next iterations.

\item We add the components for the iron templates at UV and optical
  wavelength (\S\ref{sec:comp-iron}).  The optical template is enabled
  for sources with $z \lesssim 1$, while the UV iron template is
  always enabled, although the equivalent width is fixed to 138\AA{}
  if $z < 0.4$ (\S\ref{sec:comp-ironuv}).  Then we re--run the fitting
  process by varying just the iron template parameters and we fix them
  at their best fit values for the next iterations.

\item Now we add the ``known'' emission lines, and ensure there are
  enough spectral channels to constrain the line parameters, otherwise
  the line is ignored (\S\ref{sec:prep}).  After re--running the
  minimization we fix all the emission line parameters for the next
  iterations.

\item we iteratively add the ``unknown'' emission lines
  (\S\ref{sec:comp-linesunk}), i.e. lines which are not a--priori
  associated to any specific transition.  The center wavelength of
  these lines is first set at the position of the maximum positive
  residual, i.e. where most likely there is some further emission
  component, and then left free to change during the fitting process.
  We repeat this procedure up to 10 times.  Finally we re--run the
  fitting process with all parameters, for all components, left free
  to vary.  At the end we check which ``unknown'' line component is
  actually required to provide a better fit: if the line luminosity
  uncertainty is greater than 3 time the luminosity we disable the
  component and re--run the fit.
\end{enumerate}

\subsubsection{Data reduction}
\label{sec:reduction}

After the fitting process we use the resulting best fit parameter
values to compile a list of spectral properties for the source being
analyzed.  In a few cases a further {\it reduction} is required to
provide the final quantities.  The details of the reduction process
are as follows:
\begin{itemize}
\item the AGN continuum and galaxy luminosities are estimated directly
  on the continuum component values, in at least six wavelengths, five
  of them equally spaced in the (logarithmic) wavelength range
  [\lmin{}+5\%, \lmax{}$-5$\%].  We restricted the range by 5\% on both
  sides to avoid artifacts at the edges of the available spectrum.
  The other estimates are given at the fixed rest frame wavelengths
  1450\AA, 2245\AA{}, 3000\AA{}, 4210\AA{} or 5100\AA{}, depending on
  the source redshift;

\item the ``unknown'' lines are associated to the ``known'' emission
  lines if the center wavelength of the latter falls within the FWHM
  of the former.  The association is performed only if the ``unknown''
  line has a FWHM of at least 1000 km s$^{-1}$ and the ``known'' line
  being considered is a broad line (narrow lines are always modeled
  with a single component).  If two or more emission lines
  components are associated then we compute the total line profile as the
  sum of individual line profiles, and estimate the relevant
  quantities as follows:
    \begin{itemize}
    \item the line luminosity is the sum of integrated luminosities
      of each component;
      \item the FWHM is the full--width of the total profile at half
        the maximum of the total profile peak;
      \item the velocity offset is calculated by considering the
        offset between the total profile peak position and the
        emission line reference wavelength;
    \end{itemize}
  The uncertainties on the luminosity is computed as the weighted sum
  of the luminosity uncertainty on each component.  The uncertainties
  on FWHM and velocity offset are computed as the weighted average of
  the corresponding uncertainties in the fit parameters, with the
  weight given by the line intensities.  These parameters are
  typically degenerate hence the uncertainties may be underestimated,
  see \S\ref{sec:MC-uncert} for a different approach to estimate
  parameter uncertainties;

\item beyond the spectral quantities, we also provide several
  ``quality flags'' to assess the reliability of the automatic
  results.  If a flag is raised for a given source then the
  corresponding quantity should be used cautiously, and a visual
  analysis of the spectrum is recommended.  To produce the plots in
  the following sections we considered only the sources with no
  quality flag raised (i.e. whose quality flag is 0).  The list of all
  quality flags and their meaning is discussed in
  \S\ref{sec:flattened-struc}, while the tables in
  \S\ref{sec:qualityflags} summarize the fractions of sources which
  raised a quality flag.
\end{itemize}
All quantities calculated by \qsfit{} are described in
\S\ref{sec:flattened-struc}.

The uncertainties provided by \qsfit{} are calculated using the Fisher
matrix method
\citep[e.g.][]{2009-Heavens-StatisticalTechniques,2010-Andrae-ErrorEstimation},
by assuming Gaussian uncertainties in the SDSS data, negligible
correlation among model parameters, and symmetric uncertainty
intervals.  These assumptions are not always justifiable, hence
\qsfit{} uncertainties should be considered as rough estimates of
parameter uncertainties.  A better approach, although significantly
more demanding in terms of computational time, is to use the Monte
Carlo resampling method as discussed in \S\ref{sec:MC-uncert}.  In the
following we will always consider the Fisher matrix uncertainties
provided by \qsfit{}.

\subsection{\qsfit{} plots}
\label{sec:ex_plot}

\begin{figure*}
  \includegraphics[width=.9\ww]{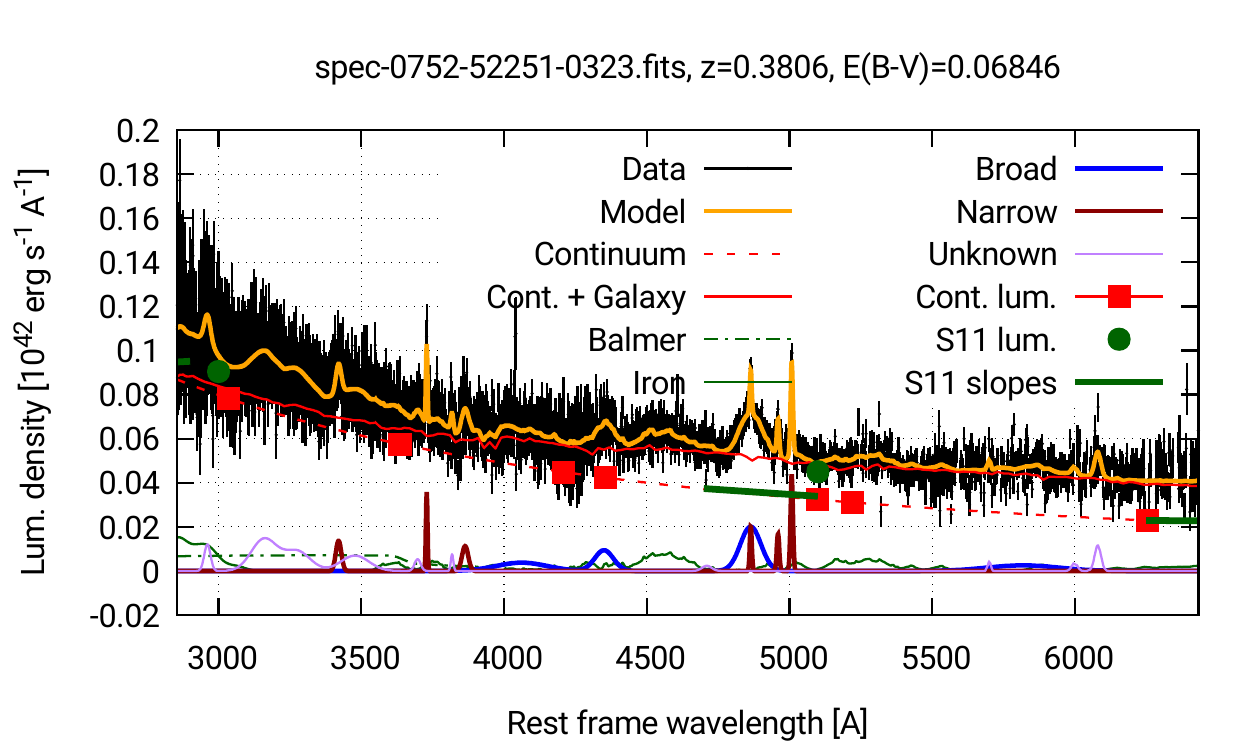}\\
  \includegraphics[width=.9\ww]{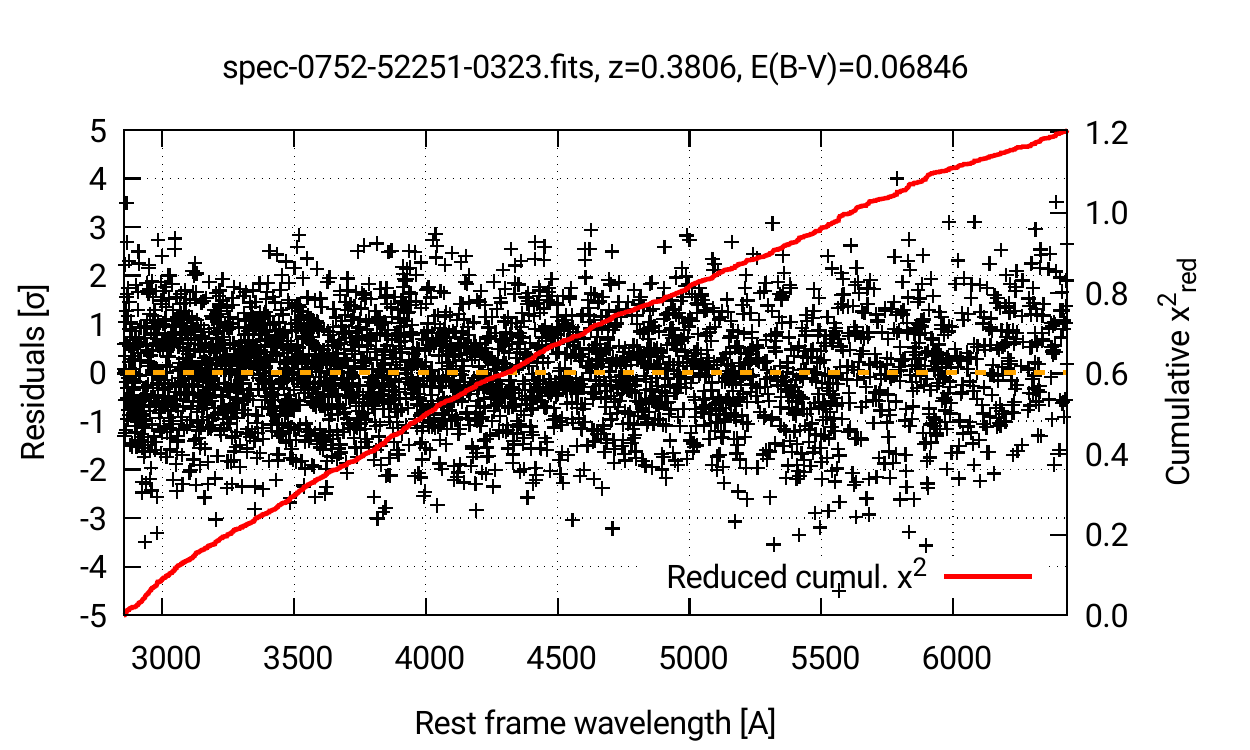}
  \caption{Comparison of the \qsfit{} model and SDSS data: the upper
    panel shows the comparison between the SDSS data with their error
    bars (black vertical lines) and the \qsfit{} model (orange line);
    the lower panel shows the residuals (data $-$ model) in units of
    $1 \sigma$ uncertainties in the data (black cross symbol) and the
    cumulative $\chi^2_{\rm red}$ (red line, values on the right
    axis).  The upper panel also shows the individual components in
    the \qsfit{} model: the dashed red line is the continuum component
    (\S\ref{sec:comp-continuum}), while the solid red line is the sum
    of continuum and host galaxy components (\S\ref{sec:comp-galaxy});
    the dot--dashed green line is the Balmer component
    (\S\ref{sec:balmer}); the solid green lines are the optical
    (\S\ref{sec:comp-ironopt}) and UV (\S\ref{sec:comp-ironuv}) iron
    templates; the sum of all broad and of all narrow emission line
    (\S\ref{sec:comp-lines}) components are shown with blue and brown
    lines respectively.  The sum of all ``unknown'' lines
    (\S\ref{sec:comp-linesunk}) is shown with a purple solid line.
    The red square and green circle symbols are the continuum
    luminosity estimated by \qsfit{} and S11 respectively.  The thick
    green line segments show the slopes estimated by S11 (anchored on
    the short wavelength side on \qsfit{} continuum component).}
  \label{fig:ex1}
\end{figure*}

After analyzing the data \qsfit{} produces two plots using {\sc
  gnuplot}.\footnote{\url{http://gnuplot.info/}} An example is shown
in Fig.~\ref{fig:ex1}: the upper panel shows the comparison between
the SDSS data with their error bars (black vertical lines) and the
\qsfit{} model (orange line); the lower panel shows the residuals
(data minus model) in unit of $1\sigma$ uncertainties in the data
(black cross symbol) and the cumulative $\chi^2_{\rm red}$ (red line,
values on the right axis).  The upper panel also shows the individual
component in the \qsfit{} model: the dashed red line is the continuum
component (\S\ref{sec:comp-continuum}), while the solid red line is
the sum of continuum and host galaxy components
(\S\ref{sec:comp-galaxy}); the dot--dashed green line is the Balmer
component (\S\ref{sec:balmer}); the solid green lines are the optical
(\S\ref{sec:comp-ironopt}) and UV (\S\ref{sec:comp-ironuv}) iron
templates; the sum of all broad and of all narrow emission line
(\S\ref{sec:comp-lines}) components are shown with blue and brown
lines respectively.  The sum of all ``unknown'' lines
(\S\ref{sec:comp-linesunk}) is shown with a purple solid line.  The
red squared and green circle symbols are the continuum luminosity
estimated by \qsfit{} and S11 respectively.  The thick green line
segments show the slopes estimated by S11 (anchored on the short
wavelength side on \qsfit{} continuum component).

In the example shown in Fig.~\ref{fig:ex1} the comparison between the
data and the \qsfit{} model yield a $\chi^2_{\rm red}$ value of 1.19,
with 3248 degrees of freedom (DOF).  Given the very high number of DOF
the probability of obtaining a worst $\chi^2_{\rm red}$ by random
fluctuations is very low ($\ll 10^{-5}$), hence the model should
actually be rejected on a statistical basis.  Nevertheless a
qualitative inspection shows that the residuals are equally
distributed on positive and negative values, and the cumulative
$\chi^2_{\rm red}$ line (red line in lower panel) do not shows
significant ``jumps'' or discontinuities.  Also, the spectral
decomposition of the \hb{} and \oiii{} region, as well as the iron
emission, appear reasonable.  Finally, the correlation among
parameters (\S\ref{sec:MC-uncert}) implies that the actual number of
DOF is larger than the quantity {\it spectrum channel $-$ free
  parameters}, hence the quoted $\chi^2$ is actually an upper limit
for the actual value.  In summary, we consider our spectral quantities
reliable, even if the model should be rejected according to the
$\chi^2_{\rm red}$ goodness of fit test.

\section{The \qsfitcat{} catalog}
\label{sec:catalog}

\begin{figure}
  \includegraphics[width=.5\ww]{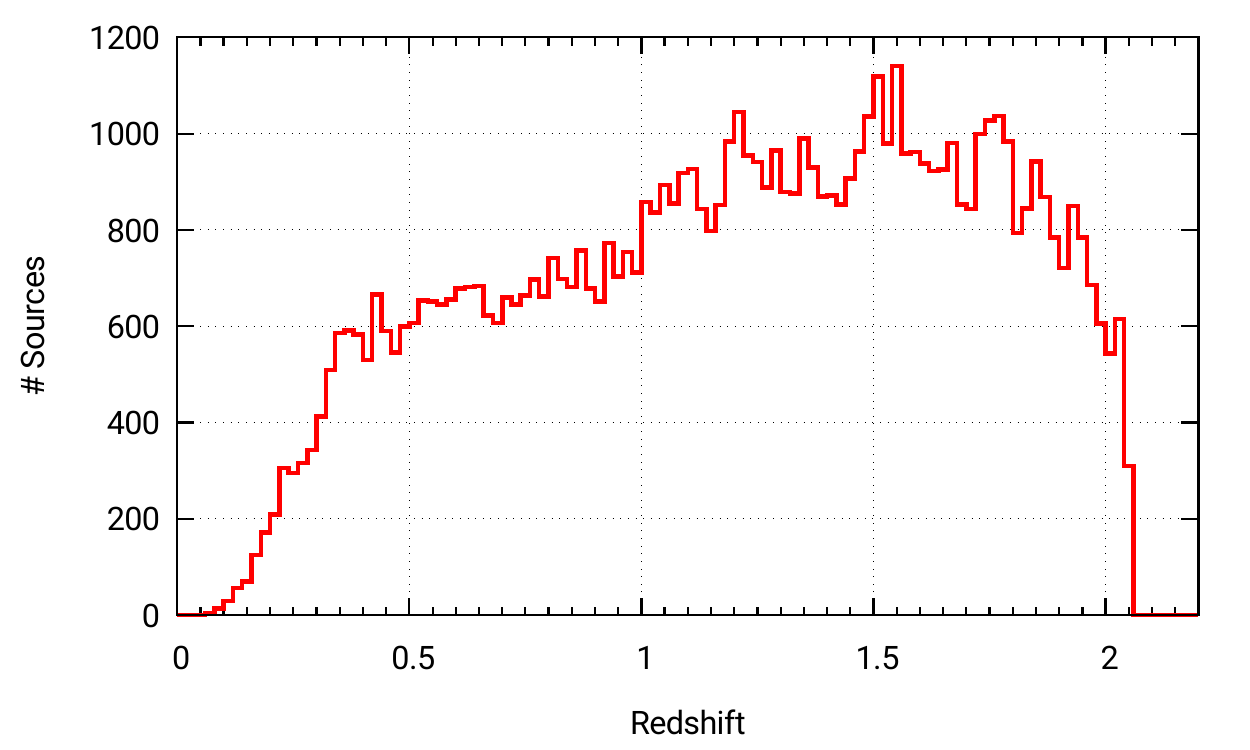}\\
  \includegraphics[width=.5\ww]{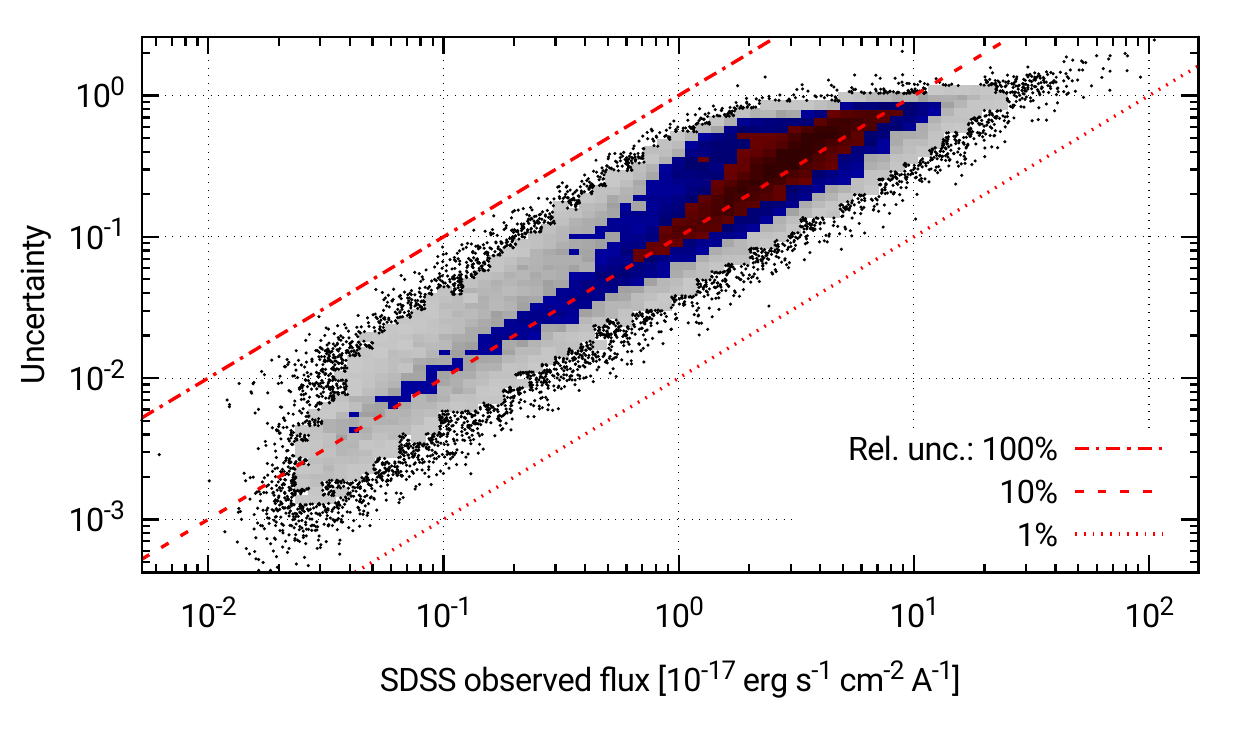}
  \caption{Distribution of redshift (upper panel) and SDSS observed
    flux density vs. its uncertainty (lower panel).  See
    \S\ref{sec:catalog} for a description of the colors used in the
    2--dimensional histogram.  The dotted, dashed and dot--dashed red
    lines are drawn in the region of points whose relative uncertainty
    is 1\%, 10\% and 100\% respectively.}
  \label{fig:redshift}
\end{figure}

We used \qsfit{} to analyze a subsample of the fifth edition of the
SDSS quasar catalog \citet{2010-Schneider-catdr7}, i.e. the same
sample analyzed by \citet[][the S11 catalog]{2011-shen-catdr7}.  We
selected all the objects in the S11 catalog whose redshift is $<2.05$
and whose {\tt BAL\_FLAG} is set to 0, i.e. they are not
broad--absorption line quasars.  This amounts to 79,986 sources.  Then
we dropped all the sources whose number of ``good'' spectral channels
(quality mask=0) was less than 75\% (\S\ref{sec:fit-proc}).  The final
sample comprises \nn{} sources, which make up the \qsfitcat{} catalog
(version \version{}) of spectral quantities.  Fig.~\ref{fig:redshift}
shows the redshift distribution in our sample (upper panel) and the
distribution of the SDSS flux density vs. its uncertainty (lower
panel).
The colors in the 2--dimensional histogram in the lower panel of
Fig.~\ref{fig:redshift} represent the count densities.  The bins drawn
with a red and blue palette contain respectively 40\% and 70\% of the
sources, the gray, red and blue ones (taken together) contain 95\% of
the sources.  The remaining 5\% are shown with individual dots.  This
color code is used for all the 2--D plots in this paper.

For each source in our subsample we performed the spectral analysis
using the \qsfit{} procedure and collected all the results in the
\qsfitcat{} catalog.  A similar work has already been done by
\citet{2011-shen-catdr7}, although with a different approach
(\S\ref{sec:cmpS11}).  In order to perform a meaningful comparison of
our results and those from S11 (\S\ref{sec:cmpS11}) we used exactly
the same redshift and E(B$-$V) values used by S11, as well as their
cosmology ($H_0=70$ km s$^{-1}$ Mpc$^{-1}$, $\Omega_{\rm M}=0.3$ and
$\Omega_{\Lambda}=0.7$).

\begin{figure}
  \includegraphics[width=.5\ww]{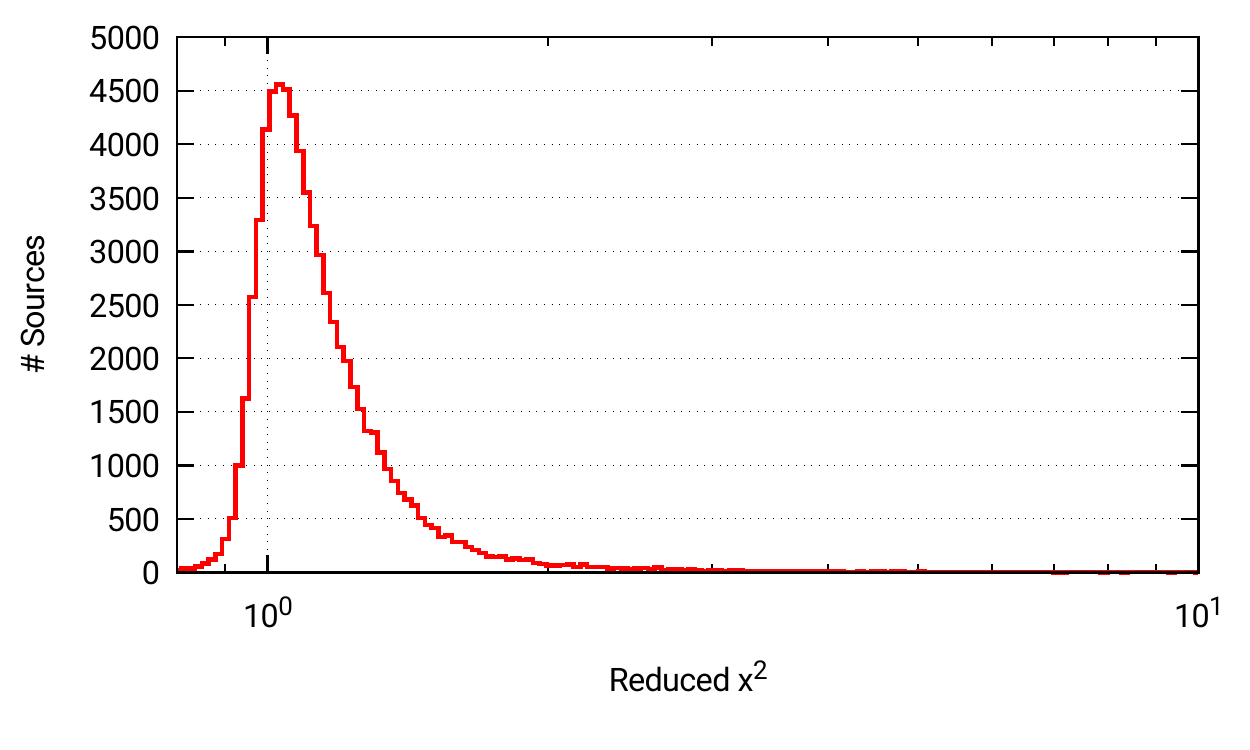}
  \includegraphics[width=.5\ww]{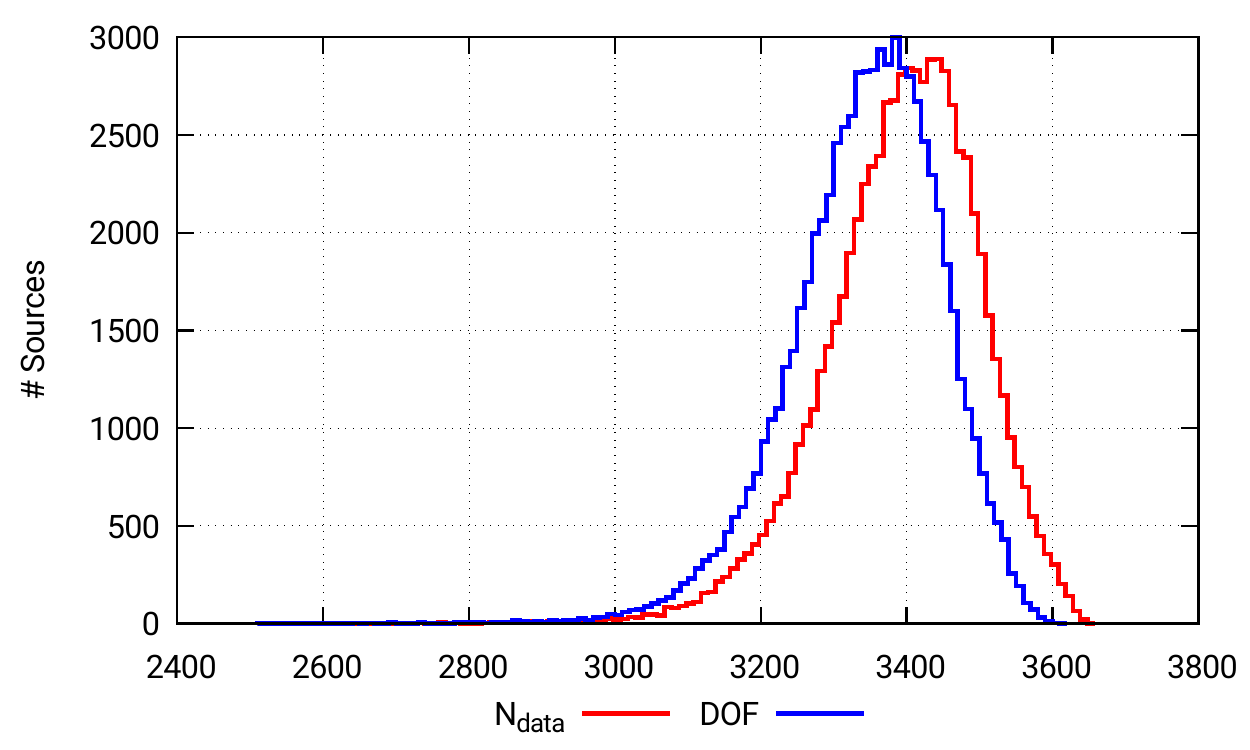}
  \caption{Upper panel: distribution of reduced $\chi^2_{\rm red}$ in
    the \qsfitcat{} catalog.  Lower panel: distribution of number of
    spectral channels in the SDSS data ($N_{\rm data}$, red line) and
    of degrees of freedom in the fit (DOF, blue line). }
  \label{fig:redchisq}
\end{figure}

In order to estimate the reliability of \qsfit{} results we performed
a qualitative visual inspection of many of the \qsfit{} plots
(\S\ref{sec:ex_plot}) and found that our model usually provides a
rather good representation of the data.  A quantitative assessment of
the agreement is given by the $\chi^2_{\rm red}$ values, whose median
value is 1.09 and is almost always smaller than 2
(Fig.~\ref{fig:redchisq}, upper panel).
The number of DOF is typically in the range [3100, 3600] with a median
value of 3380 (Fig.~\ref{fig:redchisq}, lower panel).  Despite the
rather low values of $\chi^2_{\rm red}$ the goodness of fit tests
typically fails since the number of DOF is very high.  Hence the
\qsfit{} model is too simple to account for all the details available
in the data, and further components are likely required (e.g. emission
lines, improved iron and host galaxy templates, etc.).  Nevertheless,
as discussed in (\S\ref{sec:ex_plot}), a qualitative examination of
the \qsfit{} plots, as well as a comparison with the S11 catalog
(\S\ref{sec:cmpS11}), typically ensure that the spectral quantities
can be considered reliable estimates.  The ``average'' reliability of
the \qsfit{} estimates can also be assessed by comparing the average
SDSS spectra with the average \qsfit{} components (see
Fig.~\ref{fig:arit5100_03} and the discussion in \S\ref{sec:slopes}).

Moreover, the current \qsfit{} implementation aims to provide a
reasonably good fit of SDSS data in the vast majority of the cases.
In order to provide a more accurate representation of the data, the
\qsfit{} procedure can be customized and tuned for the analysis of a
specific source (\S\ref{sec:qsfitManual}).

\begin{figure}
  \includegraphics[width=.5\ww]{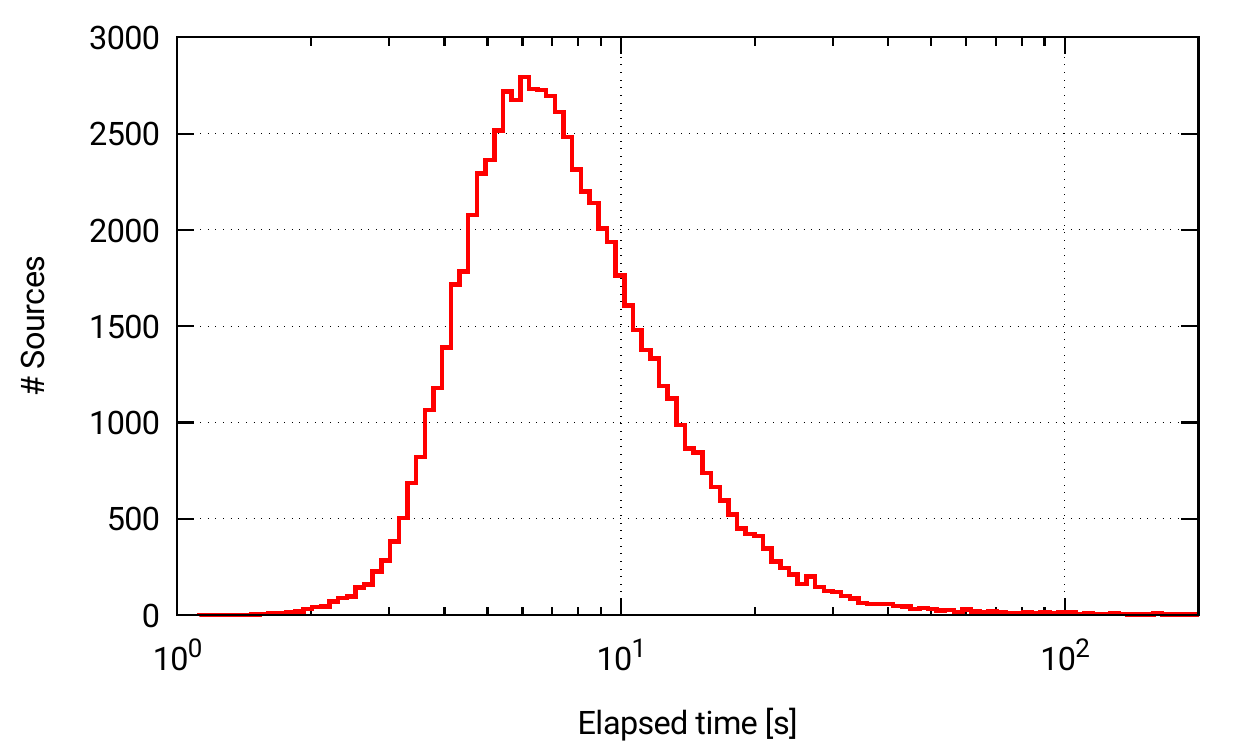}
  \caption{Distribution of elapsed time (in seconds) to perform a
    \qsfit{} analysis on a modern laptop.}
  \label{fig:elapsed}
\end{figure}

The \qsfitcat{} catalog is available as a FITS
file\footnote{\url{http://qsfit.inaf.it/cat_1.2/fits/qsfit_1.2.fits}}
where we report all the columns discussed in \S\ref{sec:reduction}, as
well as all the columns from the S11 catalog.  The \qsfitcat{} catalog
is also browsable through a web
tool\footnote{\url{http://qsfit.inaf.it/}} which allows the user to
search for a specific source, sort sources according to a parameter,
e.g. redshift, interactively browse the spectra inspecting the various
model components, eventually rebinning the spectra, and finally
download the plots (comparison of data and model, fit residuals) and
the \qsfit{} log file for a specific source.  The rather low
computational time ($\sim$~7 seconds on a modern day laptop,
Fig.~\ref{fig:elapsed}) required to analyze a source allows to re--run
the analysis of individual source, re--build the whole \qsfitcat{}
catalog, expand it, etc., without the need for expensive computational
resources.

\section{Comparison of S11 and \qsfitcat{} catalogs}
\label{sec:cmpS11}

In this section we will compare our results with those from
\citet[][S11]{2011-shen-catdr7} in order to assess the reliability of
\qsfit{} results.  Although the purpose of both the S11 and \qsfitcat{}
catalogs is similar, the adopted analysis algorithms show significant
differences.  The most relevant are:
\begin{itemize}
\item S11 evaluates the continuum underlying the emission lines over a
  rather narrow wavelength range, in the neighborhood of the line,
  whereas \qsfit{} uses a continuum component spanning the whole
  available wavelength range (\S\ref{sec:comp-continuum});

\item S11 fits the continuum and iron contributions first, then
  subtract them from the observed flux, and finally fits the emission
  lines.  \qsfit{} follows a similar approach, but at the last step
  all parameters (continuum + galaxy + iron + emission lines) are free
  to vary simultaneously;

\item S11 neglects the host galaxy contamination while \qsfit{} uses an
  elliptical galaxy template to account for such contribution
  (\S\ref{sec:comp-galaxy});

\item S11 uses the iron template from
  \citet{1992-boroson-emlineprop-irontempl} for the \ha{} and \hb{}
  emission lines, while \qsfit{} uses the template from
  \citet{2004-veron-spectra-izw1};

\item S11 models the broad emission lines with either one, two or
  three Gaussian profiles, according to which case provide the best
  fit.  Also, the \oiii{}4959\AA{} and \oiii{}5007\AA{} are both
  modeled with two Gaussian profiles. \qsfit{} always starts with a
  single Gaussian profile for each broad and narrow line.  If an
  ``unknown'' emission line (\S\ref{sec:comp-linesunk}) lies
  sufficiently close to a ``known'' line it can be associated to the
  latter (\S\ref{sec:modelFitting}) and the final profile may be
  modeled with more than one Gaussian profile.
  Tab.~\ref{tab:QualityLine} shows how many lines were fitted with
  more than one component.
\end{itemize}

Despite the differences in the adopted analysis algorithms, in several
cases the spectral estimates in the S11 and \qsfitcat{} catalogs show a
very good correlation, as is the case for, e.g., the continuum
luminosity estimates and the \civ{}~1549\AA{}, \mgii{}~2798\AA{} and
\hb{}~4861\AA{} emission line luminosities.
\begin{figure*}
  \includegraphics[width=.55\ww]{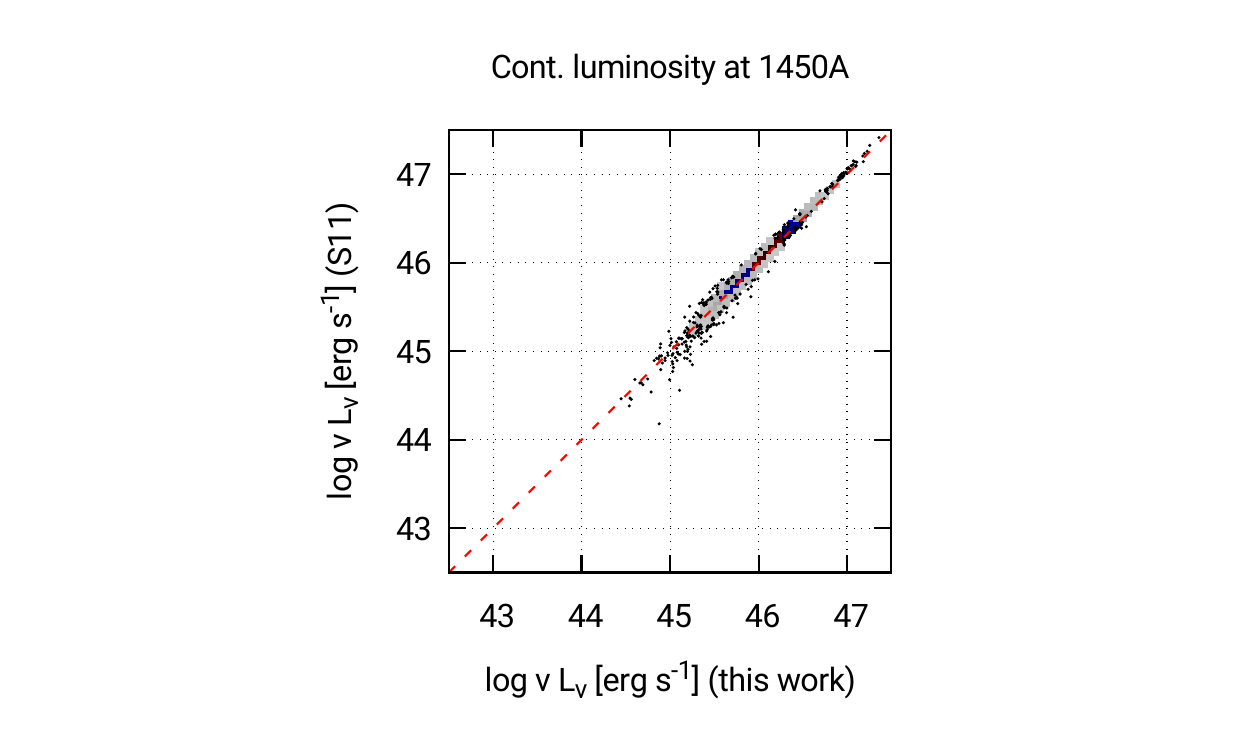}
  \hspace{-2cm}
  \includegraphics[width=.55\ww]{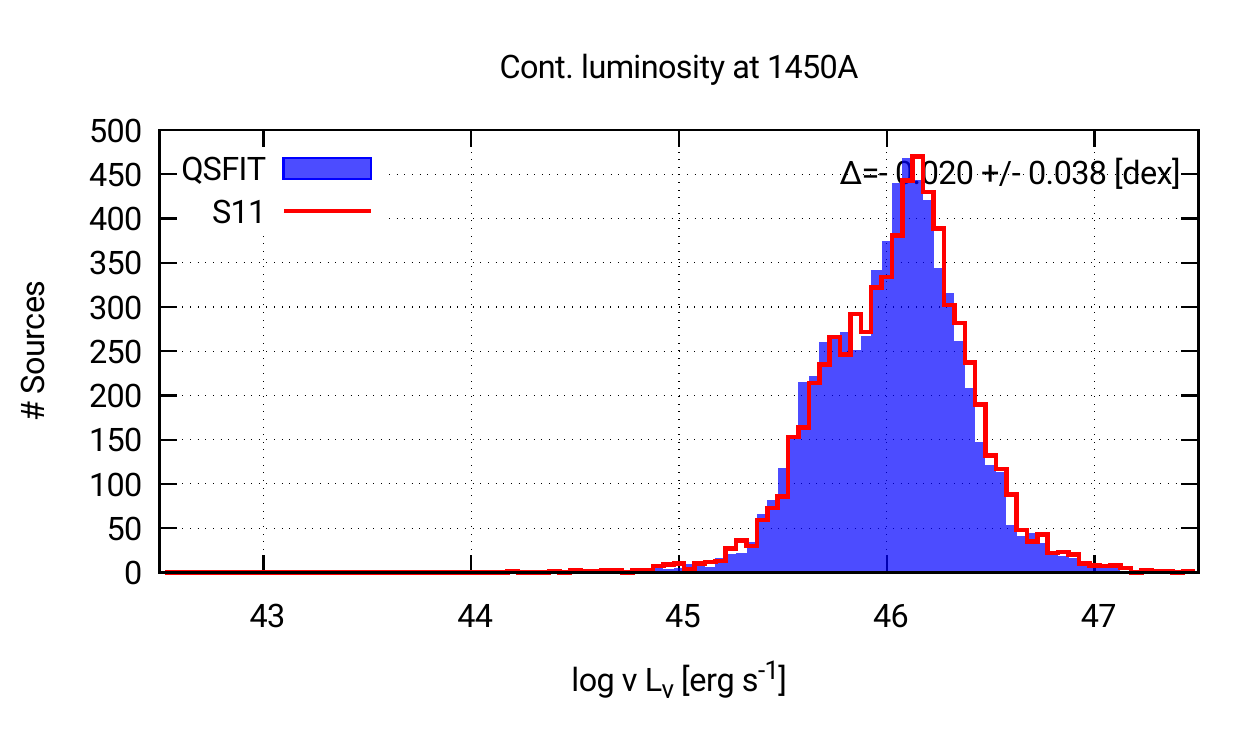}\\
  \includegraphics[width=.55\ww]{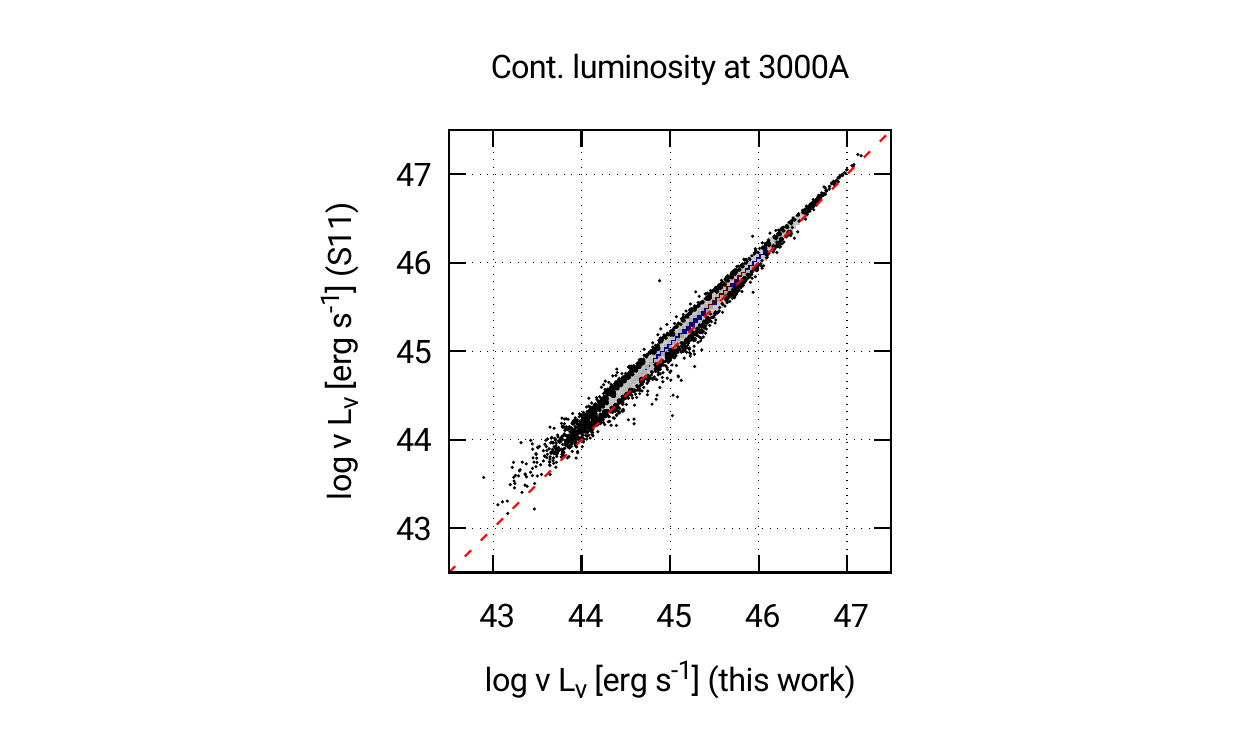}
  \hspace{-2cm}
  \includegraphics[width=.55\ww]{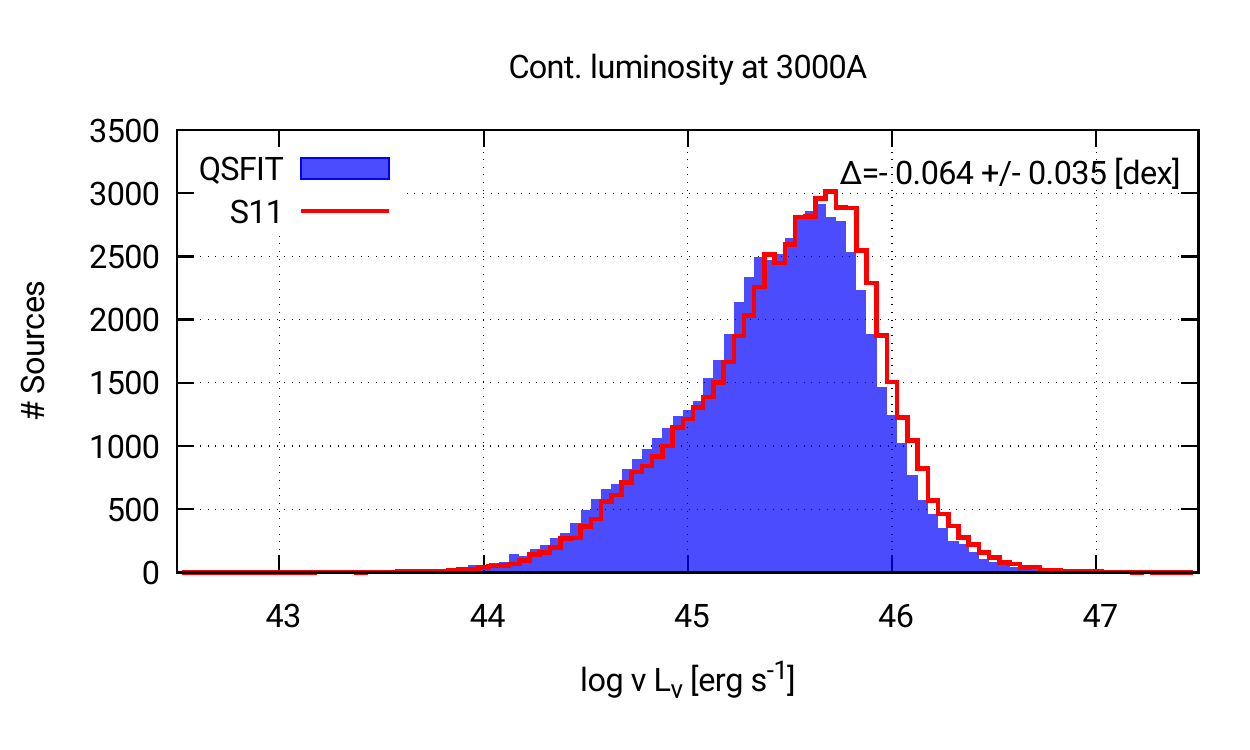}\\
  \includegraphics[width=.55\ww]{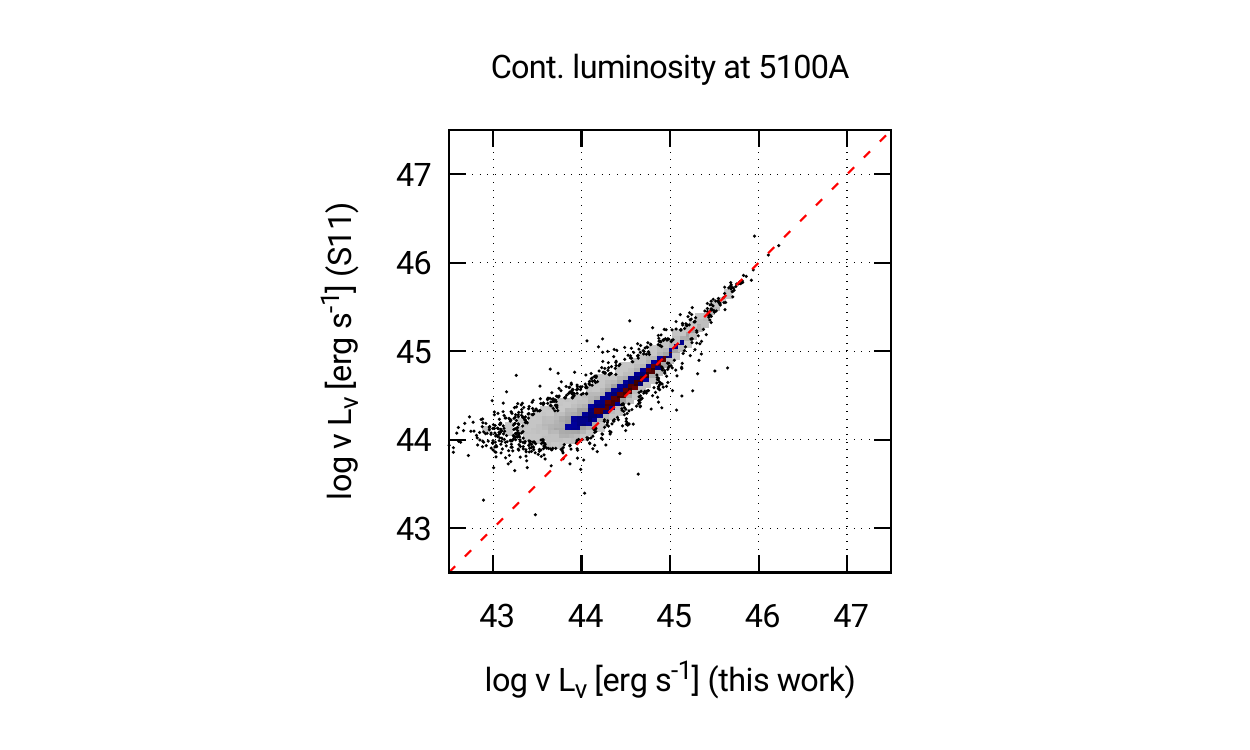}
  \hspace{-2cm}
  \includegraphics[width=.55\ww]{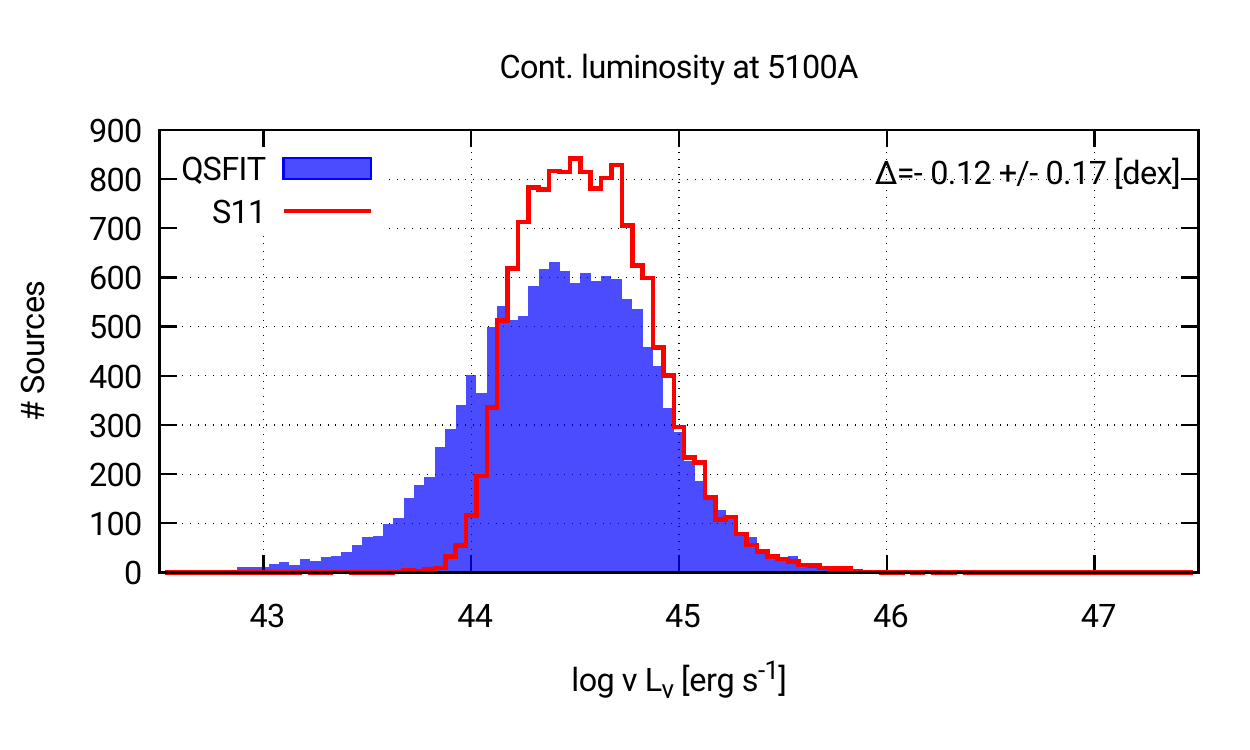}\\
  \caption{Comparison of \qsfit{} and S11 continuum luminosity
    estimates at 1350\gaa{}, 3000\gaa{} and 5100\gaa{}.  The left
    panels show the 2--D histogram plots (see \S\ref{sec:catalog} for a
    description of such plots) with \qsfit{} values on the X axis and
    S11 values on the Y axis (the red dashed line is the equality
    line).  In the right panels we compare the distribution of the two
    populations.  The average and standard deviation of the difference
    between the two catalogs is shown in the upper right corner.  The
    estimates are compatible in the two catalogs with a scatter of
    $\lesssim 0.04$ dex ($\sim 10$\%) over $\sim 2$ dex, except for
    the 5100\gaa{} case of low luminosity AGN ($L_{\rm 5100} \lesssim
    10^{44.5}$ erg s$^{-1}$) since the host galaxy contribution has
    been considered in \qsfit{} but neglected in S11 (see also
    \S\ref{sec:slopes}).}
  \label{fig:cmp_contlum}
\end{figure*}
Fig.~\ref{fig:cmp_contlum} shows the comparison of \qsfit{} and S11
continuum luminosity estimates at 1350\gaa{}, 3000\gaa{} and
5100\gaa{}.  The left panels show the 2--D histogram plots (see
\S\ref{sec:catalog} for a description of such plots) with \qsfit{}
values on the X axis and S11 values on the Y axis (the red dashed line
is the equality line).  The right panels show the comparisons between
the population distributions.  The average and standard deviation of
the difference between the two catalogs is shown in the upper right
corner.  The estimates are compatible in the two catalogs with a
scatter of $\lesssim$0.04 dex ($\sim$10\%) over $\sim$2 dex, except
for the 5100\gaa{} case of low luminosity AGN ($L_{\rm 5100} \lesssim
10^{44.5}$ erg s$^{-1}$) since the host galaxy contribution has been
considered in \qsfit{} but neglected in S11 (see also
\S\ref{sec:slopes}).

\begin{figure*}
  \includegraphics[width=.55\ww]{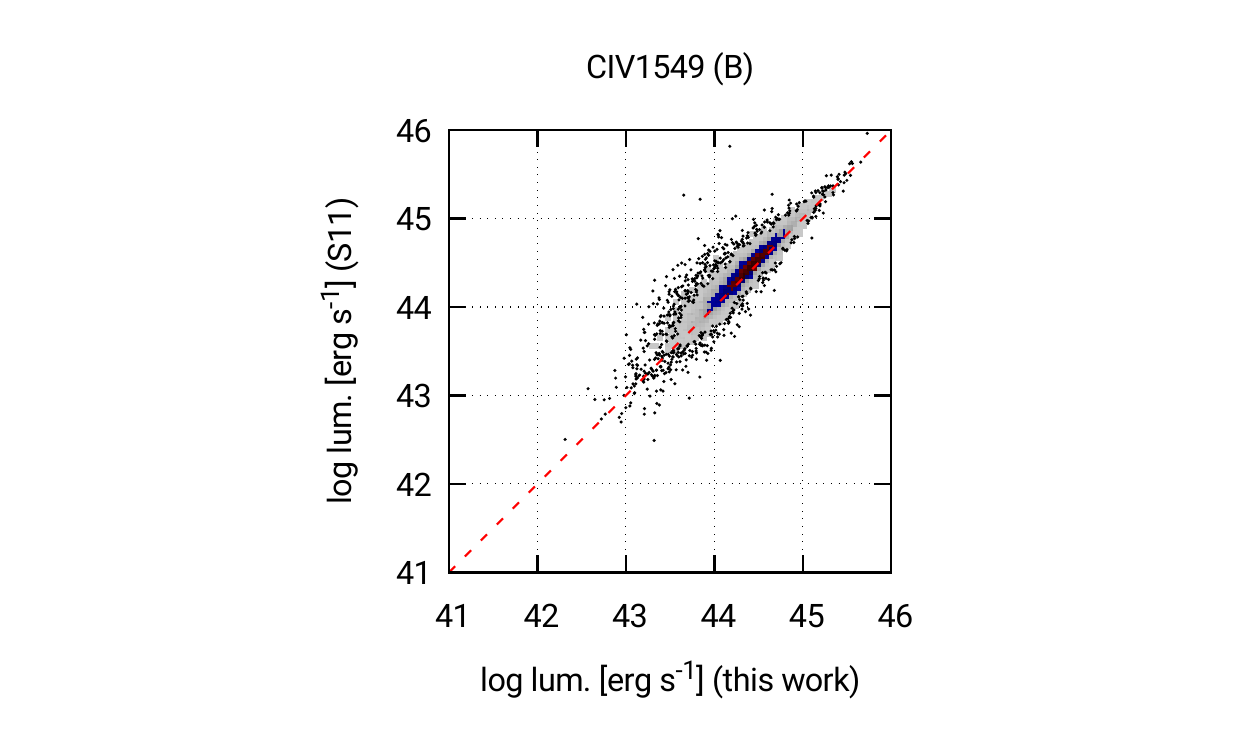}
  \hspace{-2cm}
  \includegraphics[width=.55\ww]{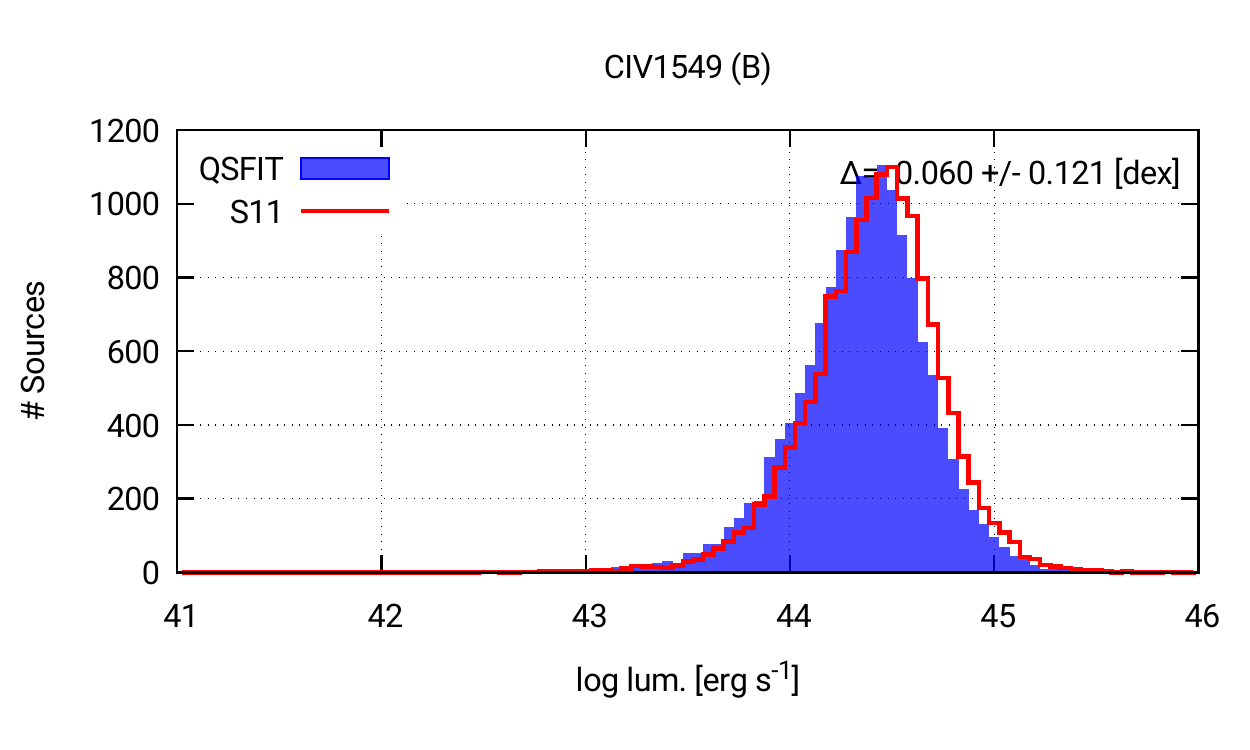}\\
  \includegraphics[width=.55\ww]{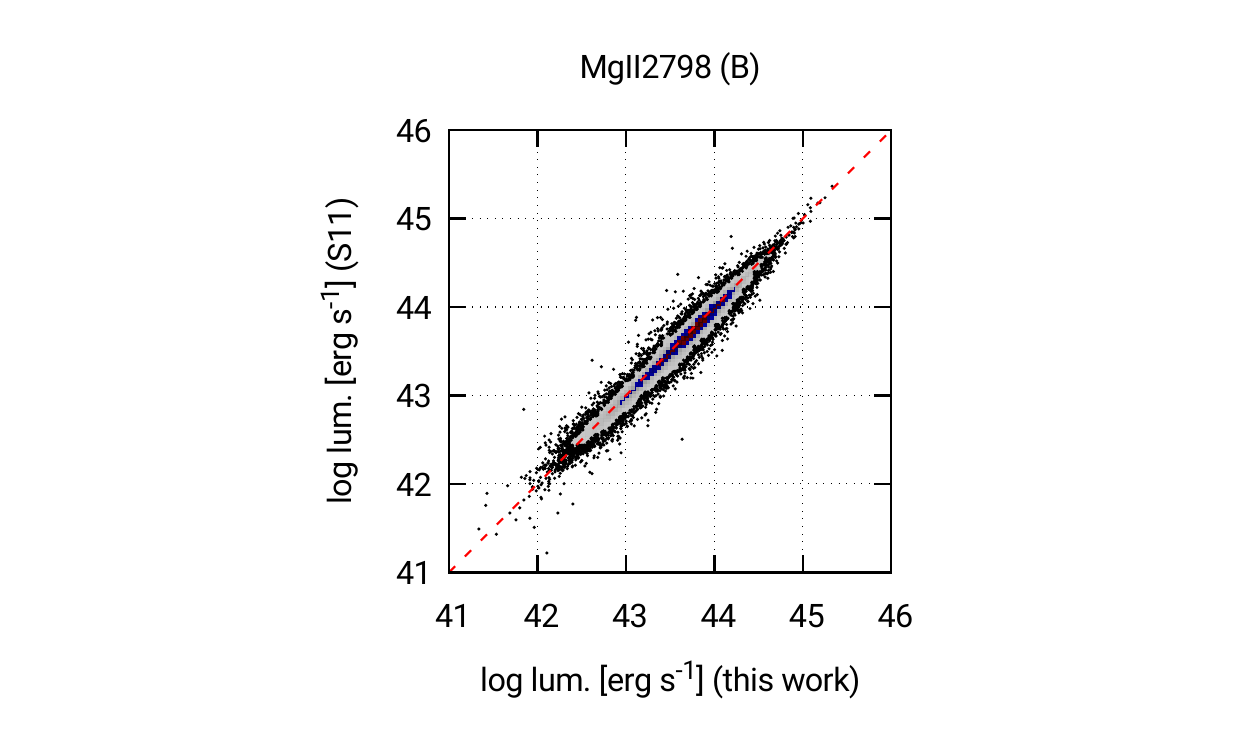}
  \hspace{-2cm}
  \includegraphics[width=.55\ww]{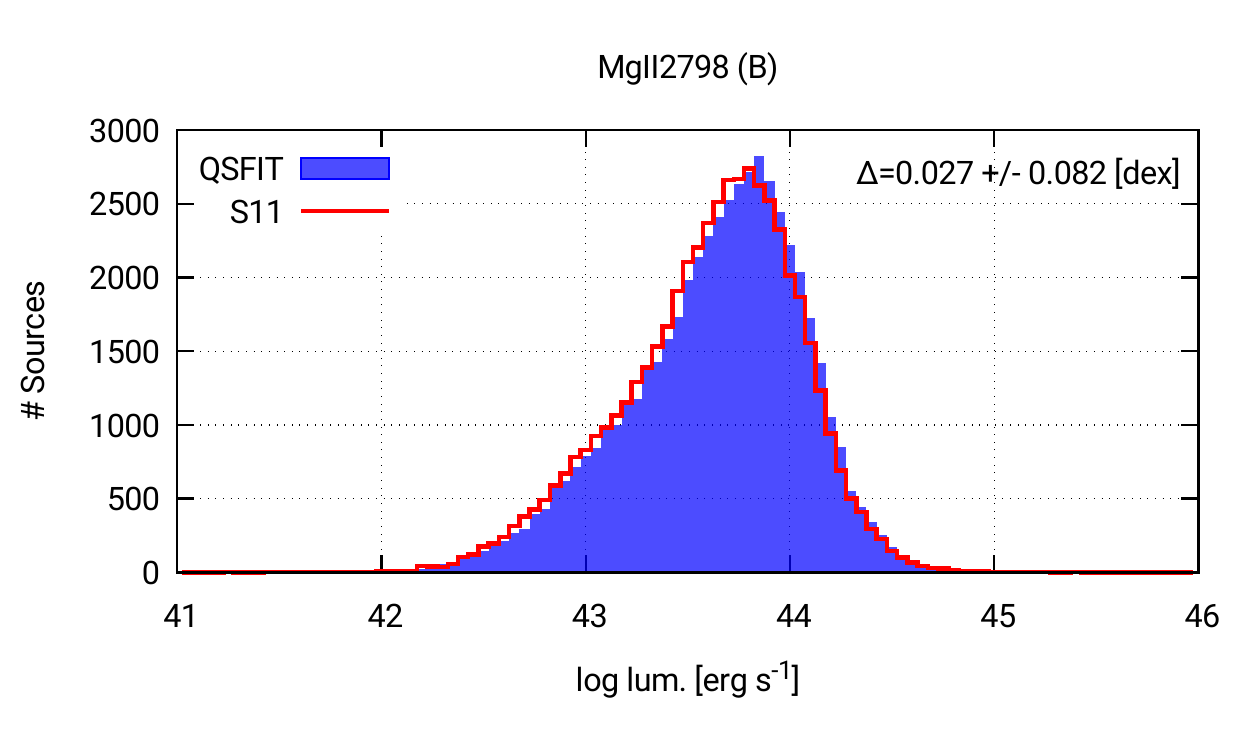}\\
  \includegraphics[width=.55\ww]{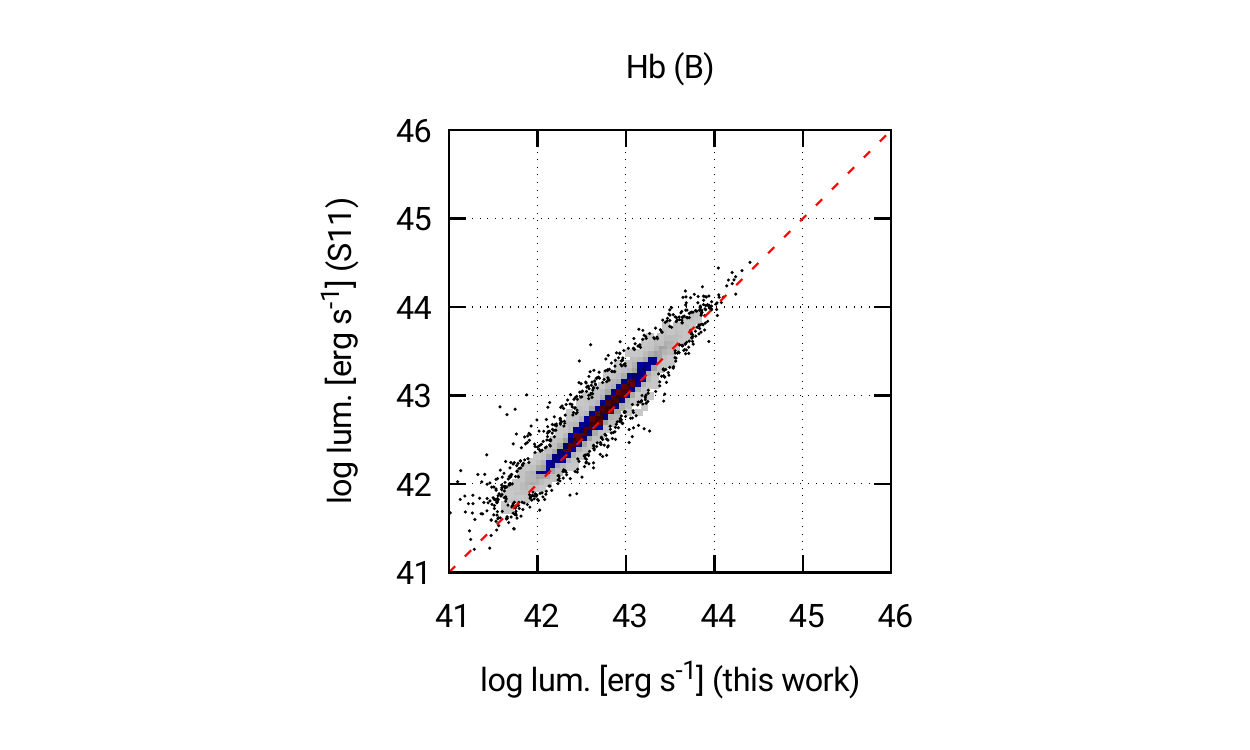}
  \hspace{-2cm}
  \includegraphics[width=.55\ww]{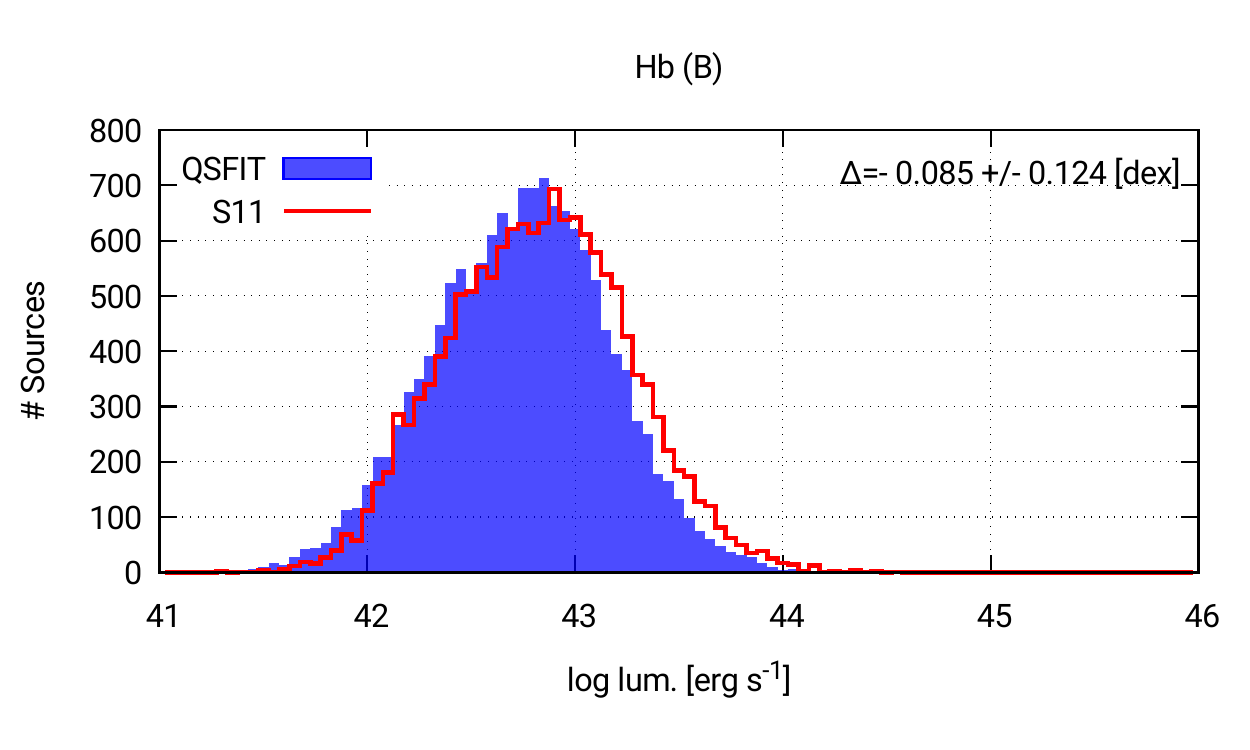}\\
  \caption{Comparison of \qsfit{} and S11 broad component luminosities
    of \civ{}~1549\AA{}, \mgii{}~2798\AA{} and \hb{}~4861\AA{}
    emission lines.  The plot meaning is the same as in
    Fig.~\ref{fig:cmp_contlum}.  The most prominent differences are in
    the \civ{}~1549\AA{} and \hb{}~4861\AA{} emission lines, mainly
    due to the different continuum subtraction methods
    (\S\ref{sec:fit-proc}).}
  \label{fig:cmp_lumLines}
\end{figure*}

\begin{figure}
  \includegraphics[width=9cm]{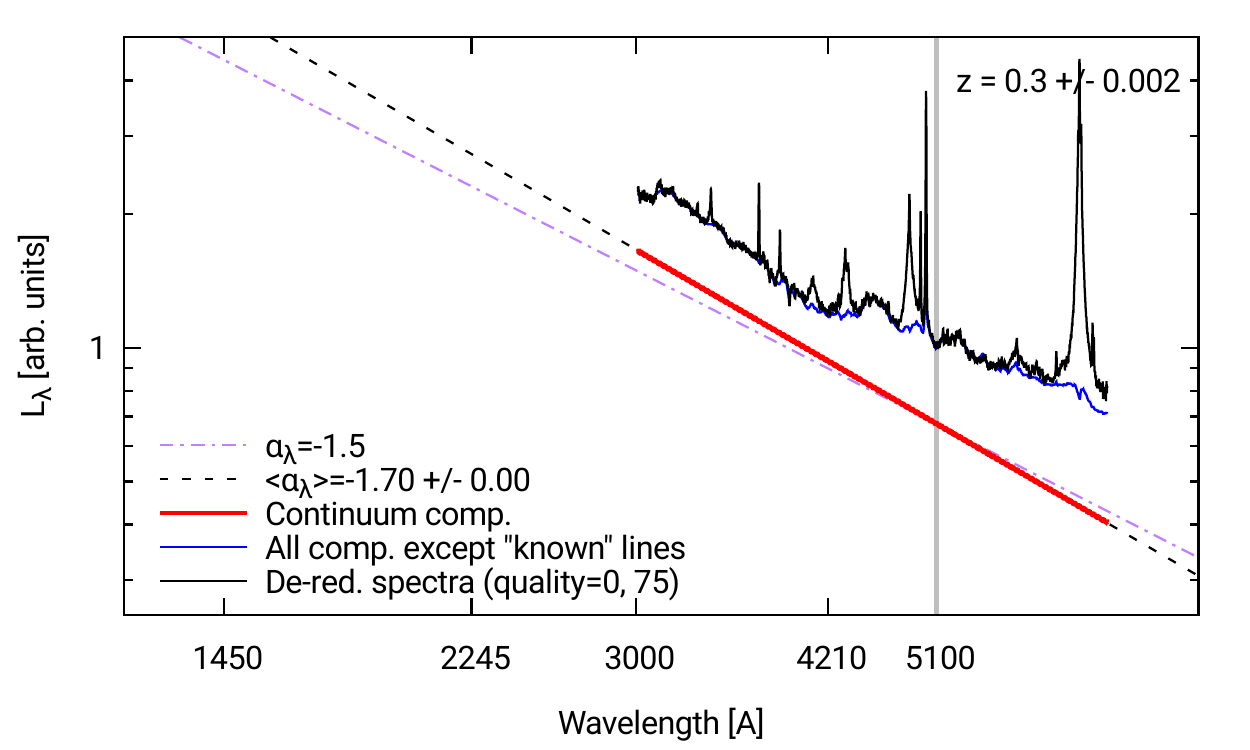}
  \caption{Comparison of the averaged spectrum of SDSS de--reddened
    spectra in the redshift bin $z = 0.3 \pm 0.002$ (black line) with
    the average spectrum of the sum of all \qsfit{} components except
    the ``known'' emission lines (blue line).  The agreement between
    the blue and black lines is remarkably good over the entire
    wavelength range (except the regions of the main emission line).
    Hence the blue line represent a reliable estimate of the continuum
    below the line (at least on average).
  }
  \label{fig:arit5100_03}
\end{figure}

Fig.~\ref{fig:cmp_lumLines} shows the same comparisons for the broad
component luminosities of \civ{}~1549\AA{}, \mgii{}~2798\AA{} and
\hb{}~4861\AA{} emission lines: also in this case the correlation is
quite good, although the scattering is larger ($\lesssim 0.18$ dex
$\sim 50$\%) over $\sim 2$ dex.  The most prominent differences are in
the \civ{}~1549\AA{} and \hb{}~4861\AA{} emission lines, and are
mainly due to the different continuum assumed below the lines
(\S\ref{sec:fit-proc}).  In \qsfit{} such continuum is the sum of all
the components except the ``known'' emission lines.
The reliability of such continuum can be inferred, at least on
average, by comparing the averaged SDSS de--reddened spectra (in a
narrow redshift bin) with the averaged sum of all \qsfit{} components
except the ``known'' emission lines.  Such comparison is shown in
Fig.~ \ref{fig:arit5100_03}, with the averaged spectrum of SDSS
de--reddened spectra in the redshift bin $z = 0.3 \pm 0.002$ shown
with a black line, and the average spectrum of the sum of all \qsfit{}
components except the ``known'' emission lines shown with a blue line.
The agreement between the blue and black lines is remarkably good over
the entire wavelength range (except the regions of the main emission
line).  Hence the blue line represent a reliable estimate of the
continuum below the line (at least on average).  The averaged (or
composite) spectra are further discussed in \S\ref{sec:slopes}.

\begin{figure*}
  \includegraphics[width=.55\ww]{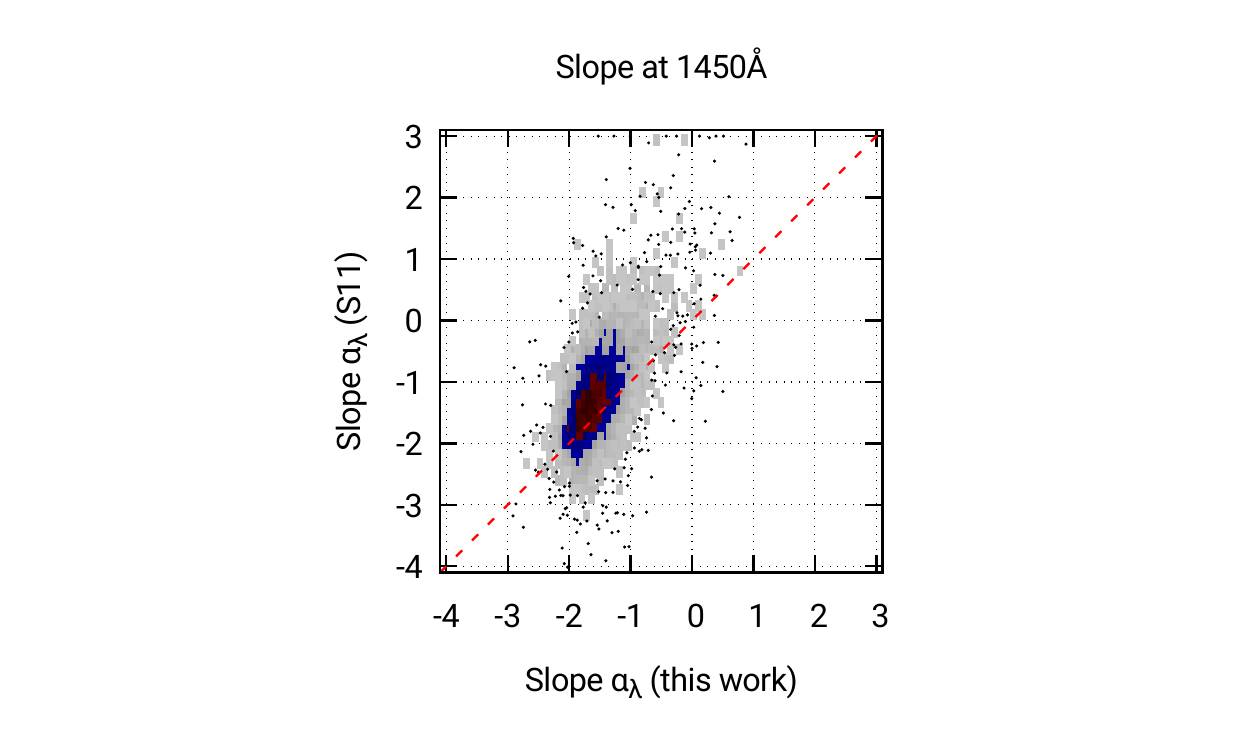}
  \hspace{-2cm}
  \includegraphics[width=.55\ww]{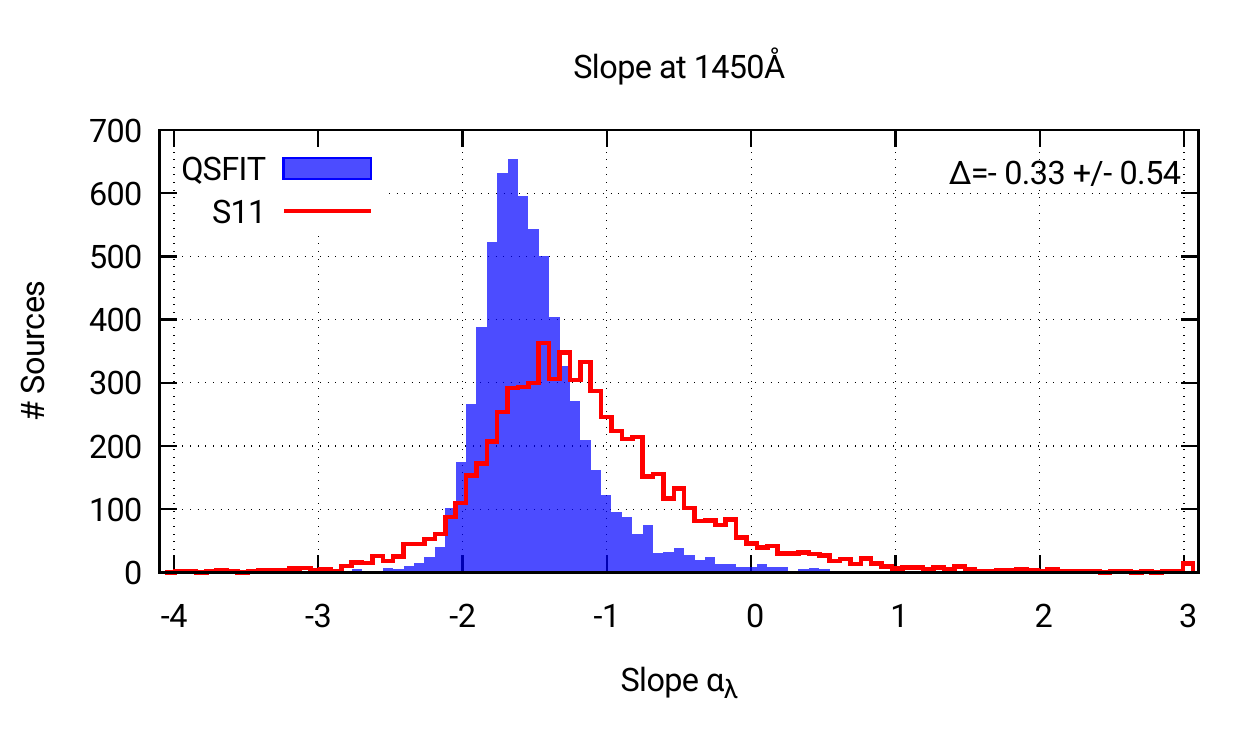}\\
  \includegraphics[width=.55\ww]{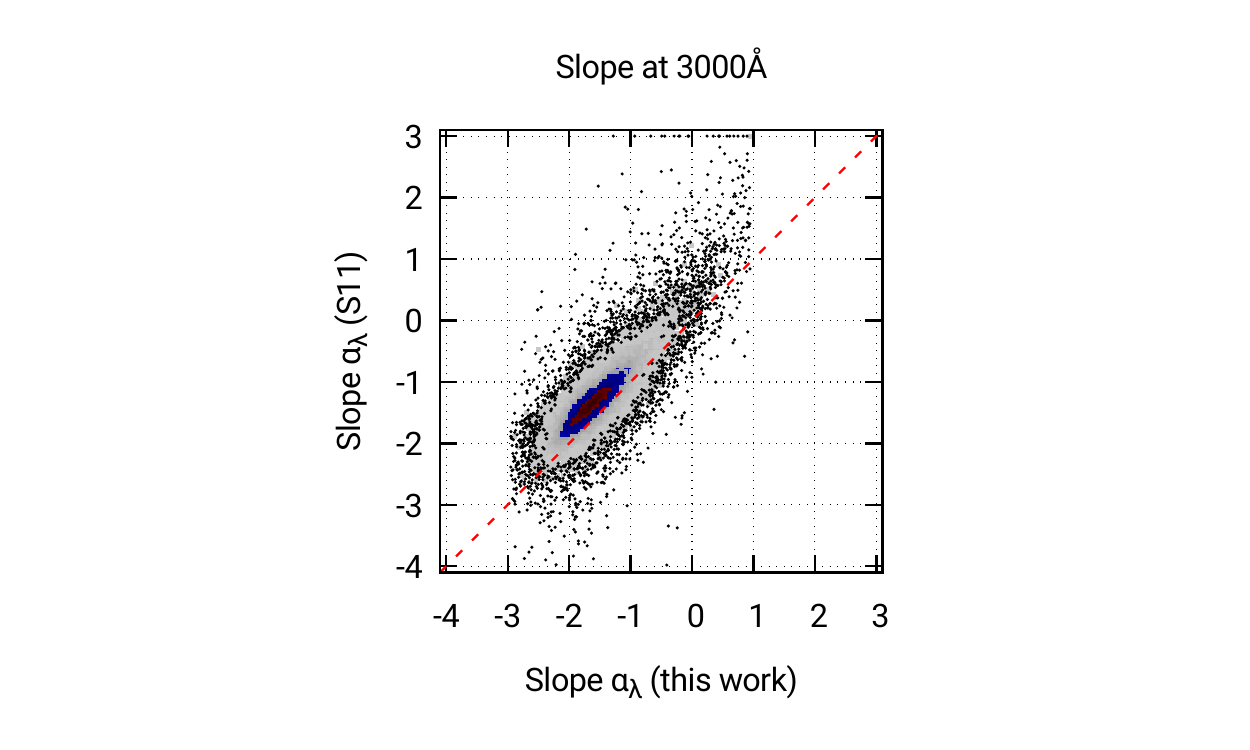}
  \hspace{-2cm}
  \includegraphics[width=.55\ww]{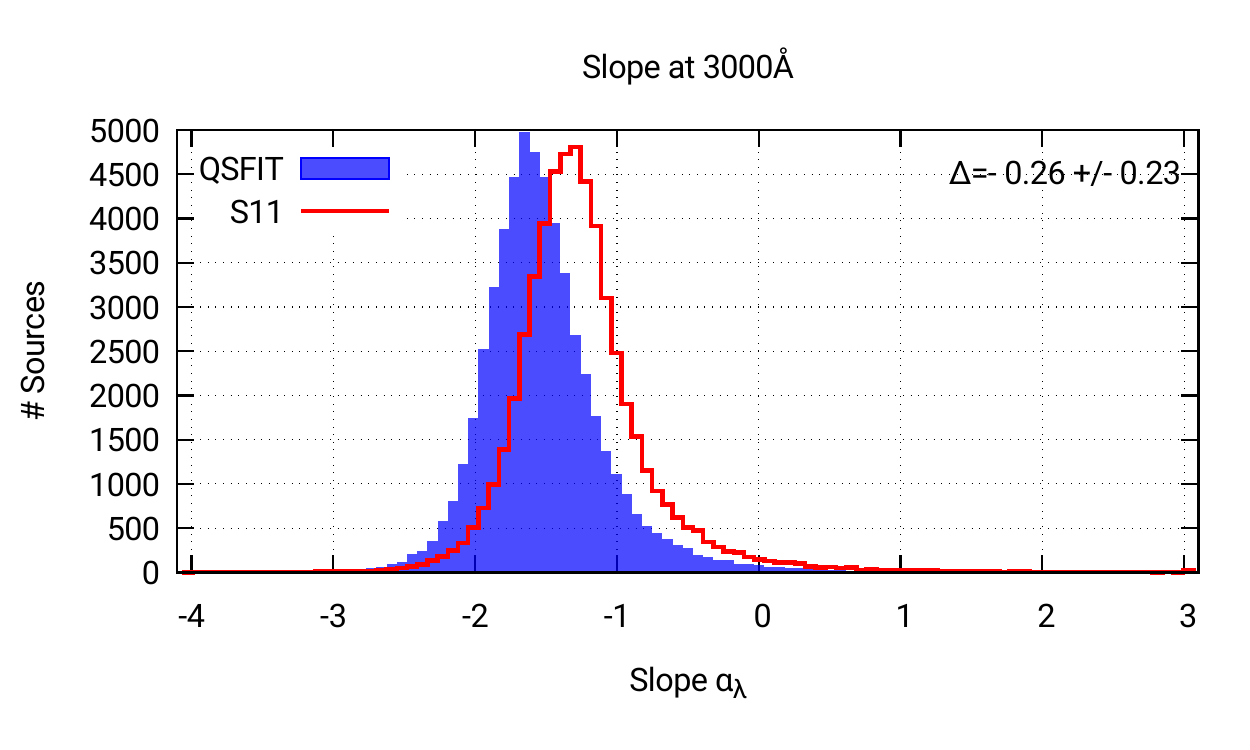}\\
  \includegraphics[width=.55\ww]{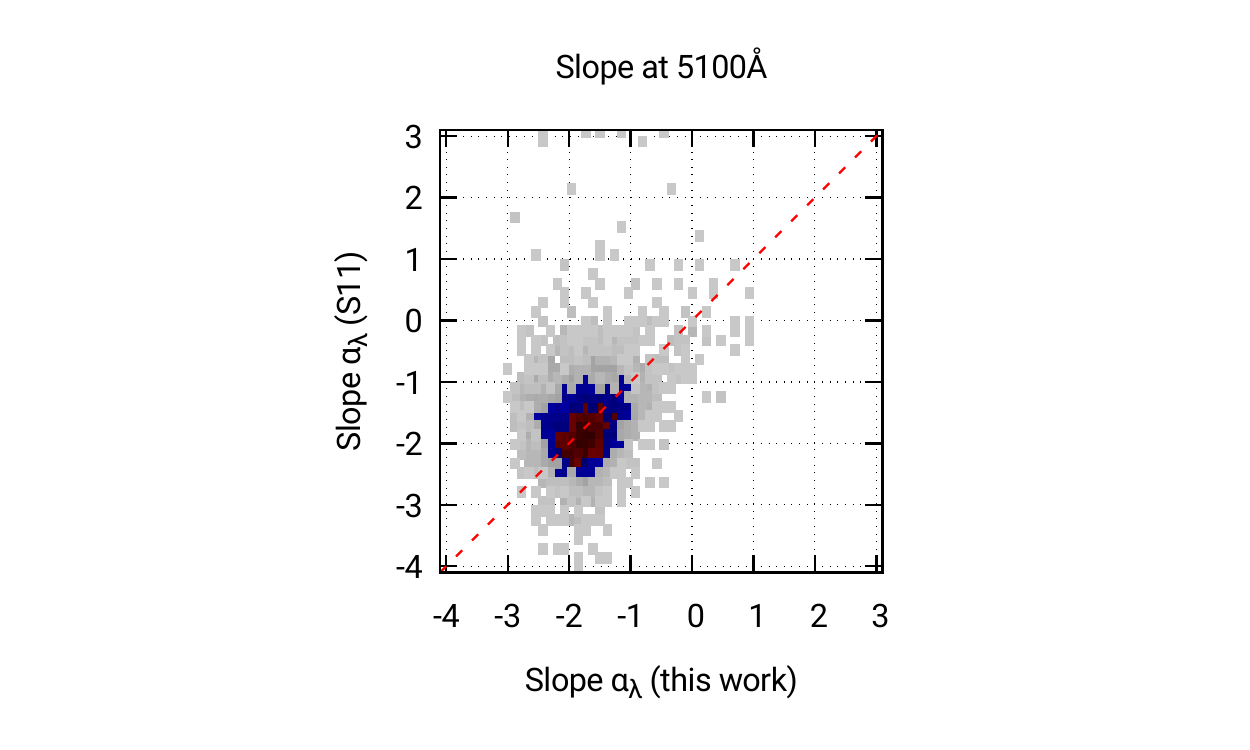}
  \hspace{-2cm}
  \includegraphics[width=.55\ww]{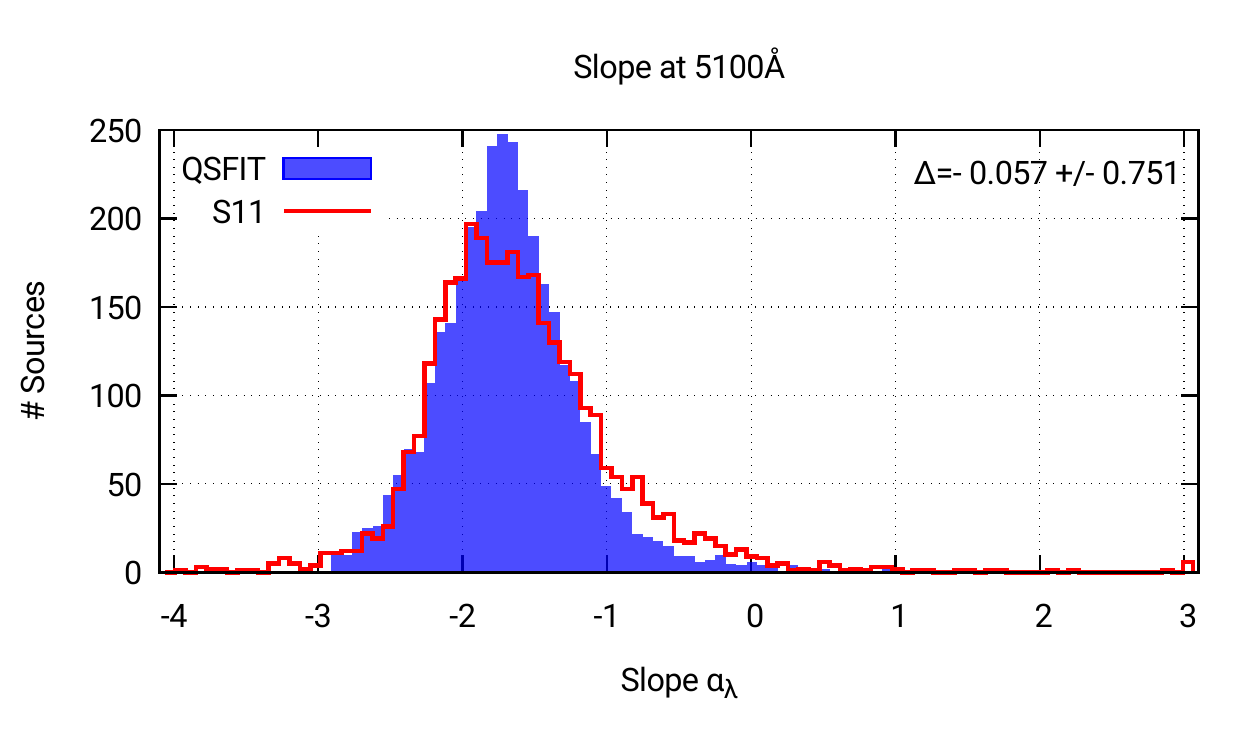}\\
  \caption{Comparison of our estimates of continuum slopes at
    1350\gaa{}, 3000\gaa{} and 5100\gaa{} with those in S11.  The
    estimates are not compatible because the continuum has been
    estimated using two different approaches: in this work we used a
    single power law extending through the entire observed wavelength
    range, while S11 used a power law in a narrow wavelength range.}
  \label{fig:cmp_contslopes}
\end{figure*}

\begin{figure*}
  \includegraphics[width=.55\ww]{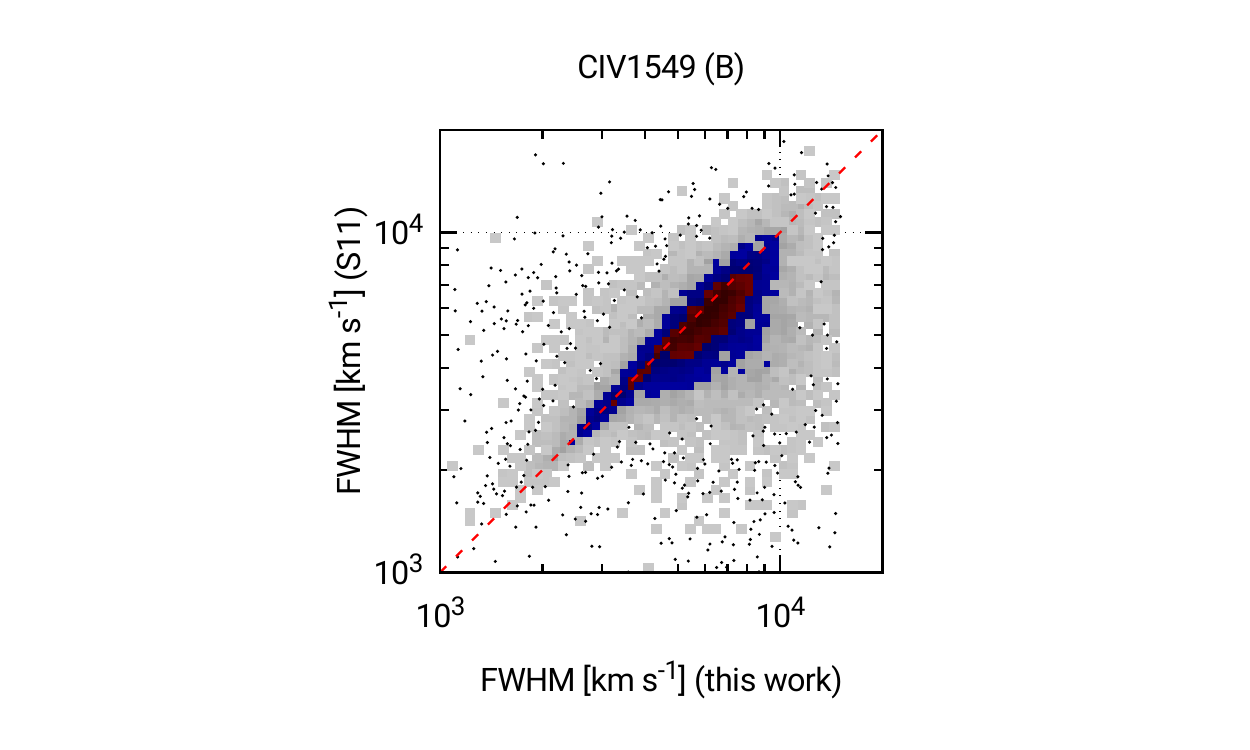}
  \hspace{-2cm}
  \includegraphics[width=.55\ww]{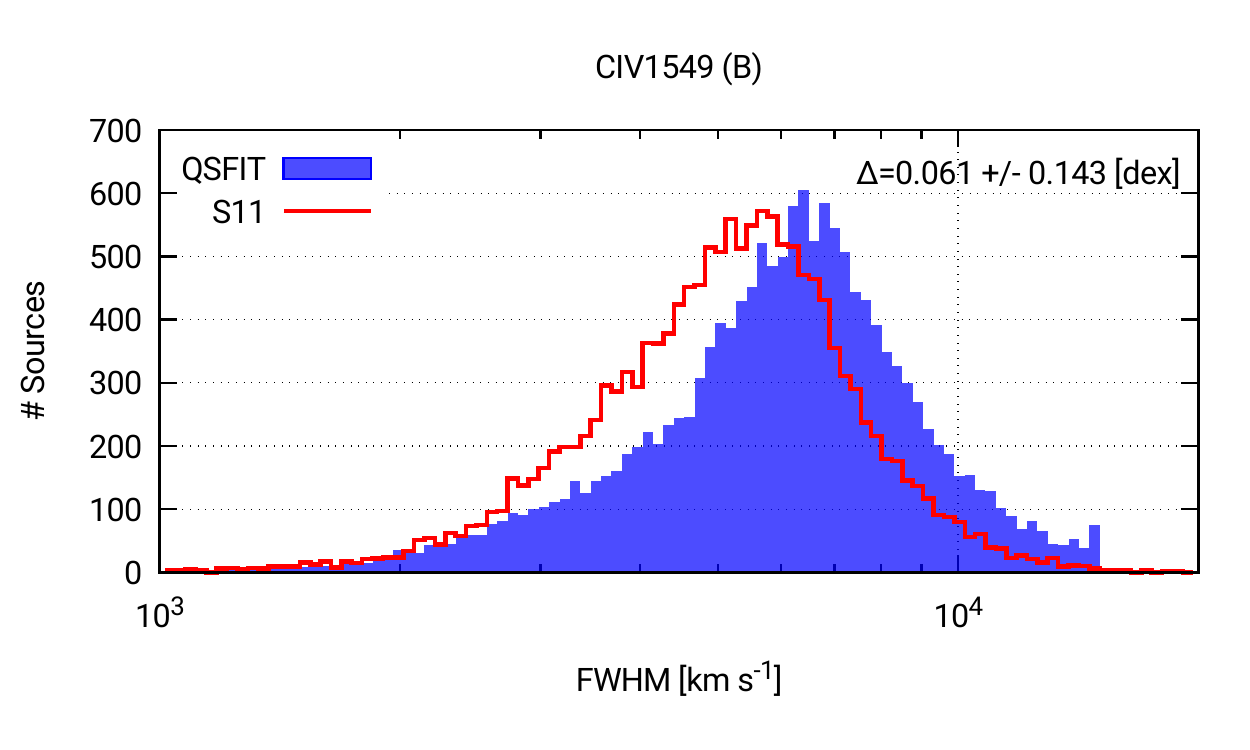}\\
  \includegraphics[width=.55\ww]{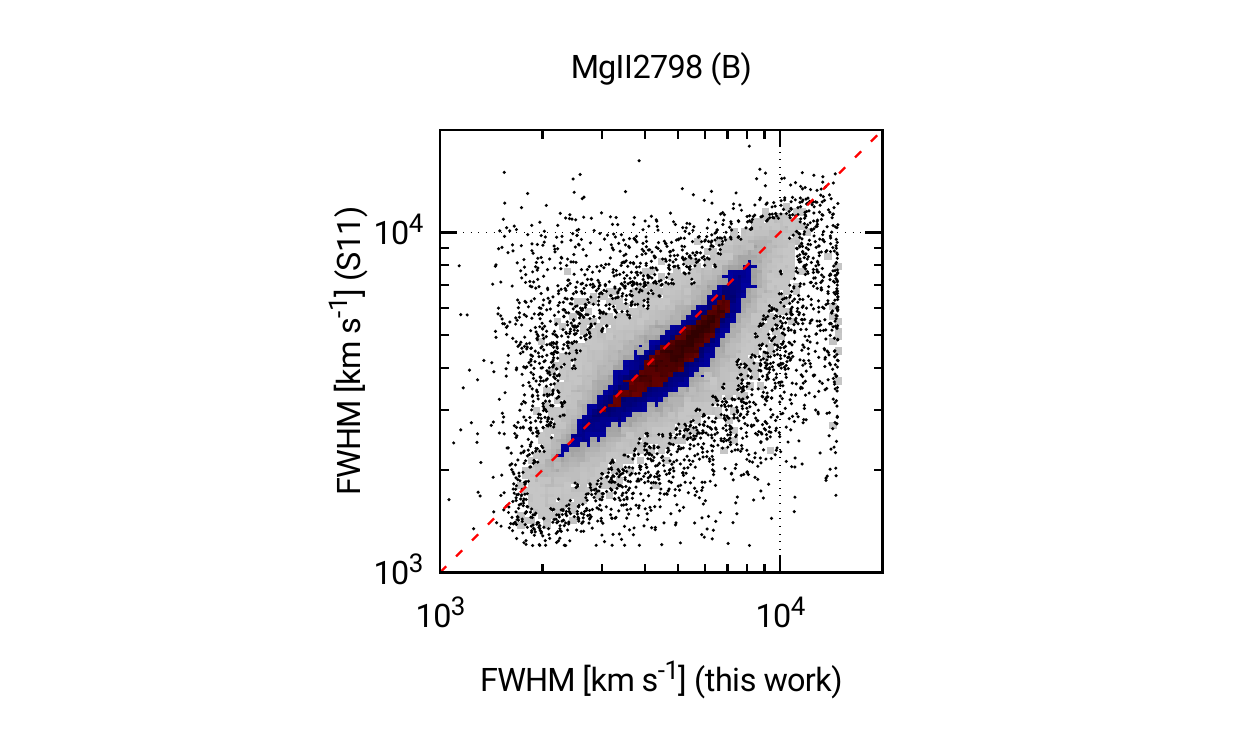}
  \hspace{-2cm}
  \includegraphics[width=.55\ww]{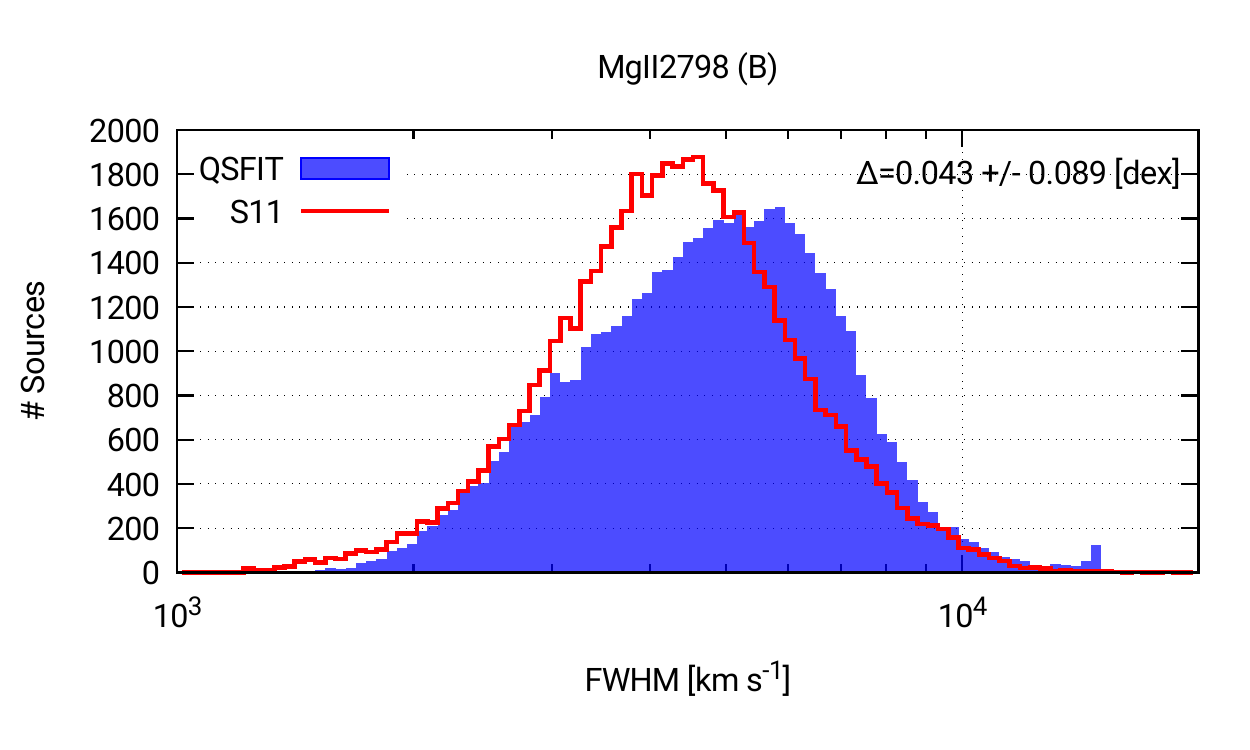}\\
  \includegraphics[width=.55\ww]{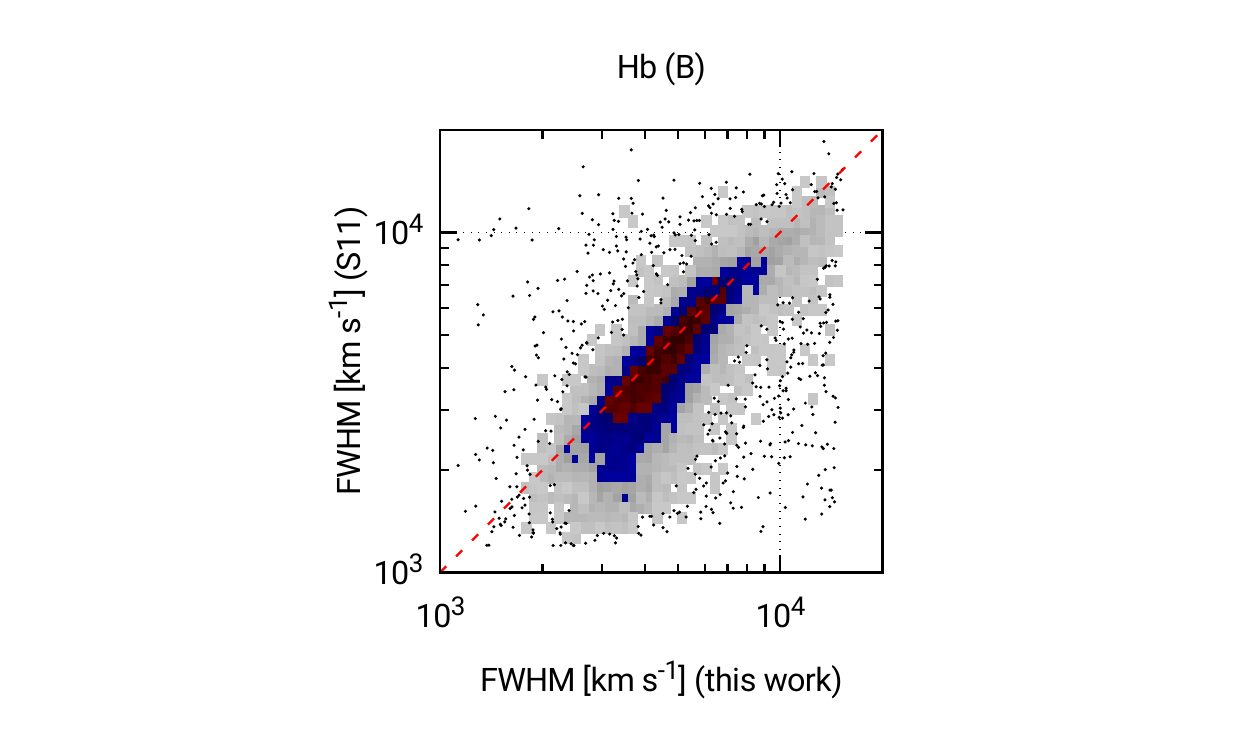}
  \hspace{-2cm}
  \includegraphics[width=.55\ww]{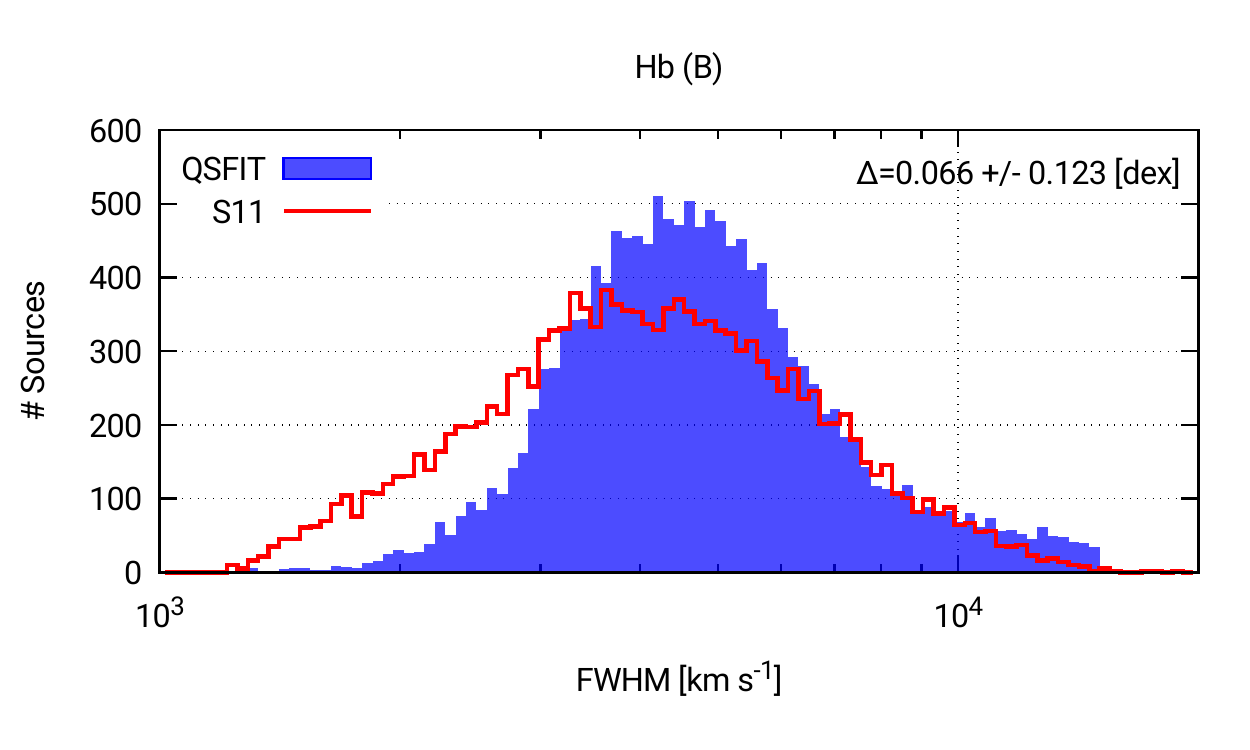}\\
  \caption{Comparison of \qsfit{} and S11 broad component widths of
    \civ{}~1549\AA{}, \mgii{}~2798\AA{} and \hb{}~4861\AA{} emission
    lines.  The plot meaning is the same as in
    Fig.~\ref{fig:cmp_contlum}.  The discrepancies are mainly due to
    the different line profile decomposition.}
  \label{fig:cmp_fwhmLines}
\end{figure*}

The very good agreement in continuum and emission line luminosities,
spanning $\sim$2 dex, stem from both an accurate spectral analysis,
and the relatively large redshift range considered.  Indeed, when
considering quantities which do not depend directly on redshift the
correlation is worse, although still present.  This is the case for,
e.g., the continuum slopes (Fig.~\ref{fig:cmp_contslopes}) and the
emission line widths (Fig.~\ref{fig:cmp_fwhmLines}).
The discrepancies are mainly due to the different adopted algorithms.
For instance, in the case of continuum slopes, we refer to the broad
band continuum slopes, while S11 refer to the local continuum in the
neighborhood of an emission line.\footnote{The plots available in the
  \qsfit{} web page allow for a quick comparison of the spectral
  slopes of both the \qsfitcat{} and S11 catalogs.}  Moreover, the
emission line widths show a significant scattering ($\lesssim 0.16$
dex, $\sim 45$\%, over less than 1 dex) since the line profile
decomposition algorithm is very different.  These findings show that
estimating the emission line widths is not a an easy and
straightforward task, and the availability of a commonly accepted
procedure would be very useful.

The rather low average differences between the populations of
luminosities and line widths discussed above allow us to conclude that,
at least on average, the \qsfit{} estimates are as reliable as the S11
ones.  Since there is not yet a complete understanding of the emission
processes generating the broad band continuum or the emission lines,
there is not a clear way to prefer either the S11, \qsfit{}, or any
other catalog.  The advantage in using the \qsfit{} one is simply that
the analysis is performed in an automatic (i.e. fast on large data
sets), replicable and shareable way.  Moreover it allows users to
customize and fine tune the analysis for specific sources.

\section{Discussion}
\label{sec:discussion}

The main purpose of the \qsfitcat{} catalog is to analyze the average
spectral properties of Type 1 AGNs by taking advantages of the large
SDSS dataset, instead of focusing on individual sources whose spectrum
may be affected by low signal to noise ratio, or be peculiar in some
aspect.  A thorough analysis of the \qsfitcat{} results is, however,
beyond the purpose of this work, and will be discussed in forthcoming
papers.  Here we will briefly explore the \qsfitcat{} data to check
whether they follow the empirical relations already known in
literature (\S\ref{sec:empirical}), discuss the reliability of the
\qsfit{} continuum slope estimates (\S\ref{sec:slopes}) and present a
few of the major results (\S\ref{sec:mainResults}).

\subsection{Empirical relations}
\label{sec:empirical}

Several empirical relations have been identified among AGN spectral
quantities \citep[e.g.][]{2000-Sulentic-PhenomenologyBLR}.  One of the
most important ones is the ``Baldwin effect''
\citep{1977-BaldwinEffect}, i.e. the correlation between the
\civ{}1549\AA{} equivalent width and the continuum luminosity.
\begin{figure}
  \includegraphics[width=.48\ww]{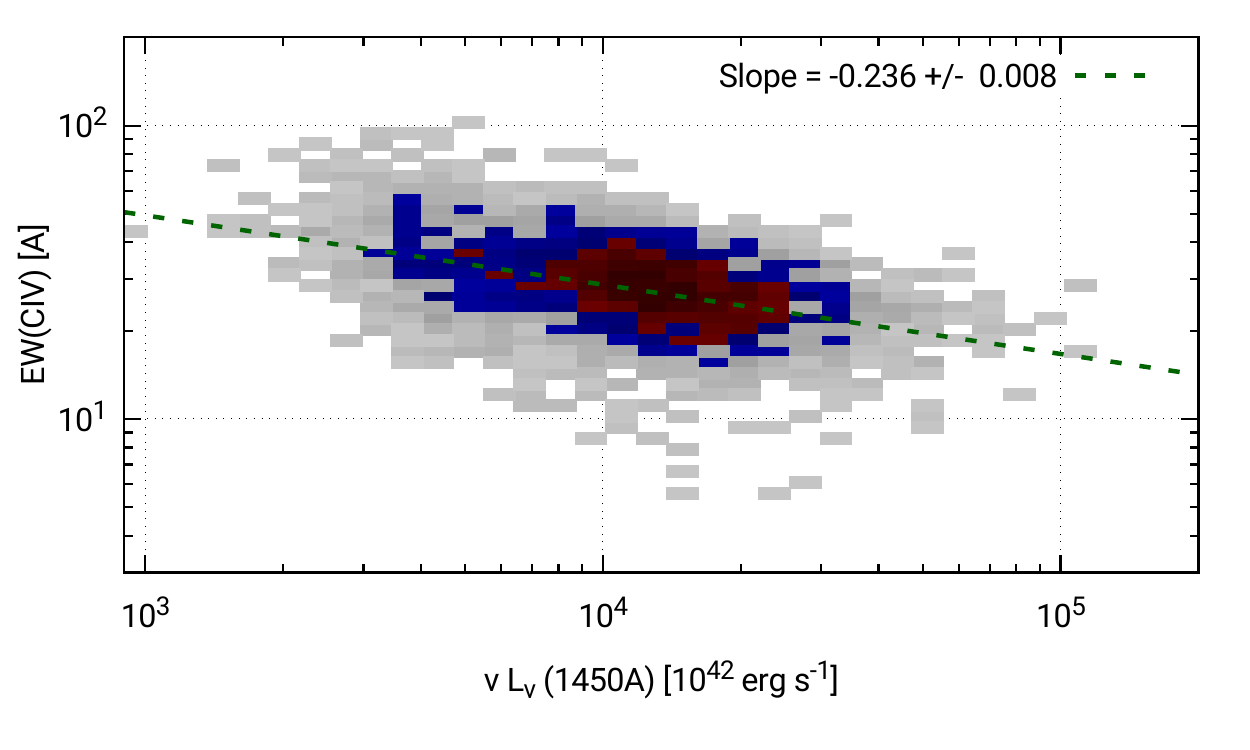}\\
  \includegraphics[width=.48\ww]{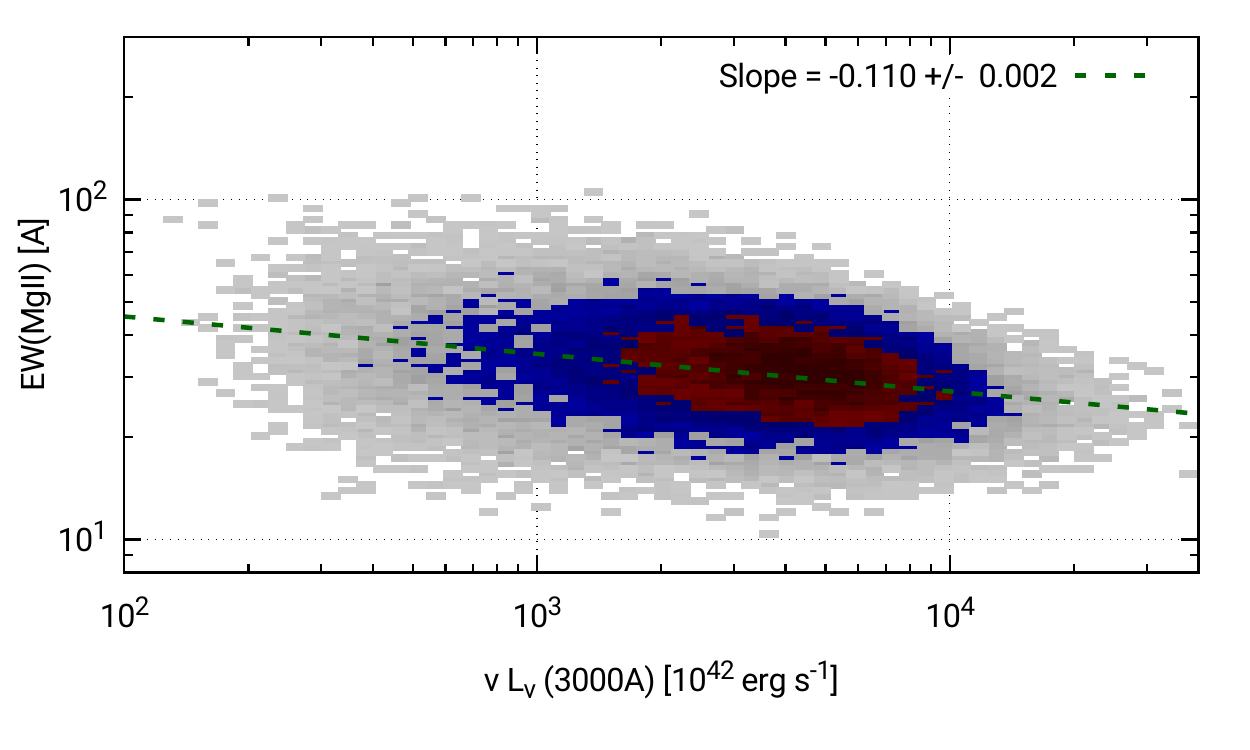}
  \caption{The Baldwin effect of the \civ{}~1549\AA{} (upper panel)
    and \mgii{}~2798\AA{} emission lines (lower panel) using the data
    from the \qsfitcat{} catalog.  The slopes are very similar to
    those reported in \citet{2001-Green-BaldwinEffect}, although the
    sample used here is significantly larger.}
  \label{fig:Baldwin}
\end{figure}
As new data became available similar correlations have been found for
other emission lines
\citep[e.g.][]{1989-Baldwin-EmissionLineProperties,
  1990-Kinney-Baldwin, 1992-Pogge-IntrinsicBaldwin,
  2001-Green-BaldwinEffect}.  In Fig.~\ref{fig:Baldwin} we show the
Baldwin effect for the \civ{}~1549\AA{} (upper panel) and
\mgii{}~2798\AA{} emission lines (lower panel) using the data from the
\qsfitcat{} catalog.  The slopes (shown in the upper right corner of
the plots) are very similar to those reported in
\citet{2001-Green-BaldwinEffect}, $-0.227\pm0.025$ and
$-0.187\pm0.017$ for \civ{}~1549\AA{} and \mgii{}~2798\AA{}
respectively, although the sample used here is significantly larger.

\begin{figure}
  \includegraphics[width=.48\ww]{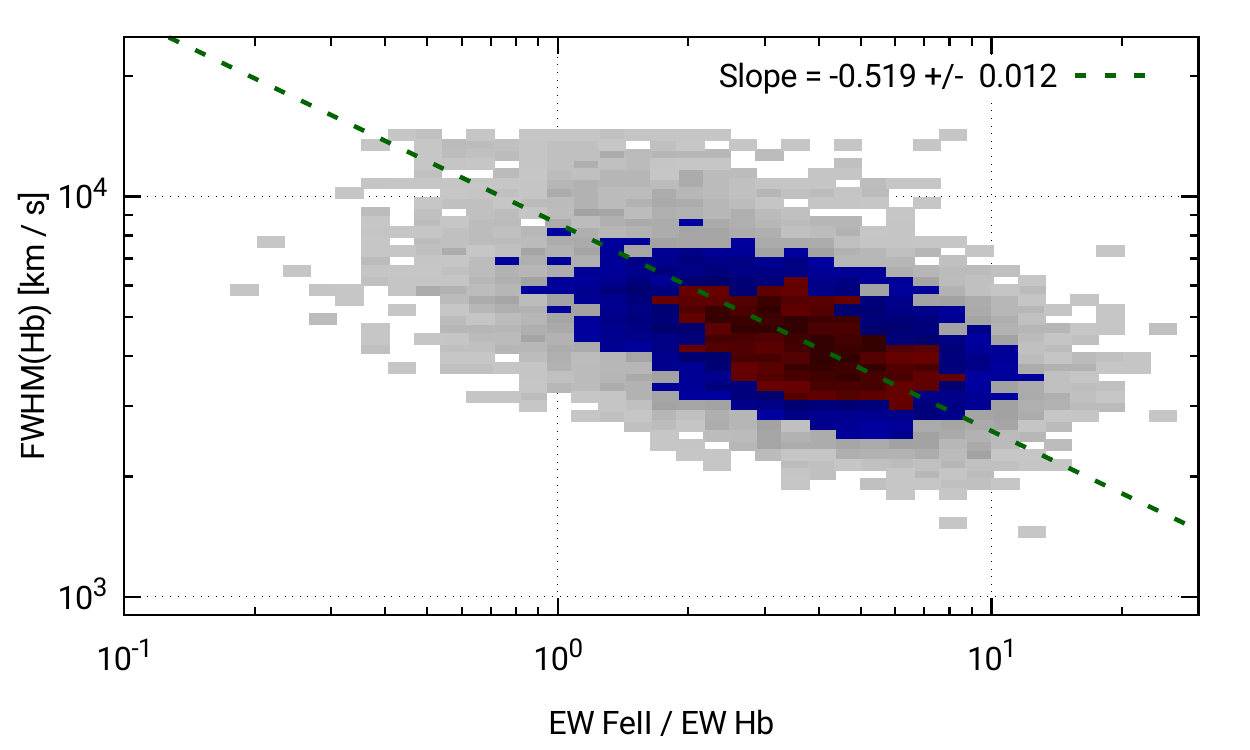}\\
  \includegraphics[width=.48\ww]{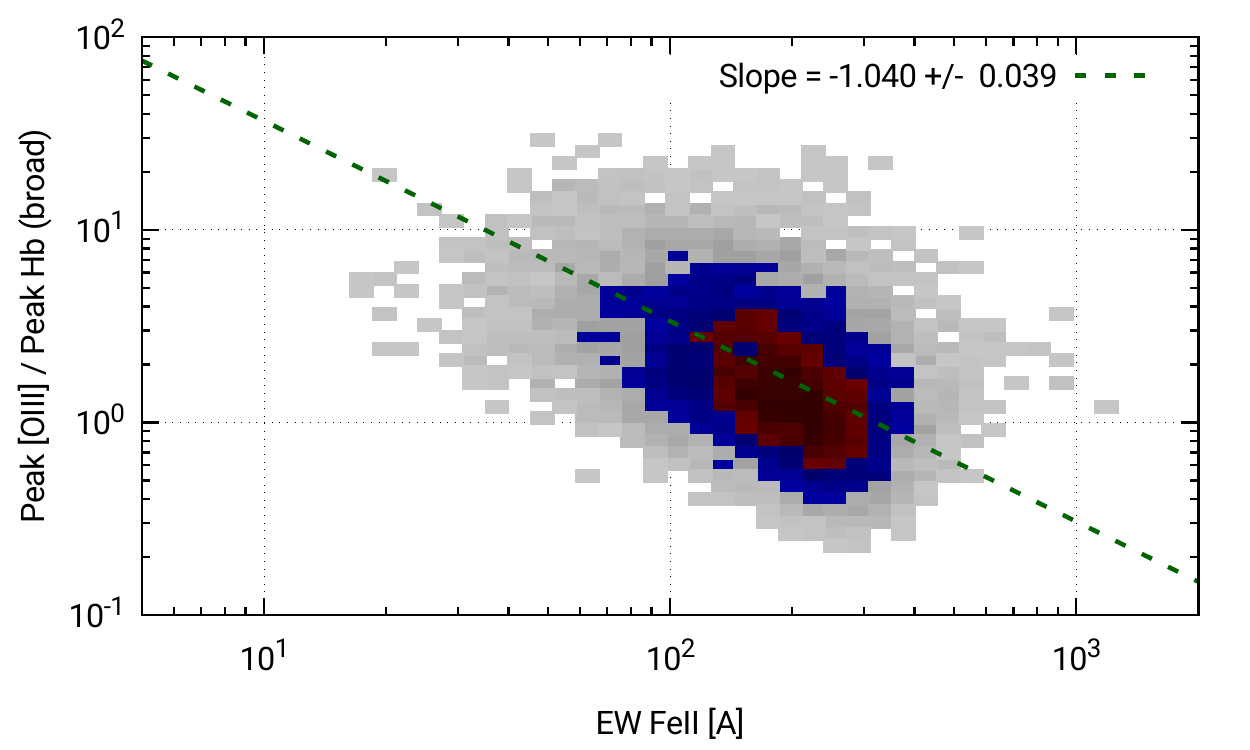}
  \caption{Correlations between the width of the broad \hb{} emission
    line (upper panel) and the peak luminosity ratios between \oiii{}
    and \hb{} (lower panel) vs. the \feii{} equivalent width, using
    the data from the \qsfitcat{} catalog.}
  \label{fig:ev1}
\end{figure}

Another important set of correlations has been highlighted by means of
the Principal Component Analysis (PCA).  The spectral quantities which
build up the ``Eigenvector 1'' \citep[i.e. the main driver for the
  variance in the observed spectral
  properties][]{1992-boroson-emlineprop-irontempl} are the peak
luminosity ratios between \oiii{} and \hb{}, the equivalent width of
the \feii{} complex at optical wavelengths, the width of the \hb{}
emission line and a few others.  These quantities present strong
correlations, which are likely the manifestation of the underlying AGN
physics \citep[e.g.][]{1996-Wang-AGNwStrongFeII, 2002-Boroson,
  2003-Sulentic-EV1-RadioLoud}.  In Fig.~\ref{fig:ev1} we show two
such correlations between the width of the broad \hb{} emission line
(upper panel) and the peak luminosity ratios between \oiii{} and \hb{}
(lower panel) vs. the \feii{} equivalent width, using the data from
the \qsfitcat{} catalog.

These examples show that the data in the \qsfitcat{} catalog allow to
recover the most common empirical relations known in literature.  By
exploiting the very large \qsfitcat{} catalog and its additional
estimates not previously available (such as the AGN continuum slopes
or the host galaxy luminosities) it will be possible to study in a
much deeper detail the correlations.  These topics will be the
subject of a forthcoming paper.

\subsection{Reliability of continuum slope estimates}
\label{sec:slopes}

\begin{figure*}
  \includegraphics[width=.48\ww]{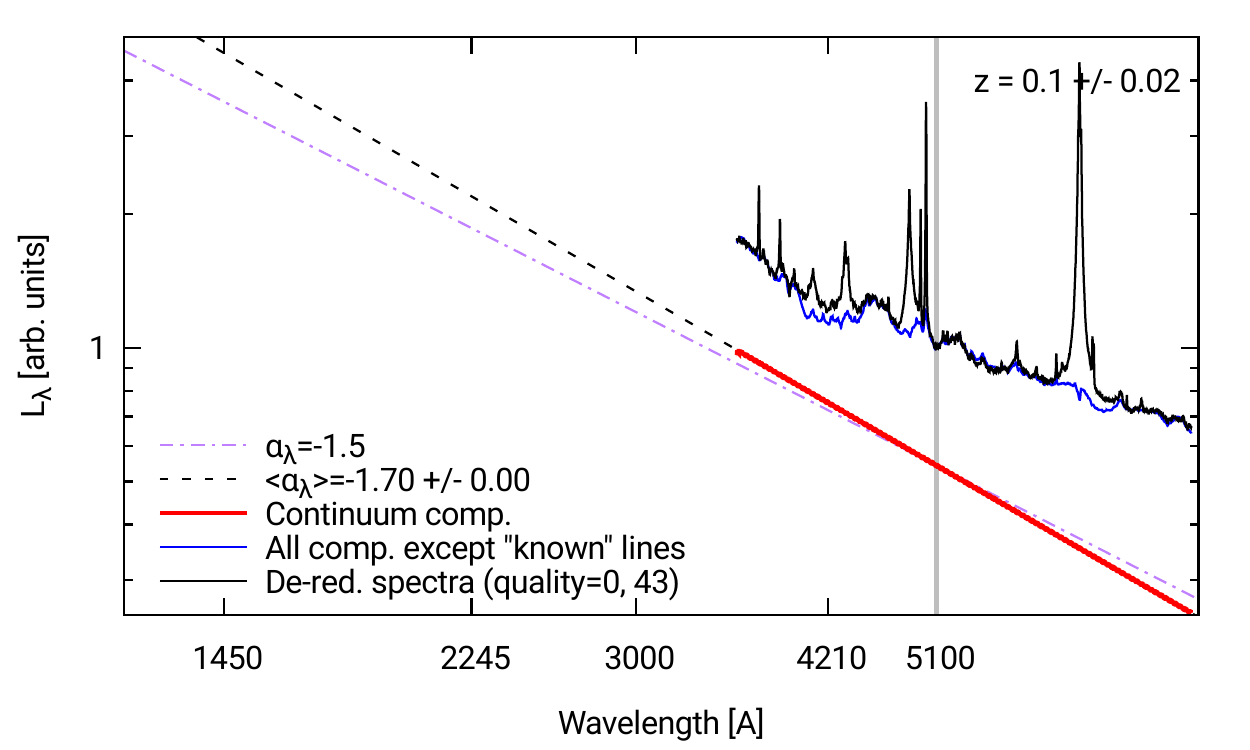}
  \hspace{-0.4cm}
  \includegraphics[width=.48\ww]{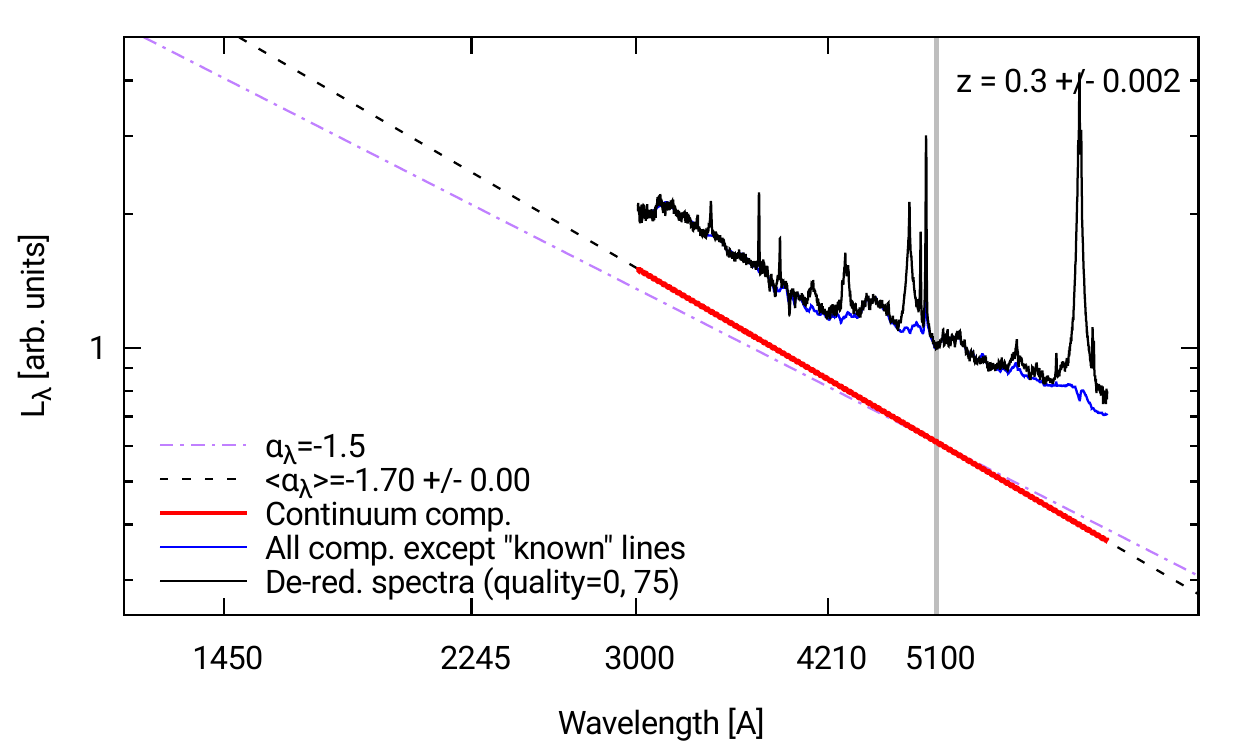}\\
  \includegraphics[width=.48\ww]{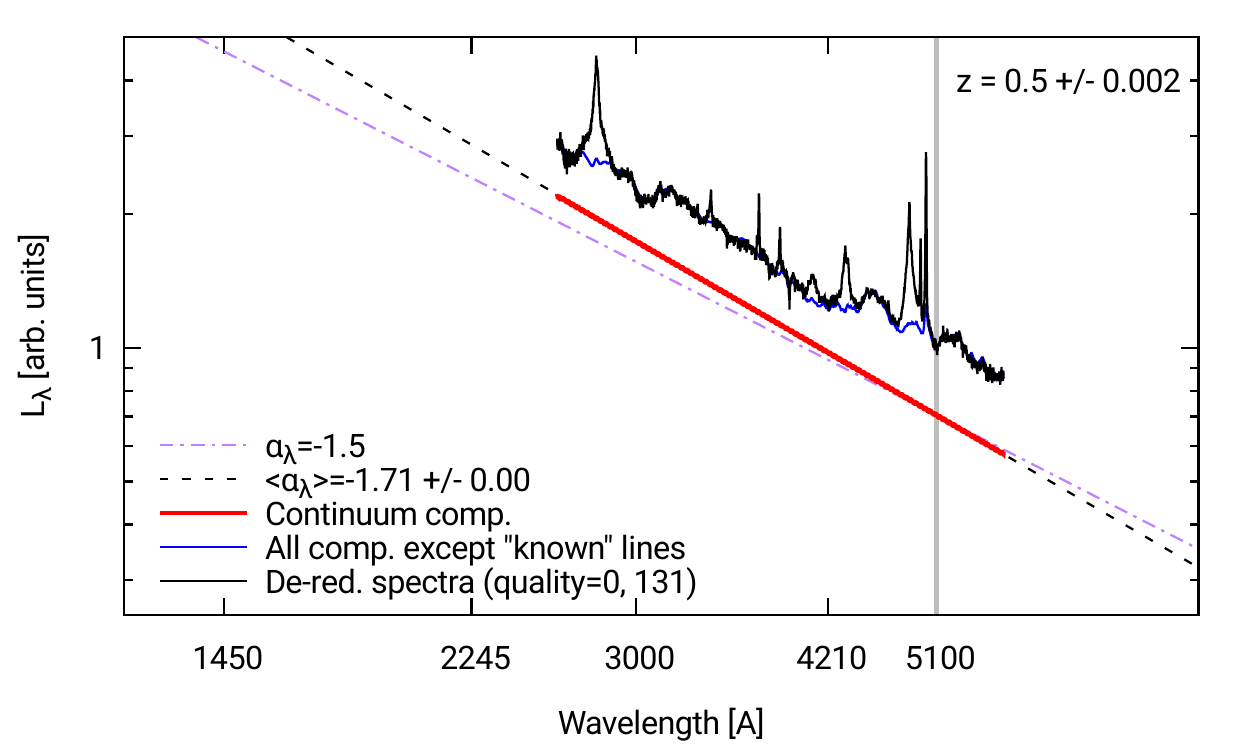}
  \hspace{-0.4cm}
  \includegraphics[width=.48\ww]{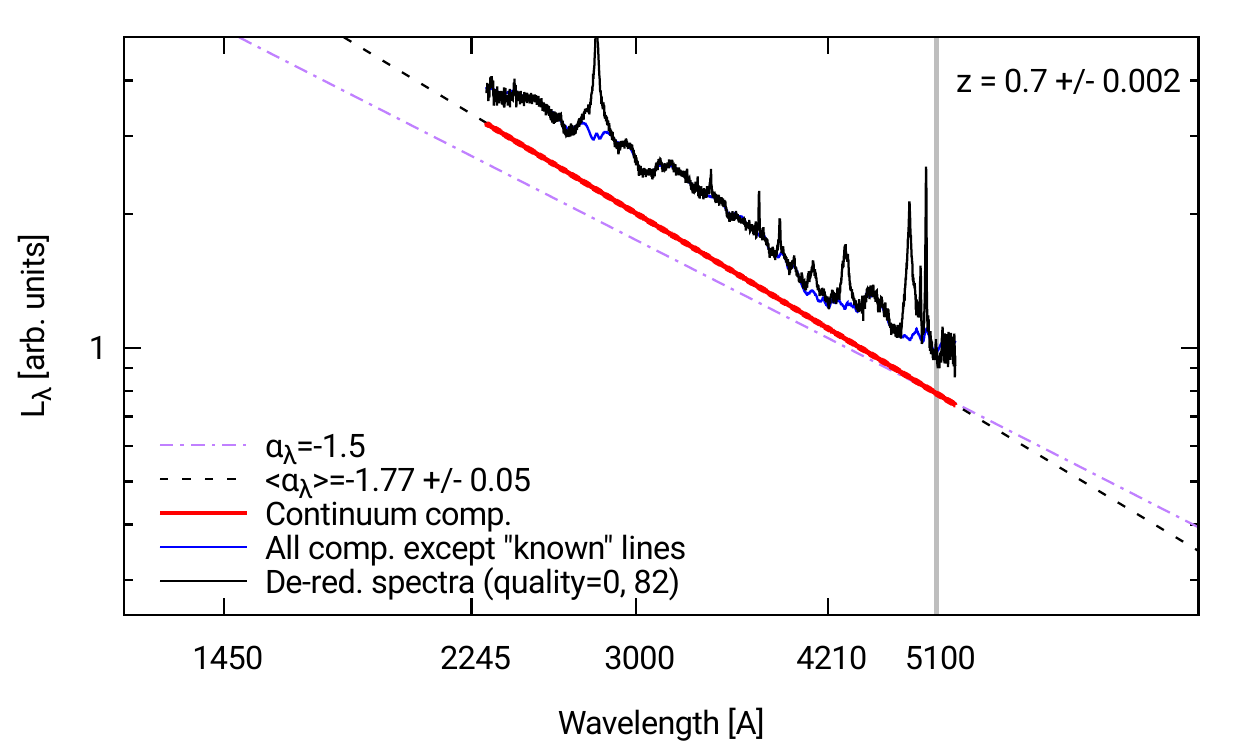}\\
  \caption{Comparison of SDSS de--reddened spectra and \qsfit{}
    components composites in very narrow redshift bins (shown in the
    upper right corner of each plot).  All composite spectra are
    calculated as the geometrical average of the individual spectra in
    the subsample.  Each spectrum is normalized by the source
    luminosity at wavelength $\lambda_{\rm norm} = 5100$\AA{}
    (vertical grey line).  The composite spectrum of SDSS de--reddened
    spectra for all sources in the redshift bin whose continuum
    quality flag at $\lambda_{\rm norm}$ is 0 ({\tt
      CONT5100\_\_QUALITY}, \S\ref{sec:flattened-struc}) is shown with
    a black solid line.  The number of sources in this subsample is
    shown in the legend.  Also shown are the composite spectrum of the
    \qsfit{} continuum component alone (red line) and the composite of
    the sum of all the \qsfit{} components except the ``known''
    emission lines (blue line).  The average \qsfit{} slope at
    $\lambda_{\rm norm}$ is shown with a black dashed line, and the
    numerical value is shown in the legend along with the associated
    standard deviation of the mean.  Finally the purple dot--dashed
    line identifies the reference slope $\alpha_\lambda=-1.5$.}
  \label{fig:stack5100}
\end{figure*}

\begin{figure*}
  \includegraphics[width=.48\ww]{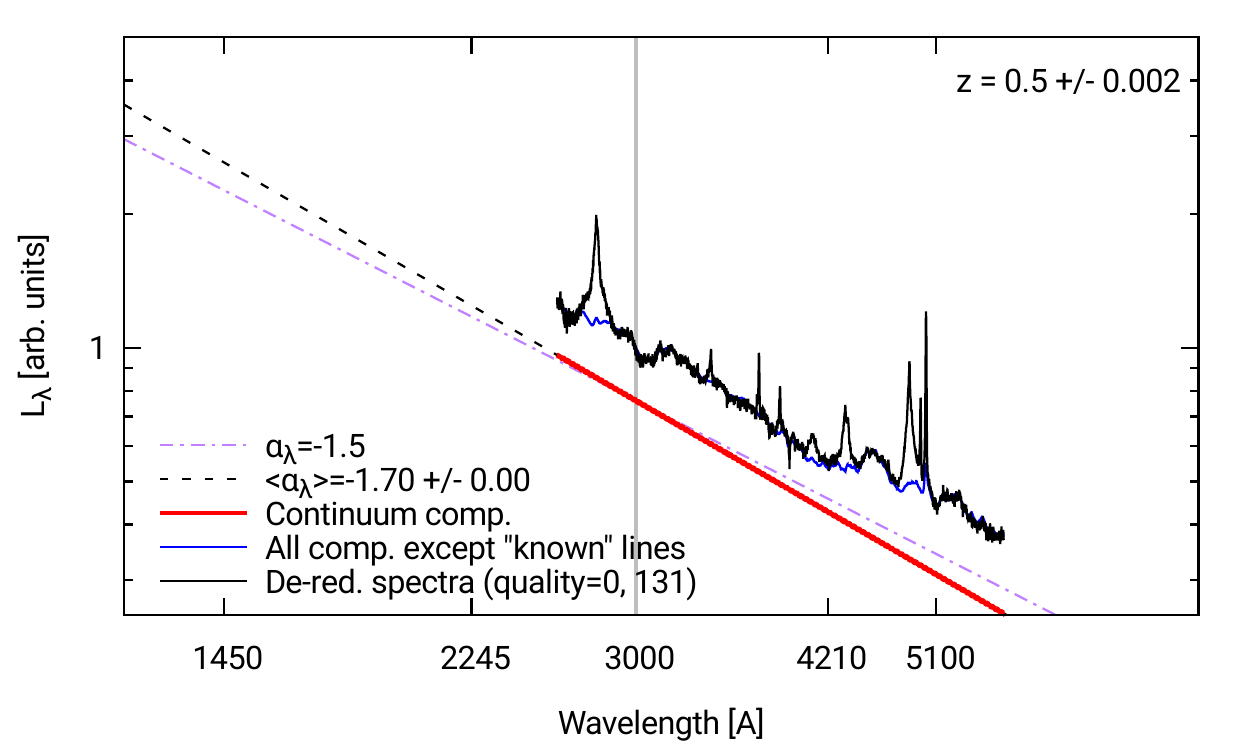}
  \hspace{-0.4cm}
  \includegraphics[width=.48\ww]{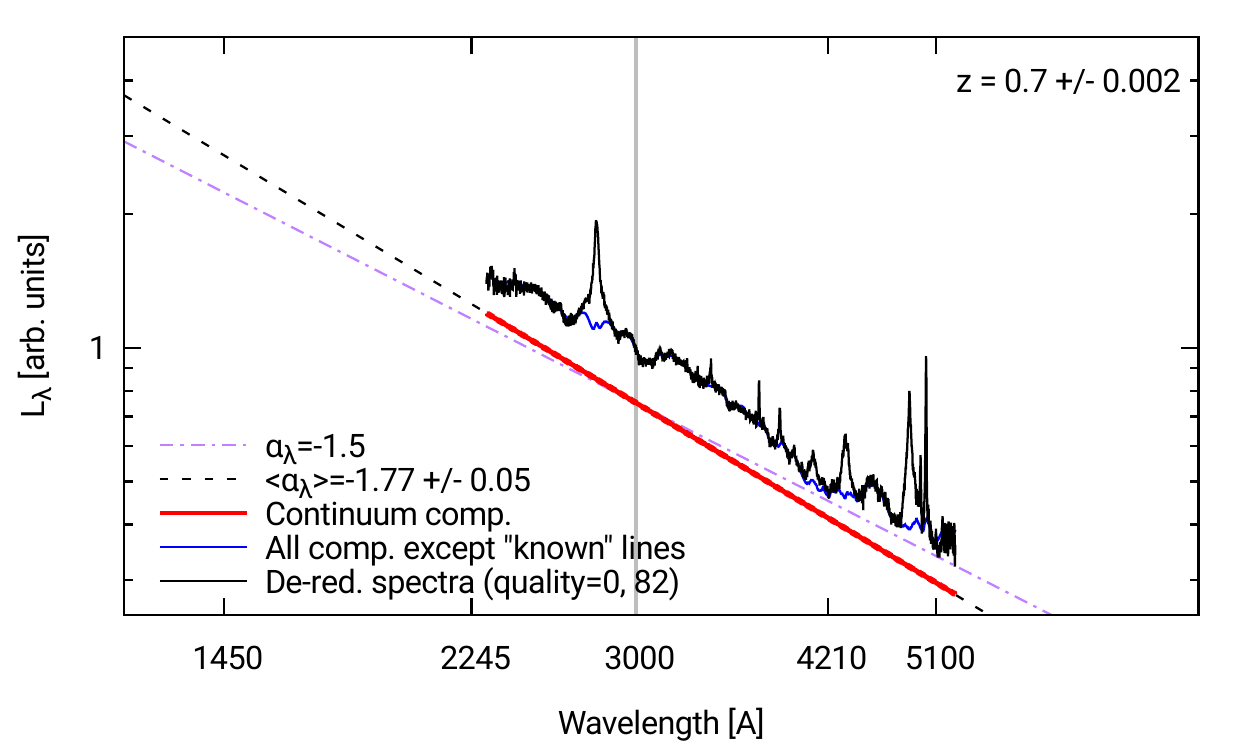}\\
  \includegraphics[width=.48\ww]{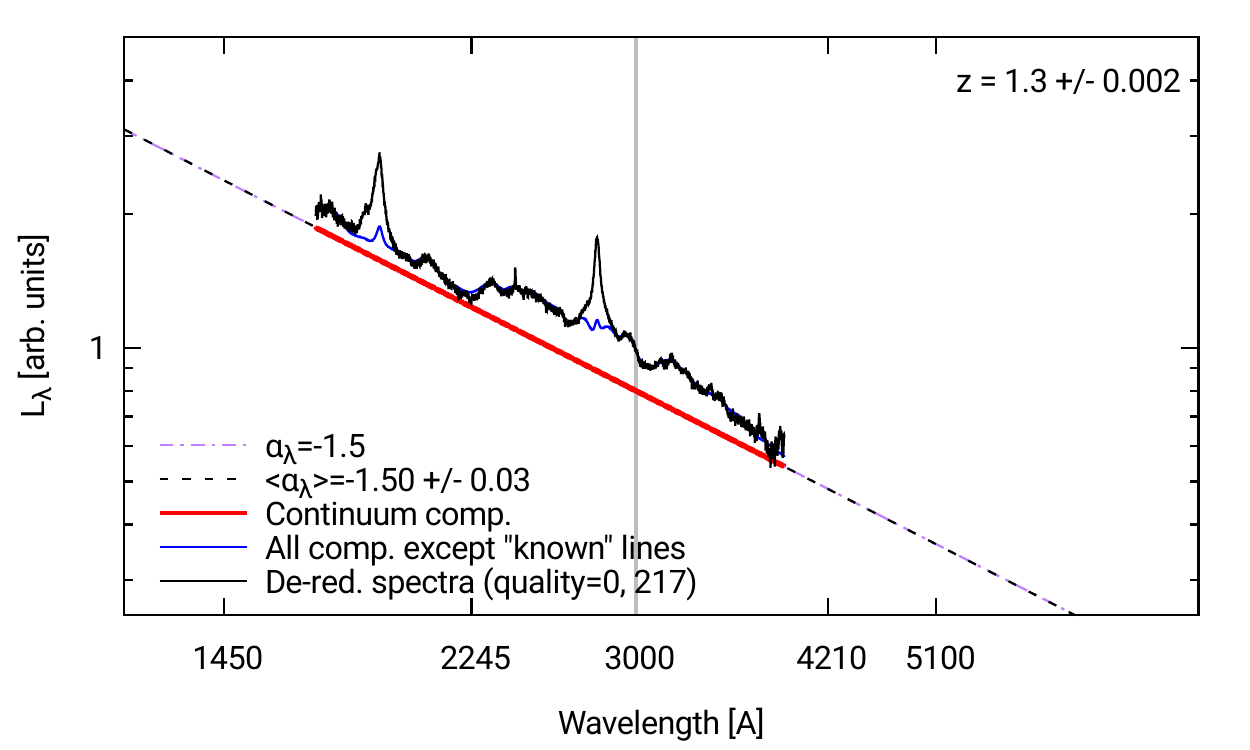}
  \hspace{-0.4cm}
  \includegraphics[width=.48\ww]{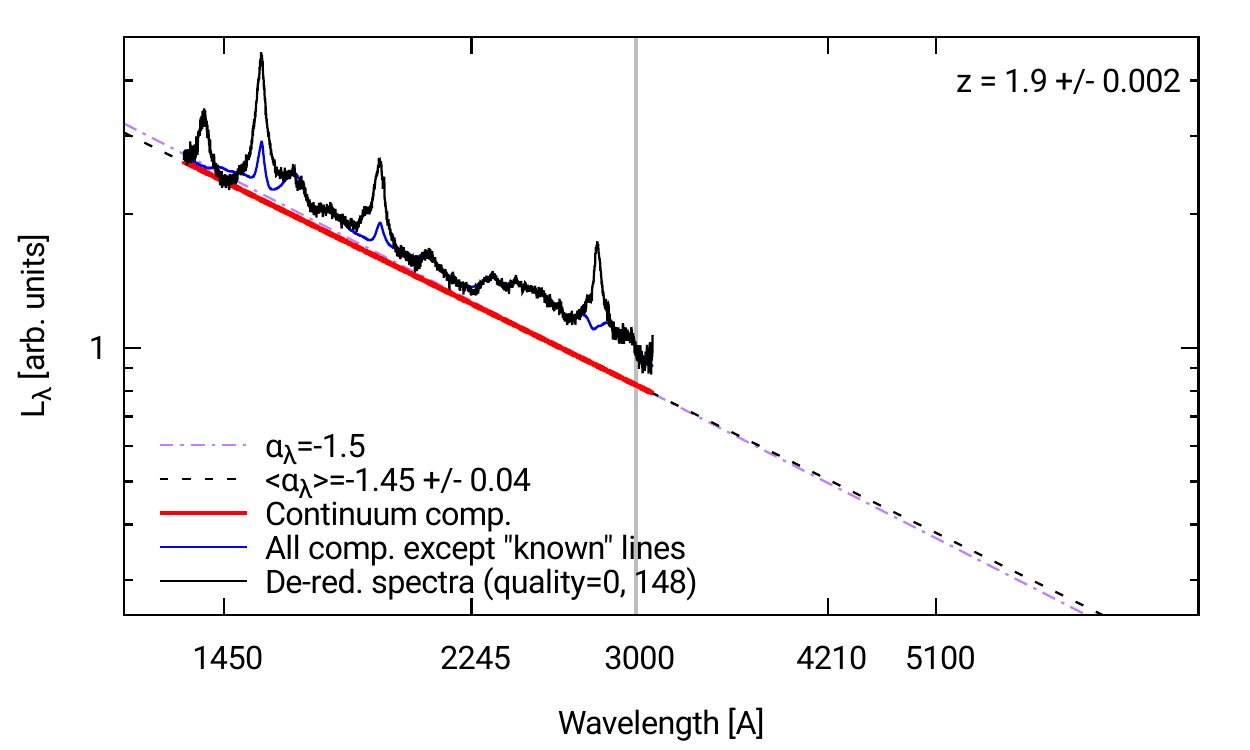}\\
  \caption{Comparison of SDSS de--reddened spectra and \qsfit{}
    components composites in very narrow redshift bins (shown in the
    upper right corner of each plot).  All composite spectra are
    calculated as the geometrical average of the individual spectra in
    the subsample.  Each spectrum is normalized by the source
    luminosity at wavelength $\lambda_{\rm norm} = 3000$\AA{}
    (vertical grey line).  The composite spectrum of SDSS de--reddened
    spectra for all sources in the redshift bin whose continuum
    quality flag at $\lambda_{\rm norm}$ is 0 ({\tt
      CONT3000\_\_QUALITY}, \S\ref{sec:flattened-struc}) is shown with
    a black solid line.  The number of sources in this subsample is
    shown in the legend.  Also shown are the composite spectrum of the
    \qsfit{} continuum component alone (red line) and the composite of
    the sum of all the \qsfit{} components except the ``known''
    emission lines (blue line).  The average \qsfit{} slope at
    $\lambda_{\rm norm}$ is shown with a black dashed line, and the
    numerical value is shown in the legend along with the associated
    standard deviation of the mean.  Finally the purple dot--dashed
    line identifies the reference slope $\alpha_\lambda=-1.5$.}
  \label{fig:stack3000}
\end{figure*}

In order to assess the reliability of \qsfit{} continuum estimates we
compared the SDSS de--reddened averaged (or composite) spectrum with
the average spectrum of the \qsfit{} continuum components in a few
subsamples spanning a very narrow redshift range.  Each composite
spectrum is calculated as the geometrical average of all spectra in
the subsample, in order to preserve the slopes in a log--log plot.
Before calculating the average we normalize each de--reddened spectrum
and the associated \qsfit{} components by the source luminosity at
wavelength $\lambda_{\rm norm}$.  Fig.~\ref{fig:stack5100} shows such
a comparison for the sources in the redshift bins $z=0.1 \pm 0.02$,
$z=0.3 \pm 0.002$, $z=0.5 \pm 0.002$ and $z=0.7 \pm 0.002$
respectively, using $\lambda_{\rm norm} = 5100$\AA{} (vertical grey
line).  The redshift bins are shown in the upper right corner.  The
bin at $z=0.1$ is larger in order to contain a sufficient number of
sources.
The composite spectrum of the sources in the redshift bin whose
continuum quality flag at $\lambda_{\rm norm}$ is 0 ({\tt
  CONT5100\_\_QUALITY}, \S\ref{sec:flattened-struc}) is shown with a
black solid line.  The number of sources in this subsample is shown in
the legend.  Also shown are the composite spectrum of the \qsfit{}
continuum component alone (shown with a red line) and the composite
spectrum obtained by summing all the \qsfit{} components except the
``known'' emission lines (blue line).  The latter represents the
``floor'' below the emission lines, i.e. the sum of the AGN, host
galaxy, iron complex and Balmer continuum contributions as estimated
by \qsfit{}.  The average \qsfit{} slope at $\lambda_{\rm norm}$ is
shown with a (black dashed line), and the numerical value is shown in
the legend along with the associated standard deviation of the mean.
Finally the purple dot--dashed line identifies the reference slope
$\alpha_\lambda=-1.5$, to be discussed below.  A similar comparison is
shown in Fig.~\ref{fig:stack3000} for the sources at $z=0.5 \pm
0.002$, $z=0.7 \pm 0.002$, $z=1.3 \pm 0.002$ and $z=1.9 \pm 0.002$
respectively, using $\lambda_{\rm norm} = 3000$\AA{}.

In all plots the blue line follows accurately the black
line.\footnote{Such agreement between the blue and black line is best
  visible when the arithmetic average is used, like in
  Fig.~\ref{fig:arit5100_03}, in place of the geometric one.  However,
  in this section the focus is on the reliability of the slopes hence
  we decided to show the geometric composites only.  The plots with
  both geometric and arithmetic composites in all the redshift bins
  between 0.1 and 2 are available in the \qsfit{} webpage.}  Hence
\qsfit{} is able to find the appropriate continuum below the main
emission lines, and the host galaxy, Balmer continuum and iron
templates appear adequate to fit the SDSS spectra.  Having considered
such components in the above discussion, the remaining contribution is
what we assumed to be the AGN continuum (\S\ref{sec:comp-continuum}),
whose composite spectrum is shown with the red line, and whose average
slope is shown with a black dashed line.

\subsection{\qsfitcat{} catalog main results}
\label{sec:mainResults}

Here we will discuss two of the most important results of the \qsfit{}
analysis, namely the lack of appreciable redshift evolution of the AGN
continuum slope and of the Balmer continuum luminosity (when
normalized with continuum luminosity).

\begin{figure}
  \includegraphics[width=.48\ww]{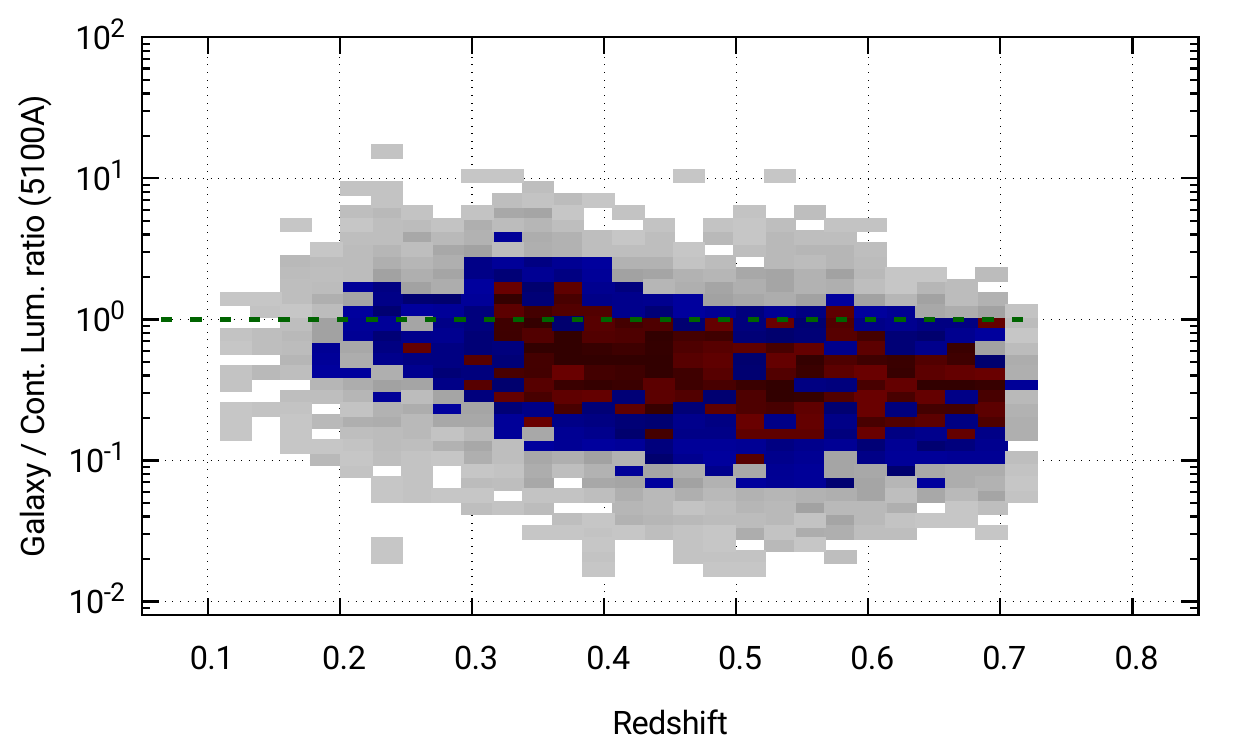}\\
  \caption{Ratio of galaxy luminosity over continuum luminosity, both
    estimated at 5100\AA{}, as a function of redshift. At $z\lesssim
    0.3$ almost half of the sources have ratio $\sim$~1.  At $z
    \gtrsim 0.5$ the galaxy contribution is typically smaller than the
    continuum one.  The galaxy contribution is fitted up to $z = 0.8$
    but this plot stops at smaller redshifts since the 5100\AA{} is
    available only up to $z \sim 0.7$.}
  \label{fig:galaxyContRatio}
\end{figure}

In Fig.~\ref{fig:galaxyContRatio} we show the ratio of galaxy
luminosity over continuum luminosity, both estimated at 5100\AA{}, as
a function of redshift.  At redshifts $\lesssim 0.3$ the AGN continuum
component contribution is smaller than, or comparable with, the
emission from the galaxy (Fig. \ref{fig:galaxyContRatio}).
This explains why our continuum luminosity estimates are smaller than
the S11 ones (see Fig.~\ref{fig:cmp_contlum}) at low
luminosities. However, for these sources we can only indirectly probe
the AGN continuum, and our estimates rely on the assumptions that the
host galaxy template and the fixed continuum slope ($\alpha_{\lambda}
= -1.7$ for the sources with $z \le 0.6$, \S\ref{sec:comp-continuum})
are suitable for the source being analyzed.  Hence, the \qsfit{}
continuum properties (especially the slopes) for sources with $z
\lesssim 0.3$ may be biased by our assumptions.  At redshifts $ z
\gtrsim 0.5$ the galaxy contribution become gradually less important
and the AGN continuum can in principle be directly probed.  However,
the presence of a very broad emission feature between $\sim$2200\AA{}
and $\sim$4000\AA{} known as ``small blue bump'' \citep[SBB, due to
  the Balmer continuum and many blended iron lines,
][]{1985-Wills-SBB}, clearly visible in the lower left panel of
Fig.~\ref{fig:stack3000}, may introduce a bias in the \qsfit{} slope
estimates.  The bias is due to the fact that at $z \lesssim 0.8$ the
small blue bump is not entirely visible using just the data at optical
wavelengths, and consequently the AGN continuum component is ``pulled
upward'' at short wavelengths to fit the regions close to the \mgii{}
emission line.  The slope $\alpha_\lambda$ would be, therefore,
systematically underestimated.  As a consequence we had to fix the
continuum slope $\alpha_{\lambda} = -1.7$ for the sources with $z \le
0.6$.  In order to remove this limit additional data at UV wavelengths
are required.

At redshifts $z \gtrsim 0.6$ the small blue bump is almost always
entirely visible (see Fig.~\ref{fig:stack3000}, upper right panel) and
the host galaxy is typically negligible, hence the \qsfit{} slope
estimates are not affected by the aforementioned bias, and the slope
parameter is free to vary in the fit.

\begin{figure}
  \includegraphics[width=.45\ww]{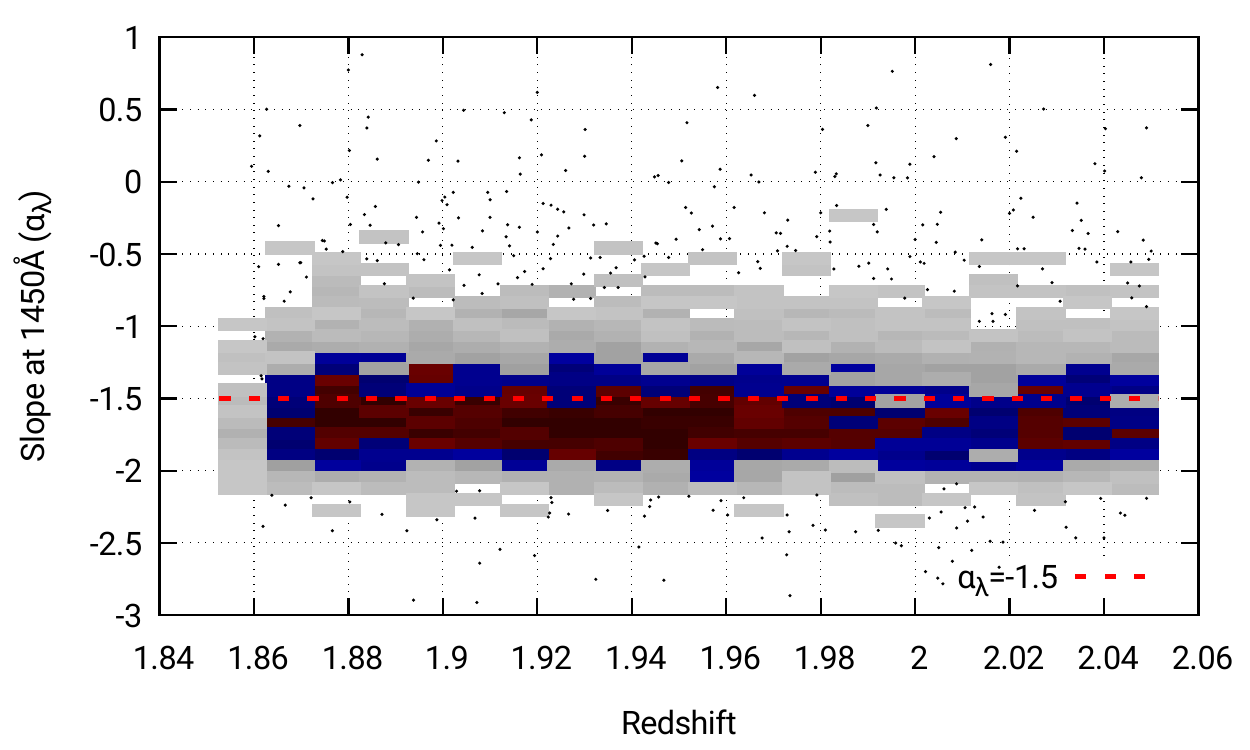}\\
  \hspace{-0.5cm}
  \includegraphics[width=.45\ww]{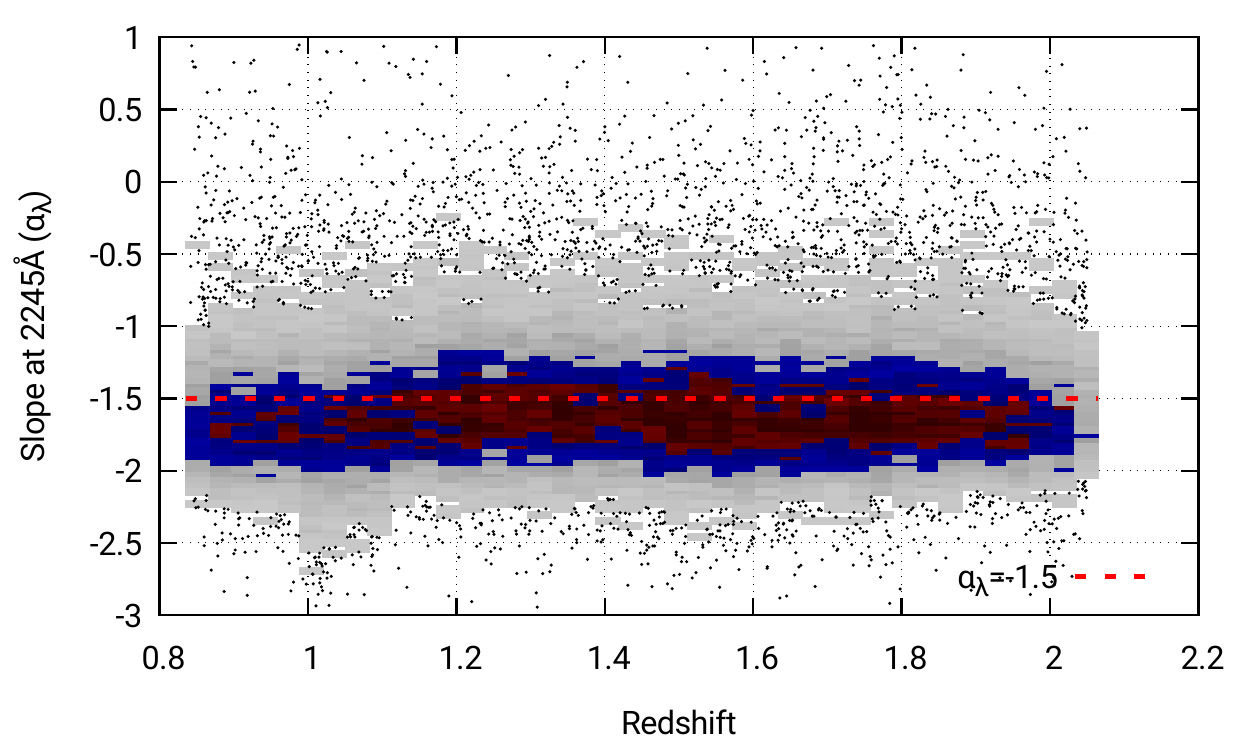}\\
  \hspace{-0.5cm}
  \includegraphics[width=.45\ww]{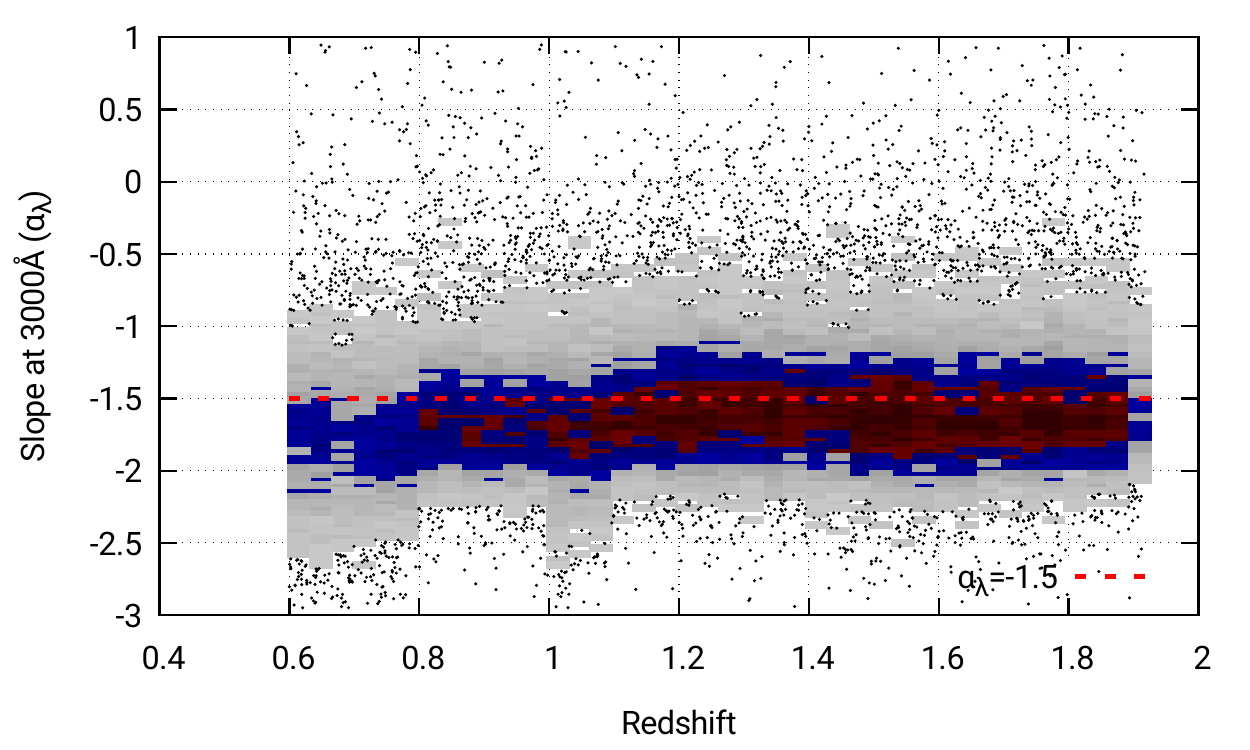}\\
  \hspace{-0.5cm}
  \includegraphics[width=.45\ww]{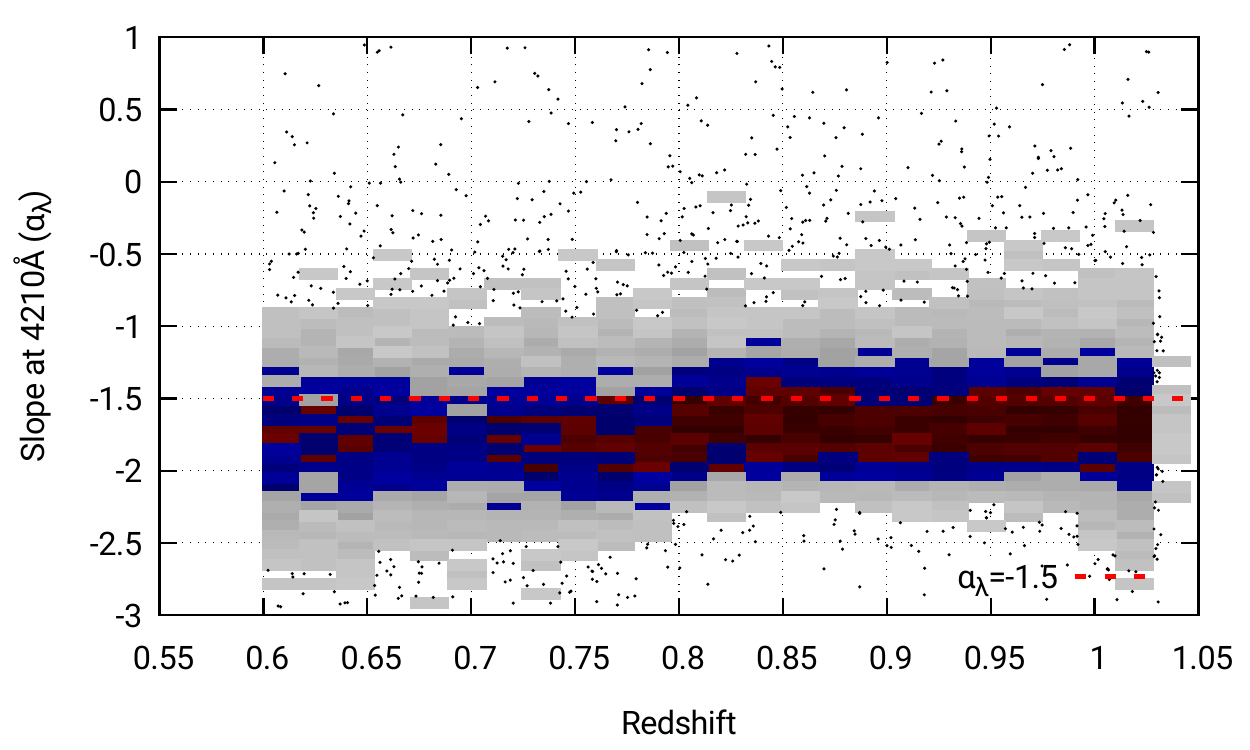}\\
  \hspace{-0.5cm}
  \includegraphics[width=.45\ww]{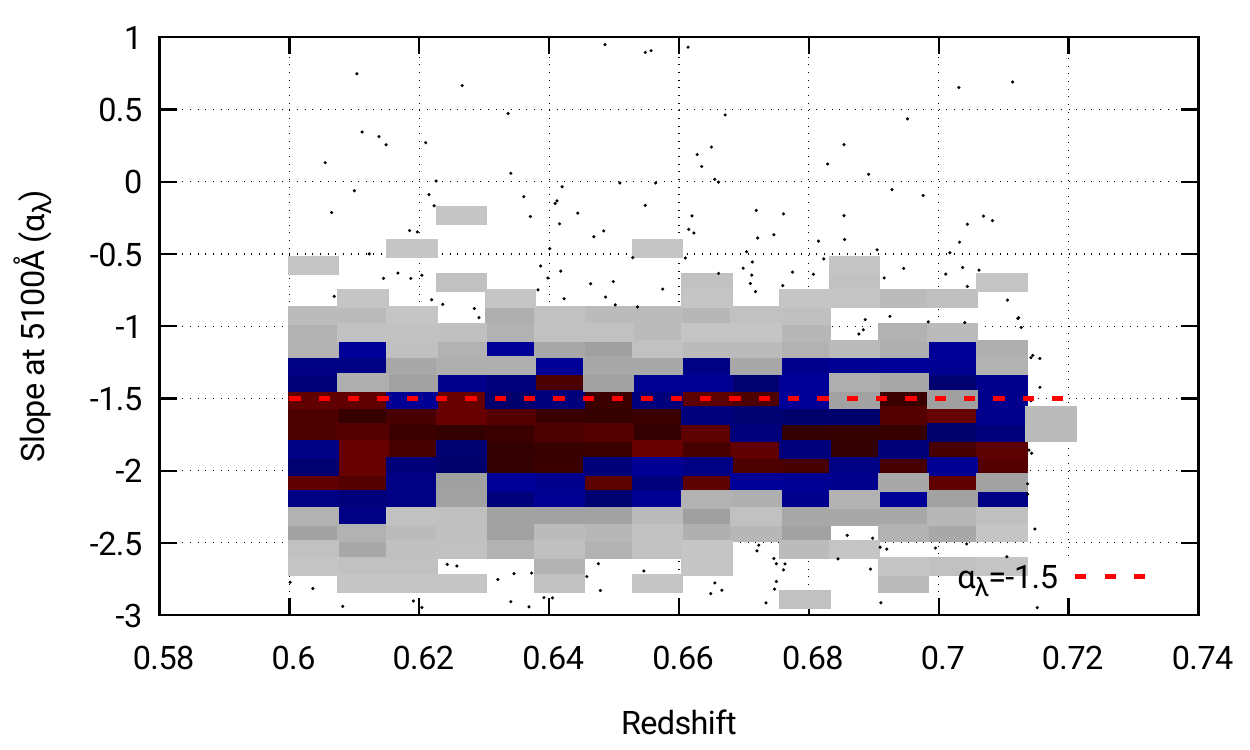}\\
  \hspace{-0.5cm}
  \caption{2--D histogram of the broad--band continuum slopes
    vs. redshift at the fixed rest frame wavelengths 1450\AA{},
    2245\AA{}, 3000\AA{}, 4210\AA{} and 5100\AA{} (the meaning of the
    2--D histogram color code is described in \S\ref{sec:catalog}).  The
    dashed red lines is at $\alpha_\lambda = -1.5$.}
  \label{fig:slope_z}
\end{figure}

\begin{figure}
  \includegraphics[width=.48\ww]{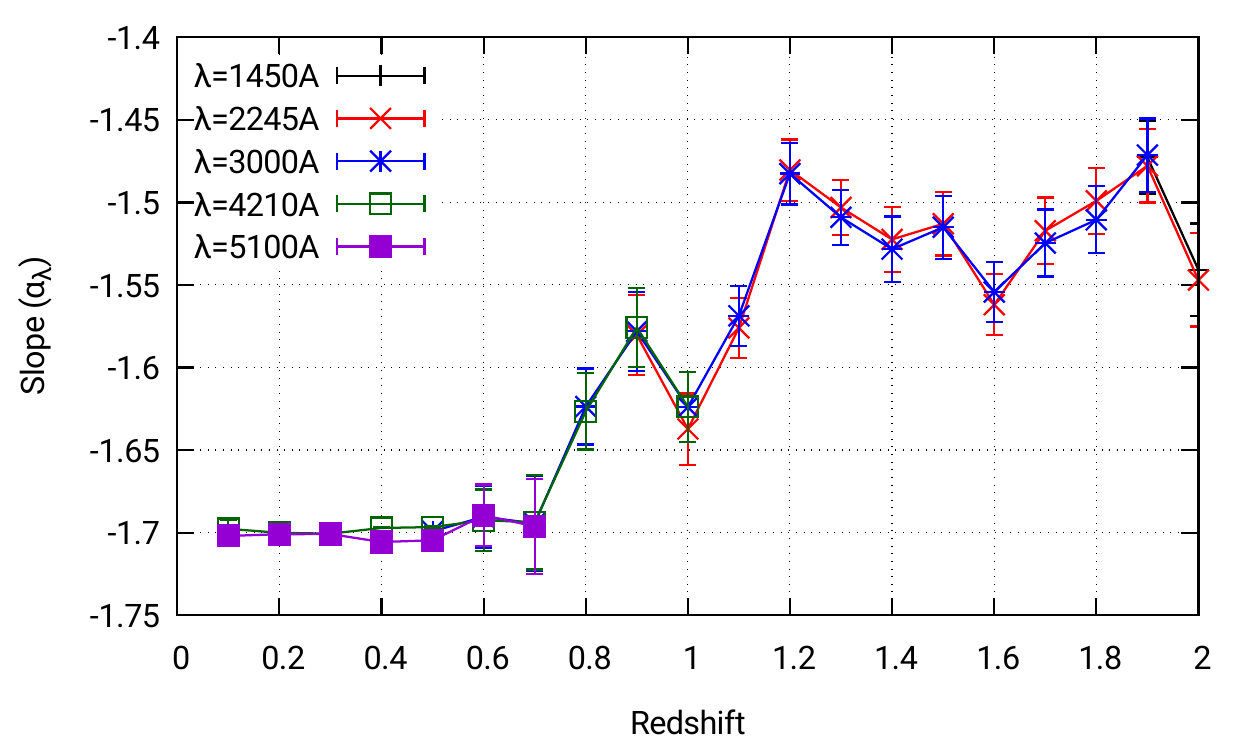}\\
  \caption{AGN continuum slopes as a function of redshift.  The slopes
    at $z\le 0.6$ at fixed at $\alpha_{\lambda} = -1.7$}
  \label{fig:slopeavg_z}
\end{figure}

The 2--D histograms of slope vs. redshift for all sources whose
\verb|CONT*__QUALITY| is 0 (\S\ref{sec:flattened-struc}) at the fixed
rest frame wavelengths 1450\AA{}, 2245\AA{}, 3000\AA{}, 4210\AA{} and
5100\AA{} is shown in Fig.~\ref{fig:slope_z} (the meaning of the 2--D
histogram color code is described in \S\ref{sec:catalog}).  The dashed
red lines is at $\alpha_\lambda = -1.5$.
Fig.~\ref{fig:slopeavg_z} shows the same data, but considering only
the sources in very narrow redshift bins ($\Delta z < 0.005$).

\begin{figure}
  \includegraphics[width=.48\ww]{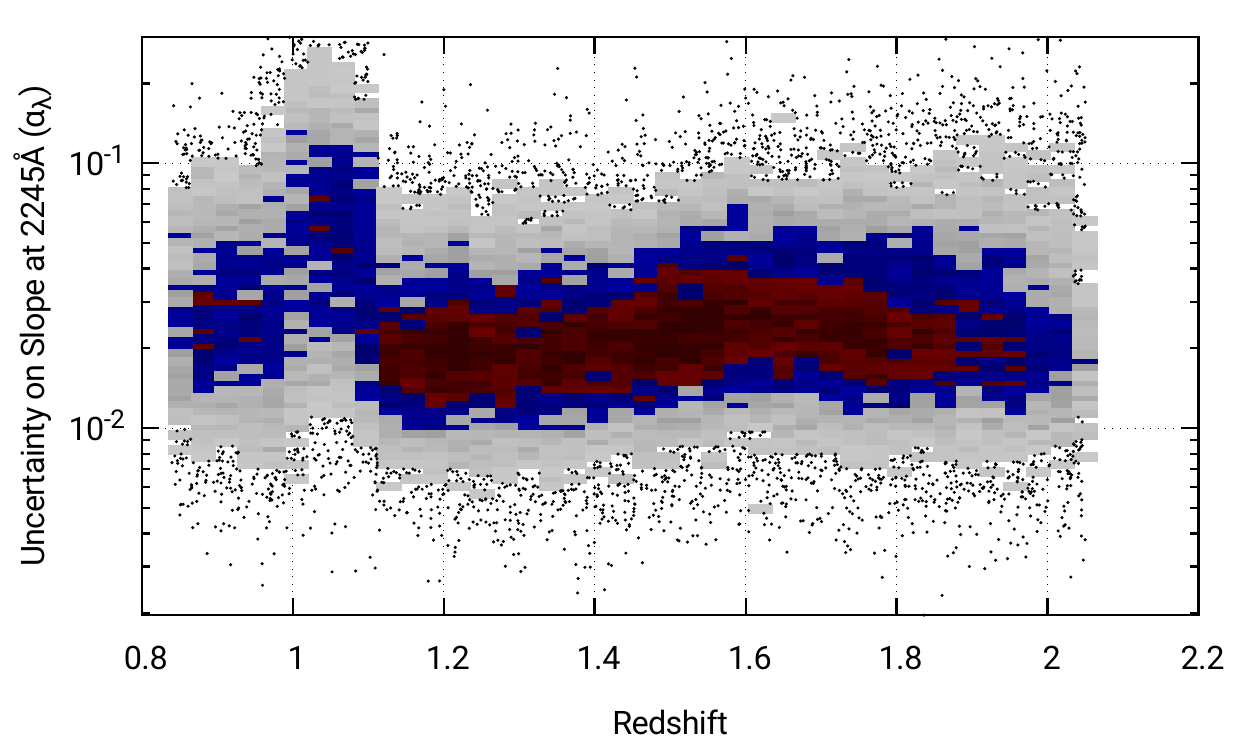}\\
  \caption{2--D histogram of the $1\sigma$ uncertainties in the
    $\alpha_\lambda(2245{\rm \AA{}})$ slope estimates vs. redshift.}
  \label{fig:uncSlope}
\end{figure}

The slopes span the very narrow range between $-$1.45 and $-$1.65 at
all redshifts between 0.8 and 2, providing remarkable evidence for the
similarity of continuum slopes across cosmic time.  This is in
agreement with previous works: the typical value of the spectral slope
at optical/UV wavelengths is $\alpha_\lambda \sim -1.5$
\citep[e.g.][]{2006-richards-meanSED, 2001-vanden-composite,
  2016-Shankar-OptUVEmissivityQSO}, however our analysis allows for
the first time to estimate the slopes on a very large sample, using a
completely automatic procedure and just the SDSS data at optical
wavelengths.  In particular, our estimates are similar to those of
\citet{2007-Davis-UVContinuum-ModelsSlopes}, who found $\alpha_\lambda
= -1.41$ (in the wavelength range 1450--2200\AA{}) and $\alpha_\lambda
= -1.63$ (in the wavelength range 2200--4000\AA{}) with rather broad
distributions (see their Fig.~2).  At $z \gtrsim 1.1$ the slopes are
remarkably constant and show no trend with redshift up to $z \sim 2$.
The scatter is larger than the standard deviation of the mean in the
redshift bins, hence the values are affected by small ($\lesssim 0.1$)
slope systematics due to the fact that at different redshift we
observe different parts of the rest frame spectrum.  Moreover, note
that the typical uncertainty associated to our estimates of
$\alpha_\lambda(2245{\rm \AA{}})$ ranges between 0.02 and 0.1
(Fig.~\ref{fig:uncSlope}), while the widths in the distributions of
Fig.~\ref{fig:slope_z} are significantly larger ($\sigma=0.4$).  Hence
these distributions are not significantly broadened by the random
uncertainties in the fitting process.  We conclude that the average
continue slope is $\alpha_\lambda\sim -1.55$, but there are AGNs whose
slope is significantly smaller or larger than the average.

At $0.6 \lesssim z \lesssim 1.1$ the slopes are slightly smaller.
Currently we are not able to discriminate whether the SBB bias is the
only cause of such low slope estimates, or rather if the slopes
actually are smaller at $z \lesssim 0.8$ with respect to higher
redshifts.  Note that both the putative galaxy bias and the systematic
SBB bias are not due to limitations of \qsfit{}: they are a
consequence of having considered just the spectra at optical
wavelengths rather than considering data at shorter wavelengths.  By
enlarging the spectral coverage we could get rid of such biases and
provide stronger constraints on the actual continuum shape at
$z\lesssim 0.8$.  On the other hand using just the SDSS data allowed
us to analyze a very large and homogeneous sample.

\begin{figure}
  \includegraphics[width=.48\ww]{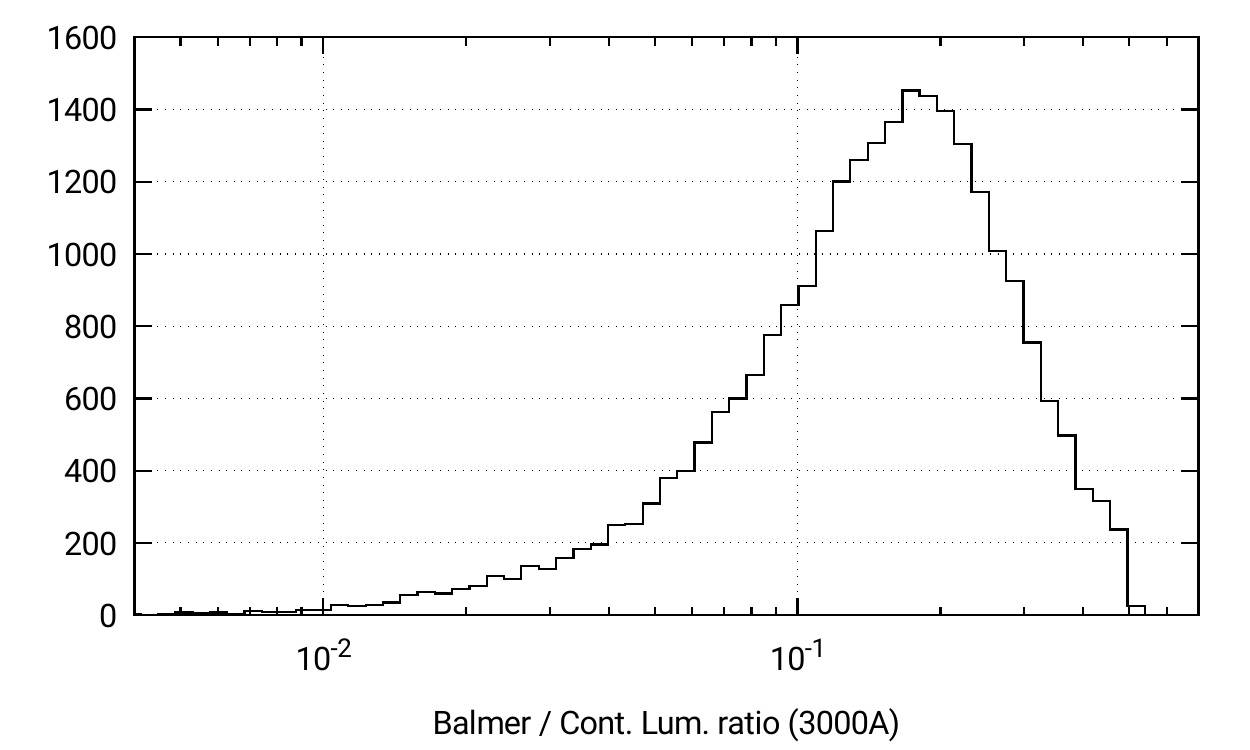}\\
  \includegraphics[width=.48\ww]{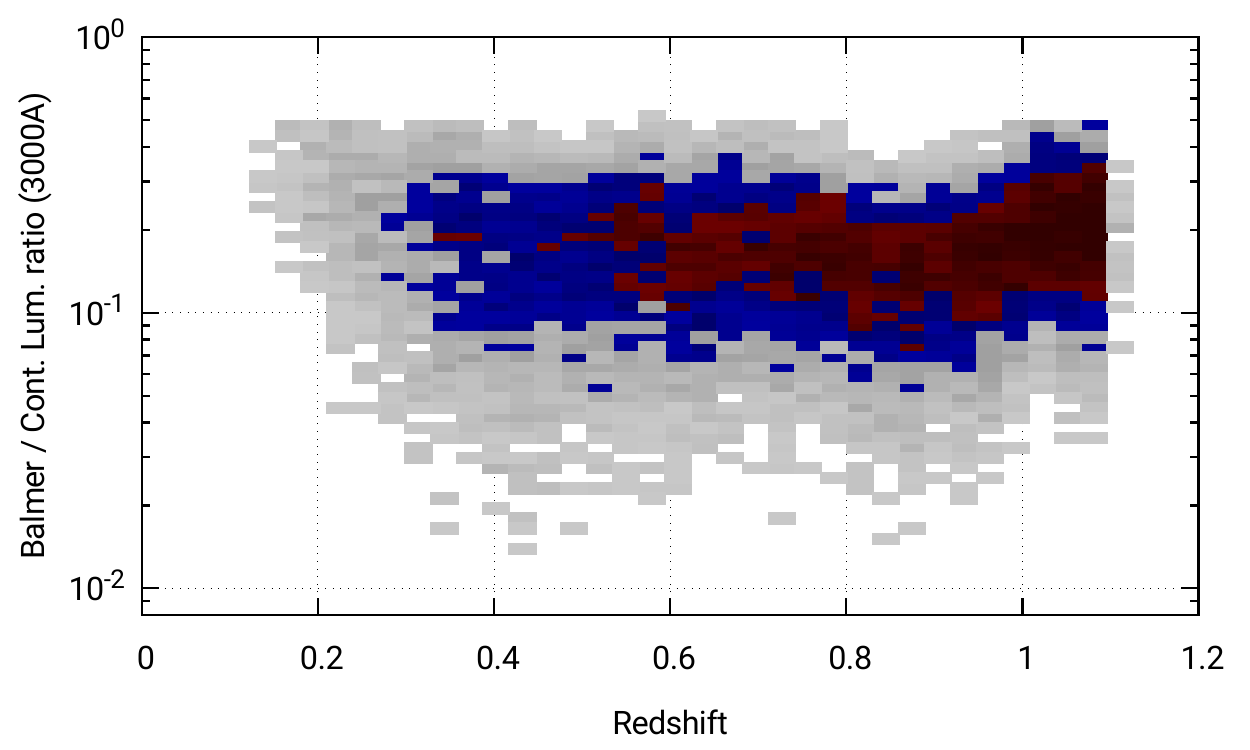}
  \caption{Upper panel: distribution of the Balmer continuum
    normalization parameter (\S\ref{sec:balmer}).  Lower panel: same
    distribution as a function of redshift.  The distribution is quite
    flat at all redshifts and has a median value of 0.15 and a width
    of 0.3 dex.}
  \label{fig:BalmerContRatio}
\end{figure}

Similar conclusions applies to the ratio of the Balmer luminosity over
the AGN continuum luminosity, both estimated at 3000\AA{}.
Fig.~\ref{fig:BalmerContRatio} shows distribution of the Balmer
continuum normalization parameter (\S\ref{sec:balmer}) in the upper
panel, and the distribution as a function of redshift in the lower
panel.  The distribution is quite flat at all redshifts and has a
median value of 0.15 and a width of 0.3 dex.

\section{Conclusions}
\label{sec:conclusion}

We presented \qsfit{} (version \version{}), an IDL package able to
automatically analyze the SDSS optical spectra of Type 1 AGNs.
\qsfit{} is released as free software, and can be customized to
analyze the spectra from other sources, or to customize the fitting
recipe.  \qsfit{} provides luminosity estimates for the AGN continuum,
the Balmer continuum, both optical and UV iron blended complex, host
galaxy and emission lines, as well as width, velocity offset and
equivalent width of 20 emission lines, as well as several ``quality
flags'' to assess the reliability of the results.

We used \qsfit{} to analyze \nn{} SDSS spectra of Type 1 AGN at $z<2$
and to produce a publicly available catalog of AGN spectral properties
(available on the \qsfit{} website, along with additional data and
plots).  Also, we compared our results with those from the S11 catalog
\citep{2011-shen-catdr7}.  The quantities are in very good agreement,
except for: (i) the continuum luminosity at 5100\AA{} for sources at
low redshift, where we also considered the host galaxy contribution;
(ii) the emission line widths whose values depend on the adopted
fitting algorithm for line decomposition.

We also checked the reliability of the \qsfit{} continuum slope
estimates by a comparison with composite spectra of the observed SDSS
data, and found that \qsfit{} performs particularly well at $z \gtrsim
0.8$, providing (for the first time) the AGN continuum slope estimates
on a very large sample, using just the data at optical wavelengths and
a completely automatic procedure.  Also, we have shown that the
continuum slope and the Balmer continuum luminosity distributions do
not show any evident trend with redshift.

The release of the \qsfit{} code and the associated catalog discussed
in this work are the first attempt towards a new approach in analyzing
AGN spectral data, where all interested astronomers are given the
possibility to study and compare the analysis methods, easily
replicate the data analysis (even on very large samples), and possibly
improve the recipes in a collaborative effort.

\bigskip
\noindent
{\bf ACKNOWLEDGEMENTS}\\

\smallskip

\noindent
We gratefully acknowledge our referee, Yue Shen, for useful suggestions
and comments that greatly improved the presentation of this work.

We acknowledge M. Vestergaard and B.J. Wilkes for having provided the
UV iron templates, and M.-P. V\'eron-Cetty, M. Joly and P. V\'eron for
the optical iron templates used in the fitting process.

We warmly thank a number of undergraduate students at the University
of Southampton who have widely tested QSFit on thousands of different
sources and under different input assumptions.

We also thank Chiara Moretti for her contribution to the development
of the \qsfit{} website, and Guido Cupani for insightful discussions
on fitting algorithms.

Funding for the SDSS and SDSS-II has been provided by the Alfred
P. Sloan Foundation, the Participating Institutions, the National
Science Foundation, the U.S. Department of Energy, the National
Aeronautics and Space Administration, the Japanese Monbukagakusho, the
Max Planck Society, and the Higher Education Funding Council for
England. The SDSS Web Site is \url{http://www.sdss.org/}.

The SDSS is managed by the Astrophysical Research Consortium for the
Participating Institutions. The Participating Institutions are the
American Museum of Natural History, Astrophysical Institute Potsdam,
University of Basel, University of Cambridge, Case Western Reserve
University, University of Chicago, Drexel University, Fermilab, the
Institute for Advanced Study, the Japan Participation Group, Johns
Hopkins University, the Joint Institute for Nuclear Astrophysics, the
Kavli Institute for Particle Astrophysics and Cosmology, the Korean
Scientist Group, the Chinese Academy of Sciences (LAMOST), Los Alamos
National Laboratory, the Max-Planck-Institute for Astronomy (MPIA),
the Max-Planck-Institute for Astrophysics (MPA), New Mexico State
University, Ohio State University, University of Pittsburgh,
University of Portsmouth, Princeton University, the United States
Naval Observatory, and the University of Washington.

\appendix

\section{Description of the \qsfitcat{} catalog}
\label{sec:flattened-struc}

The \qsfitcat {} (version \version{}) catalog is deployed as a FITS
table with 279 columns and \nn{} rows, available at the following
address:
\begin{center}
  {\url{http://qsfit.inaf.it/cat_1.0/fits/qsfit_1.0.fits}}
\end{center}
On the \qsfit{} website you will also find a second version of the
catalog where we added, for each analyzed source, the corresponding
data from the S11 catalog to allow an easy comparison.  This latter
FITS file has 434 columns.

In this section we will list the name and meaning of each column in
the \qsfitcat{} catalog.  The columns whose name has a
\verb|__QUALITY| suffix require a preliminary explanation: they store
the ``quality flags'', whose purpose is to characterize the
reliability of the associated quantity.  If a quality flag is raised
for a given source then the corresponding quantity should be used
cautiously, and a further analysis is typically required.  To save
space in the final catalog these flags are coded as ``bits'',
i.e. quantities which can be either 0 (flag not raised) or 1 (flag
raised), and packed into a binary representation of an integer number
(the bitmask).  For instance, a quality flag equal to 13 means that
the bits 0, 2, 3 are raised, since $2^0 + 2^2 + 2^3 = 13$.  A quality
flag equal to 0 means that no flag has been raised and the associated
quantity is considered reliable (within the capabilities of \qsfit{}).
The meaning of each bit in the quality flag is discussed below.

The list of columns provided in the \qsfitcat{} catalog is:
\begin{itemize}
\item \verb|SPEC|: name of the SDSS--DR10 FITS file containing the
  spectrum;
 
\item \verb|SDSS_NAME|: the SDSS name of the source (J2000.0);

\item \verb|RA|: the right ascension (J2000.0) of the source;
\item \verb|DEC|: the declination (J2000.0) of the source;
\item \verb|REDSHIFT|: the redshift of the source
\item \verb|PLATE|: spectroscopic plate number;
\item \verb|FIBER|: spectroscopic fiber number;
\item \verb|MJD|: Modified Julian Day of spectroscopic observation;
\item \verb|E_BV|: the color excess used to de--redden the SDSS
  spectrum;

\item \verb|NDATA|: the number of channels in the spectrum used in the
  analysis (\S\ref{sec:prep});

\item \verb|GOOD_FRACTION|: the fraction of ``good'' channels (whose SDSS
  mask is 0) over the total number of channels.  This quantity is
  $\ge$~0.75 for all sources analyzed in this work (\S\ref{sec:prep});

\item \verb|MEDIAN_FLUX, MEDIAN_ERR|: the median flux and median
  uncertainty in the input spectrum, in units of 10$^{-17}$ erg
  s$^{-1}$ cm$^{-2}$ \AA$^{-1}$;

\item \verb|GALAXY__LUM, GALAXY__LUM_ERR|: the host galaxy $\nu
  L_{\nu}$ luminosity at 5500\AA{} and its uncertainty, in units of
  $10^{42}$~erg~s$^{-1}$ (\S\ref{sec:comp-galaxy});

\item \verb|GALAXY__QUALITY|: quality flag for the galaxy estimates.
  The meaning of the bits are:
  \begin{itemize}
  \item Bit 0: fit of galaxy template is not sensible at this redshift;
  \item Bit 1: either the luminosity or its uncertainty are NaN or equal to zero;
  \item Bit 2: luminosity relative uncertainty $>$ 1.5;
  \end{itemize}

\item \verb|CONT*__WAVE| where ``\verb|*|'' is a number in the range
  1--5: wavelengths in \AA{} where the continuum luminosity and slopes
  (see below) has been computed. There are five such quantities whose
  values are equally spaced in the logarithmic wavelength range
  [\lmin{}, \lmax{}];

\item \verb|CONT*__LUM|, \verb|CONT*__LUM_ERR| where ``\verb|*|'' is
  in the range 1--5, or equal to 1450, 2245, 3000, 4210 or 5100: AGN
  continuum (\S\ref{sec:comp-continuum}) $\nu L_{\nu}$ luminosities
  and their uncertainties, in units of $10^{42}$~erg~s$^{-1}$,
  estimated at wavelengths \verb|CONT*__WAVE| or at the fixed rest
  frame wavelengths 1450\AA, 2245\AA{}, 3000\AA, 4210\AA{} or 5100\AA;

\item \verb|CONT*__SLOPE|, \verb|CONT*__SLOPE_err| where ``\verb|*|''
  is in the range 1--5, or equal to 1450, 2245, 3000, 4210 or 5100:
  AGN continuum slopes and their uncertainties estimated at
  wavelengths \verb|CONT*__WAVE| or at the fixed rest frame
  wavelengths 1450\AA, 2245\AA{}, 3000\AA, 4210\AA{} or 5100\AA.  The
  slopes are calculated as the derivative of the continuum component
  in the $\log \lambda$--$\log L_{\lambda}$ plane, this approach
  allows to consider the most general case of the continuum modeled as
  a smoothly broken power law (\S\ref{sec:sbpl}).  All the
  uncertainties are identical since they are given by the
  uncertainties of the $\alpha_{\lambda}$ and $\Delta
  \alpha_{\lambda}$ parameters of the AGN continuum component
  \S\ref{sec:comp-continuum}.  Note that for sources with $z < 0.6$
  the \verb|CONT*__SLOPE| is always $-1.7$ and the
  \verb|CONT*__SLOPE_err| is always 0;

\item \verb|CONT*__GALAXY| where ``\verb|*|'' is in the range 1--5, or
  equal to 1450, 2245, 3000, 4210 or 5100: host galaxy $\nu L_{\nu}$
  luminosities, in units of $10^{42}$~erg~s$^{-1}$, estimated at
  wavelengths \verb|CONT*__WAVE| or at the fixed rest frame
  wavelengths 1450\AA, 2245\AA{}, 3000\AA, 4210\AA{} or 5100\AA.
  These quantities are provided for a quick comparison with the AGN
  continuum luminosities.  The relative uncertainty of
  \verb|CONT*__GALAXY| can be estimated as
  \verb|GALAXY__LUM_ERR / GALAXY__LUM|;

\item \verb|CONT*__QUALITY| where ``\verb|*|'' is in the range 1--5,
  or equal to 1450, 2245, 3000, 4210 or 5100: quality flag for the AGN
  continuum estimates.  The meaning of the bits are:
  \begin{itemize}
    \item Bit 0: wavelength is outside the observed range;
    \item Bit 1: either the luminosity or its uncertainty are NaN or equal to zero;
    \item Bit 2: luminosity relative uncertainty $>$ 1.5;
    \item Bit 3: either the slope or its uncertainty are NaN or equal to zero;
    \item Bit 4: slope hits a limit in the fit;
    \item Bit 5: slope uncertainty $>$ 0.3;
  \end{itemize}

\item \verb|IRONUV__LUM|, \verb|IRONUV__LUM_ERR|: the integrated
  luminosity and its uncertainty for the iron template at UV
  wavelengths, in units of $10^{42}$~erg~s$^{-1}$
  (\S\ref{sec:comp-ironuv});

\item \verb|IRONUV__FWHM|, \verb|IRONUV__FWHM_ERR|: the FWHM and its
  uncertainty of the Gaussian profile used to broaden the template, in
  units of km s$^{-1}$.  The SDSS data do not allow to reliably
  constrain this quantity, hence \verb|IRONUV__FWHM| is always 3000
  km s$^{-1}$ and \verb|IRONUV__FWHM_ERR| is always 0;

\item \verb|IRONUV__EW|, \verb|IRONUV__EW_ERR|: the equivalent width
  and its uncertainty for the iron template at UV wavelengths, in
  units of \AA{};

\item \verb|IRONUV__QUALITY|:
  quality flag for the UV iron template.  The meaning of the bits are:
  \begin{itemize}
    \item Bit 0: fit of iron template is not sensible at this redshift;
    \item Bit 1: either the luminosity or its uncertainty are NaN or equal to zero;
    \item Bit 2: luminosity relative uncertainty $>$ 1.5;
  \end{itemize}

\item \verb|IRONOPT_BR__LUM|, \verb|IRONOPT_BR__LUM_ERR|: the
  integrated luminosity and its uncertainty for the ``broad'' iron
  template at optical wavelengths, in units of $10^{42}$~erg~s$^{-1}$
  (\S\ref{sec:comp-ironopt});

\item \verb|IRONOPT_BR__FWHM|, \verb|IRONOPT_BR__FWHM_ERR|: the FWHM
  and its uncertainty of the Gaussian profile used to build the
  template, in units of km s$^{-1}$.  The SDSS data do not allow to
  reliably constrain this quantity, hence \verb|IRONOPT_BR__FWHM| is
  always 3000 km s$^{-1}$ and \verb|IRONOPT_BR__FWHM_ERR| is always 0;

\item \verb|IRONOPT_BR__UNK_COUNT|: number of ``unknown'' components
  (\S\ref{sec:comp-linesunk}) whose center lies in the range
  4460--4680\AA{} or 5150--5520\AA{}. i.e. which overlaps to the
  optical iron template;

\item \verb|IRONOPT_BR__UNK_LUM|, \verb|IRONOPT_BR__UNK_LUM_ERR|: the
  integrated luminosity and its uncertainty for the ``unknown''
  components (\S\ref{sec:comp-linesunk}) overlapping the optical iron
  template, in units of $10^{42}$~erg~s$^{-1}$;

\item \verb|IRONOPT_BR__EW|, \verb|IRONOPT_BR__EW_ERR|: the equivalent
  width and its uncertainty for the ``broad'' iron template at optical
  wavelengths, in units of \AA{};

\item \verb|IRONOPT_BR__QUALITY|: quality flag for the broad optical
  iron template.  The meaning of the bits are:
  \begin{itemize}
  \item Bit 0: fit of iron template is not sensible at this redshift;
  \item Bit 1: either the luminosity or its uncertainty are NaN or equal to zero;
  \item Bit 2: luminosity relative uncertainty $>$ 1.5;
  \end{itemize}

\item \verb|IRONOPT_NA__LUM|, \verb|IRONOPT_NA__LUM_ERR|,
  \verb|IRONOPT_NA__FWHM|, \verb|IRONOPT_NA__FWHM_ERR|,
  \verb|IRONOPT_NA__QUALITY|, \verb|IRONOPT_BR__EW|,
  \verb|IRONOPT_BR__EW_ERR|: same quantities as above, but for the
  ``narrow'' optical template (\S\ref{sec:comp-ironopt});

\item \verb|*__NCOMP| where ``\verb|*|'' is one of the prefix listed
  in Tab.~\ref{tab:knownline}: number of components used to account for
  the emission line.  In the current \qsfit{} implementation narrow
  lines are always modeled with a single component;

\item \verb|*__LUM|, \verb|*__LUM_ERR| where ``\verb|*|'' is one of
  the prefix listed in Tab.~\ref{tab:knownline}: emission line integrated
  luminosities and their uncertainties, in units of
  $10^{42}$~erg~s$^{-1}$;

\item \verb|*__FWHM|, \verb|*__FWHM_ERR| where ``\verb|*|'' is one of
  the prefix listed in Tab.~\ref{tab:knownline}: emission line FWHM and
  their uncertainties, in units of km s$^{-1}$;

\item \verb|*__VOFF|, \verb|*__VOFF_ERR| where ``\verb|*|'' is one of
  the prefix listed in Tab.~\ref{tab:knownline}: emission line velocity
  offsets and their uncertainties, in units of km s$^{-1}$.  A
  positive value means the line is blue--shifted, a negative value
  means the line is red--shifted;

\item \verb|*__EW|, \verb|*__EW_ERR| where ``\verb|*|'' is one of the
  prefix listed in Tab.~\ref{tab:knownline}: the equivalent width and
  its uncertainty in units of \AA{};

\item \verb|*__QUALITY| where ``\verb|*|'' is one of the prefix listed
  in Tab.~\ref{tab:knownline}: quality flag for the emission line.  The
  meaning of the bits are:
  \begin{itemize}
  \item Bit 0: either the luminosity or its uncertainty are NaN or equal to zero;
  \item Bit 1: luminosity relative uncertainty $>$ 1.5;
  \item Bit 2: either the FWHM or its uncertainty are NaN or equal to zero;
  \item Bit 3: FWHM value hits a limit in the fit;
  \item Bit 4: FWHM relative uncertainty $>$ 2;
  \item Bit 5: either the V$_{\rm off}$ or its uncertainty are NaN or equal to zero;
  \item Bit 6: V$_{\rm off}$ value hits a limit in the fit;
  \item Bit 7: V$_{\rm off}$ uncertainty $>$ 500 km s$^{-1}$.
  \end{itemize}

\item \verb|LINE_HA_BASE__NCOMP|, \verb|LINE_HA_BASE__LUM|,
  \verb|LINE_HA_BASE__LUM_ERR|, \verb|LINE_HA_BASE__FWHM|,
  \verb|LINE_HA_BASE__FWHM_ERR|, \verb|LINE_HA_BASE__VOFF|,
  \verb|LINE_HA_BASE__VOFF_ERR|, \verb|LINE_HA_BASE__EW|,
  \verb|LINE_HA_BASE__EW_ERR|, \verb|LINE_HA_BASE__QUALITY|: same
  quantities as above, but for the H$\alpha$ ``base" emission line
  (\S\ref{sec:comp-lines});

\item \verb|BALMER__LUM|, \verb|BALMER__LUM_ERR|: $\nu L_{\nu}$
  luminosity and its uncertainty for the Balmer continuum component
  (\S\ref{sec:balmer}) estimated at 3000\AA{}, in units of
  $10^{42}$~erg~s$^{-1}$;

\item \verb|BALMER__RATIO|, \verb|BALMER__RATIO_ERR|:
  luminosity ratio of the high order blended line complex to the
  Balmer continuum (see Fig.~\ref{fig:balmer});

\item \verb|BALMER__LOGT|, \verb|BALMER__LOGT_ERR|:
  logarithm of electron temperature and its uncertainty for the Balmer
  component in units of log K.  The SDSS data do not allow to reliably
  constrain this quantity, hence \verb|BALMER__LOGT| is always
  $\log(15,000)$ K and \verb|BALMER__LOGT_ERR| is always 0;

\item \verb|BALMER__LOGNE|, \verb|BALMER__LOGNE_ERR|:
  logarithm of electron density and its uncertainty for the Balmer
  component in units of log cm$^{-3}$.  The SDSS data do not allow to
  reliably constrain this quantity, hence \verb|BALMER__LOGNE| is
  always 9 and \verb|BALMER__LOGNE_ERR| is always 0;

\item \verb|BALMER__LOGTAU|, \verb|BALMER__LOGTAU_ERR|: logarithm of
  the optical depth, and its uncertainty, at the Balmer edge.  The
  SDSS data do not allow to reliably constrain this quantity, hence
  \verb|BALMER__LOGTAU| and \verb|BALMER__LOGTAU_ERR| are always 0;

\item \verb|BALMER__FWHM|, \verb|BALMER__FWHM_ERR|: FWHM
  and its uncertainty of the Gaussian profile used to broaden the
  Balmer component, in units of km s$^{-1}$.  The SDSS data do not
  allow to reliably constrain this quantity, hence \verb|BALMER__FWHM|
  is always 5000 km s$^{-1}$ and \verb|BALMER__LOGNE_ERR| is always
  0;

\item \verb|BALMER__QUALITY|: quality flag for the Balmer
  component.  The meaning of the bits are:
  \begin{itemize}
  \item Bit 0: fit of Balmer template is not sensible at this
    redshift;
  \item Bit 1: either the luminosity or its uncertainty are NaN;
  \item Bit 2: luminosity relative uncertainty $>$ 1.5.
  \end{itemize}

\item \verb|CHISQ|: the value of the $\chi^2_{\rm red}$, calculated
  using the data uncertainties provided in the SDSS FITS files;

\item \verb|DOF|: the number of degrees of freedom (number
  of data - number of free model parameters) in the fit;

\item \verb|ELAPSED_TIME|: the elapsed time in seconds required
  to analyze the source.
\end{itemize}
All uncertainties in the \qsfit{} estimates are quoted at the 68\%
level, i.e. 1$\sigma$;

\section{qsfit{} installation, usage and customization}
\label{sec:qsfitManual}

\subsection{QSFit installation, compilation and test}
\label{sec:qsfit-install}

The only prerequisite to run \qsfit{} is IDL (ver $>= 8.1$).  If you
need to view/generate the plots you should also install Gnuplot
(ver. $>= 5.0$).  The IDL code of \qsfit{} (version \version{}) can be
downloaded at the following address:
\begin{center}
  \url{https://github.com/gcalderone/qsfit/releases}\\
\end{center}
Once downloaded the \qsfit{} package should be unpacked and placed in
a directory.  Then you should start an IDL session, change to the
directory where you stored the unpacked files, e.g (on Windows):
\begin{verbatim}
IDL> CD, 'C:\path\where\qsfit\is\located'
\end{verbatim}
or (on Linux)
\begin{verbatim}
IDL> CD, '/path/where/qsfit/is/located'
\end{verbatim}
and compile all the \qsfit{} procedures and dependencies by calling
the \verb|compile| procedure:
\begin{verbatim}
IDL> compile
\end{verbatim}
Note that all the dependencies (the Astrolib and MPFIT routines) are
included in the \verb|IDL/Contrib| directory of the \qsfit{} package
and are automatically compiled in the step above.  Also note that the
only required path in the \verb|IDL_PATH| variable is the IDL default
path to the \verb|lib| directory, you do not have to add any other
path to use \qsfit{}.  Finally, you should avoid calling the
\verb|compile| procedure and another \qsfit{} routine from the same
program.  The best practice is to call \verb|compile| manually at the
very beginning of the IDL session, before any other call.

The \qsfit{} package already comes with a SDSS DR-10 FITS file,
suitable to immediately test the code using the following command:
\begin{verbatim}
IDL> res = qsfit('data/spec-0752-52251-0323.fits',$
                 z=0.3806, ebv=0.06846)
\end{verbatim}
Once the analysis is finished (it will require a few seconds) you can
plot the results with the following command (Gnuplot is required to
run this step):
\begin{verbatim}
IDL> qsfit_plot, res
\end{verbatim}
If you went through all the steps in this section without errors then
\qsfit{} is installed properly, and you can start the data analysis as
discussed in the next section.

\subsection{QSFit usage}
\label{sec:qsfit-usage}

In order to use \qsfit{} to perform spectral analysis you should first
download all the necessary spectra.  In its current implementation
(version \version{}) \qsfit{} supports only SDSS--DR10 FITS files of
sources with $z<2$.  You can find these files using the ``Object
explorer'' facility on the SDSS--DR10 website, at the address:
\begin{center}
  \url{http://skyserver.sdss.org/dr10/en/tools/explore/obj.aspx}\\
\end{center}
Alternatively you may download a FITS file using a URL similar to the
following:
\begin{center}
  \url{http://dr10.sdss3.org/sas/dr10/sdss/spectro/redux/26/spectra/0752/spec-0752-52251-0323.fits}\\
\end{center}
The numbers to be changed are:
\begin{itemize}
\item the spectroscopic plate number (0752 in the example above. Note
  that: this number appears twice in the URL);
\item the MJD of observation (52251 in the example above);
\item the spectroscopic fiber number (0323 in the example above).
\end{itemize}
The numbers for a specific object may be found in the SDSS-DR10
website or in the \qsfitcat{} catalog (columns \verb|PLATE|,
\verb|MJD| and \verb|FIBER|).  The \qsfit{} package already comes with
one such FITS file in the ``data'' directory.  This can be used to
quickly test the \qsfit{} functionalities (see
\S\ref{sec:qsfit-install}).

To run the \qsfit{} analysis you should simply call the \verb|qsfit|
function, with a command similar to the following:
\begin{verbatim}
IDL> res = qsfit('data/spec-0752-52251-0323.fits',$
                 z=0.3806, ebv=0.06846)
\end{verbatim}
The parameters are:
\begin{itemize}
\item the path to a SDSS-DR10 FITS file;
\item the redshift of the source;
\item the E(B-V) color excess (in magnitudes).
\end{itemize}
With this command you may easily replicate all the data analysis
discussed in this paper.  The FITS file names, the redshifts and the
color excess used in the analysis are all available in the \qsfitcat{}
catalog (in the columns: \verb|SPEC|, \verb|REDSHIFT| and
\verb|E_BV|).

The \qsfit{} analysis will require a few seconds (see
Fig.~\ref{fig:elapsed}), and all the results will be returned in the
\verb|res| variable, as a nested structure.  Such structure is not
suitable to be collected in a catalog like the \qsfitcat{} one
(\S\ref{sec:catalog}), hence we provide a procedure to re--arrange the
nested structure into a ``flattened'' one, i.e. a structure whose tags
are all scalar values.  To obtain a flattened structure you should use
a command like the following:
\begin{verbatim}
IDL> flat = qsfit_flatten_results(res)
\end{verbatim}
The tags in the \verb|flat| structure are described in
\S\ref{sec:flattened-struc}, and the \qsfitcat{} catalog
(\S\ref{sec:catalog}) is actually a collection of \qsfit{} flattened
structures.  The nested structure is described in further detail in
\S\ref{sec:nested}.

If you want the results to be permanently saved in a directory (which
must have been previously created) use the \verb|OUTDIR| keyword,
e.g.:
\begin{verbatim}
IDL> res = qsfit('data/spec-0752-52251-0323.fits',$
                 z=0.3806, ebv=0.06846, $
                 outdir='output')
\end{verbatim}

This will save all the logs in a file named
\verb|output/spec-0752-52251-0323_QSFIT.log| and the results in
\verb|output/spec-0752-52251-0323_QSFIT.dat|.  The latter contains a
binary representation of the results, and can be used to restore the
data in IDL (using the \verb|restore| procedure).  Note that if you
run again the above command the program will stop immediately with the
message \texttt{File output/spec-0752-52251-0323\_QSFIT.dat already
  exists.}  This is not an error, it simply says that the analysis has
already been performed and there is no need to run it again.  To
actually re--run the analysis you should either manually delete that
file or avoid using the \verb|OUTDIR| keyword.

Once the analysis is finished you can plot the results with the
following command (Gnuplot is required to run this step):
\begin{verbatim}
IDL> qsfit_plot, res
\end{verbatim}
This will open two Gnuplot windows with the plots shown in
Fig.~\ref{fig:ex1}.  To show the data re--binned by a factor of, say,
5, you can use the following command:
\begin{verbatim}
IDL> res.gfit.plot.(0).main.rebin = 5
IDL> qsfit_plot, res
\end{verbatim}
Note that in the commands above you should use the ``nested''
structure (\verb|res|), not the ``flattened'' one (\verb|flat|) since
the latter does not contain all the necessary data to draw the plots.

You may save all the plot data in a file suitable to be later loaded
in Gnuplot, and re--directed into one of the many terminals supported
by Gnuplot.  This step is accomplished by the following command:
\begin{verbatim}
IDL> qsfit_plot, res, $
       filename='plot/spec-0752-52251-0323'
\end{verbatim}
This command will create two files in the \verb|plot| directory (which
must have been previously created): one with the \verb|.gp| extension
and one with the \verb|_resid.gp| suffix.  The former is the actual
plot, to compare the data and the model (like the one in
Fig.~\ref{fig:ex1}, upper panel).  The latter is the plot of the
residuals (Fig.~\ref{fig:ex1}, lower panel).  These are just text
files containing the data and the Gnuplot command to actually show the
plots and are completely decoupled from IDL.  You can load them
in Gnuplot using the following commands:
\begin{verbatim}
gnuplot> load 'plot/spec-0752-52251-0323.gp'
gnuplot> load 'plot/spec-0752-52251-0323_resid.gp'
\end{verbatim}
You may even modify them with a text editor before loading into
Gnuplot, in order to change the way the plots are displayed
(e.g. colors, format, legends, etc.) or the terminal being used.
The default Gnuplot terminal is \verb|wxt|, which opens an interactive
window showing the plot.  To obtain a PDF copy you may simply use the
``Export to plot file'' button.  Alternatively you may modify the
\verb|.gp| files by changing the terminal.  E.g. replace the line
\verb|set term wxt| with:
\begin{verbatim}
  set term pdf
  set output 'filename.pdf'
\end{verbatim}
and reload the file in Gnuplot.

\subsection{The nested structure returned by \qsfit{}}
\label{sec:nested} 

The nested structure returned by the \verb|qsfit| function contains
much more information than the flattened one used to build the
\qsfitcat{} catalog.  These further information are stored in the
following tags:
\begin{itemize}
\item \verb|QSFIT_VERSION|: a string containing the version of
  \qsfit{} which generated the results;

\item \verb|LOG|: an array of strings containing all the log messages
  generated during the analysis.  To print them you may use the
  command \verb|IDL> gprint, res.log|;

\item \verb|CONT|: an array of structures containing all the results
  related to the continuum component.  These informations are also
  contained in the flattened structure, although in a different
  format.  The \verb|CONT| tag is very useful to quickly print all the
  continuum quantities with the command \verb|IDL> gps, res.cont|;

\item \verb|LINES|: an array of structures containing all the results
  related to the emission line components.  These informations are
  also contained in the flattened structure, although in a different
  format.  The \verb|LINES| tag is very useful to quickly print all
  the emission lines quantities with the command:
  \verb|IDL> gps, res.lines|;

\item \verb|GFIT|: this structure contains the whole state of the
  \gfit{} framework (see \S\ref{sec:gfit}) used to actually perform
  the fitting: i.e. the input data, the component settings, the best
  fit values and uncertainties, the plotting settings, etc.  For
  instance, if you want to manually plot the data and the best fit
  model (using the IDL plotting facilities) you may use the following
  commands:
\begin{verbatim}
IDL> plot , $
     res.gfit.cmp.(0).x, res.gfit.cmp.(0).y, psym=3
IDL> oplot, $
     res.gfit.cmp.(0).x, res.gfit.cmp.(0).m, col=255
\end{verbatim}
Or, if you want to print the best fit value and uncertainty of the
$\alpha_{\lambda}$ parameter in the continuum component (see
Eq.~\ref{eq:sbpl}):
\begin{verbatim}
IDL> print, a.gfit.comp.continuum.alpha1.val, $
            a.gfit.comp.continuum.alpha1.err
\end{verbatim}
A complete description of the \verb|GFIT| structure is beyond the
purpose of this paper.  Interested readers can explore the structure
using the \verb|help| command, or read the documentation in the
\verb|IDL/Glib/gfit| directory.

\end{itemize}

\subsection{Parameter uncertainty estimates through Monte Carlo resampling}
\label{sec:MC-uncert}

As discussed in \S\ref{sec:reduction} the \qsfit{} uncertainties
provided by the Fisher matrix method can be considered as rough
estimates, mainly because of correlation among model parameters
(e.g. between continuum and galaxy luminosities, or between emission
line luminosities and widths) and asymmetric uncertainty intervals.  A
better and more reliable estimate for the uncertainty of each
parameter is provided by the Monte Carlo resampling method
\citep{2009-Burtscher-MC-Resampling,2010-Andrae-ErrorEstimation},
which only relies on the assumptions that the \qsfit{} model provide a
reliable representation of the data and that the uncertainties on the
observed spectrum are Gaussian distributed with known standard
deviation.  To apply the resampling method we consider the best fit
model parameters provided by \qsfit{} (\S\ref{sec:reduction}) but
neglect the associated uncertainties.  Instead, we assume the best fit
model to be the ``true'' one, with no uncertainties, and generate a
set of mock spectra whose flux, in each spectral channel, is a
Gaussian random variable with standard deviation equal to the
uncertainty of input data.  Then we re--run the \qsfit{} analysis on
the mock spectra and collect all the results (again, neglecting the
Fisher uncertainties).  Finally, we consider the distributions of best
fit values for each model parameter, whose width provide an estimate
for the associated uncertainties.  For example to estimate the
1$\sigma$ uncertainty we may consider the interval between the 15.8
and 84.2 percentile of the whole distribution.

\begin{figure*}
  \includegraphics[width=.45\ww]{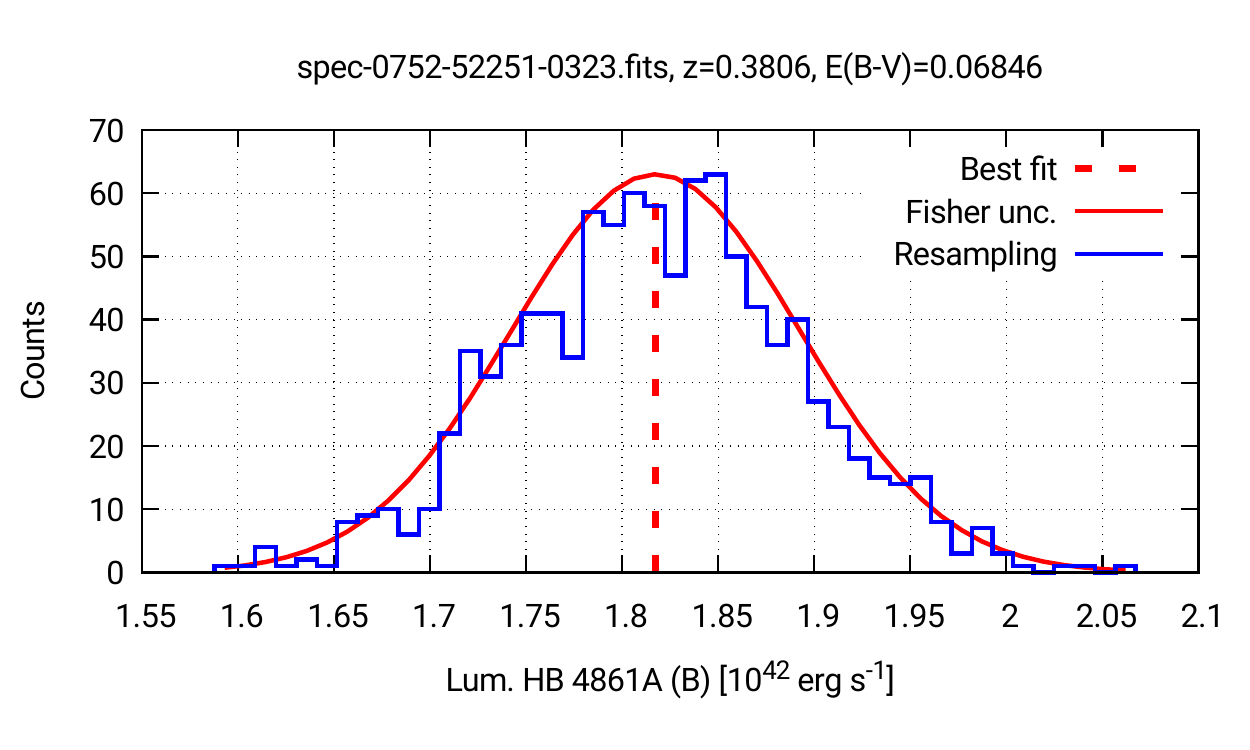}
  \includegraphics[width=.45\ww]{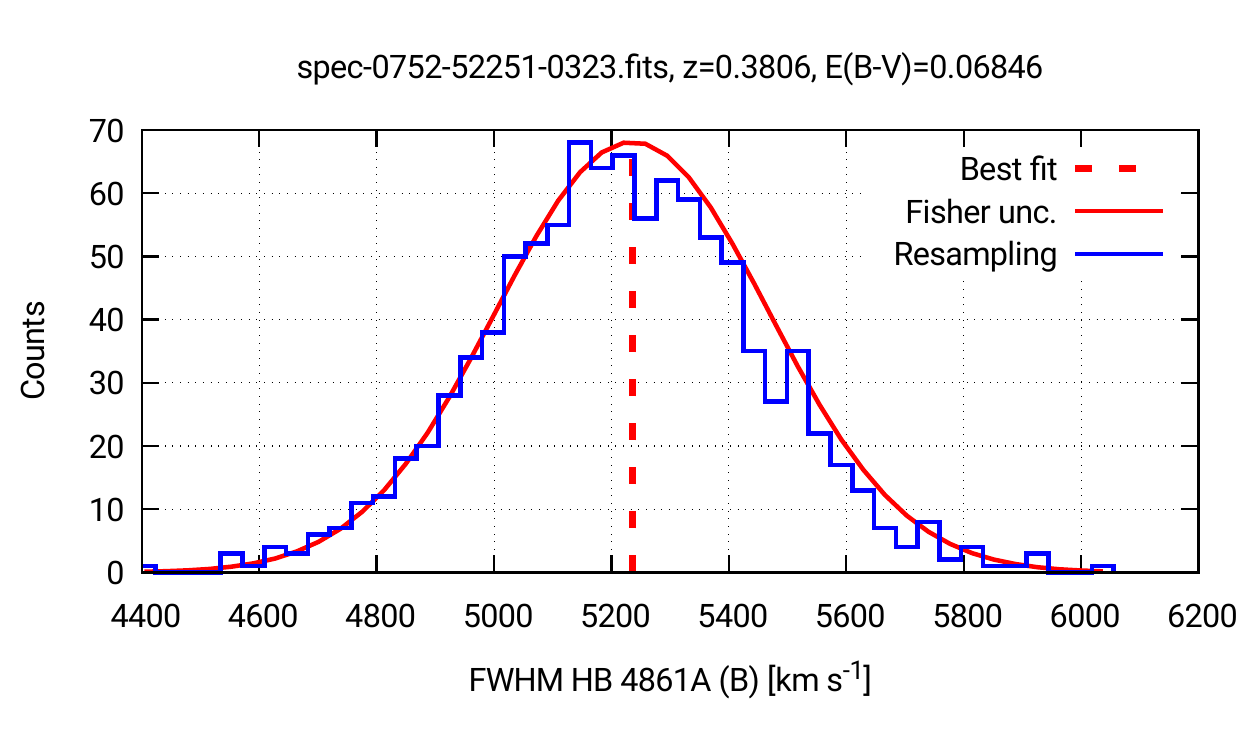}\\
  \includegraphics[width=.45\ww]{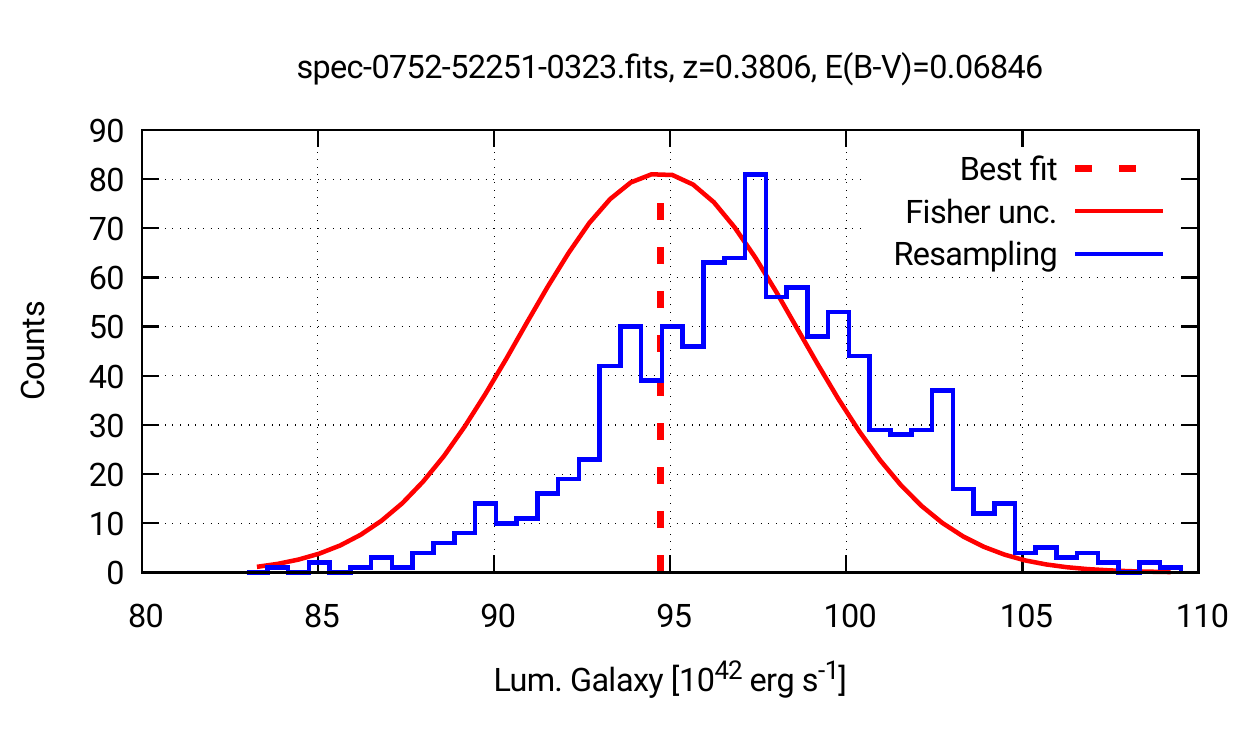}
  \includegraphics[width=.45\ww]{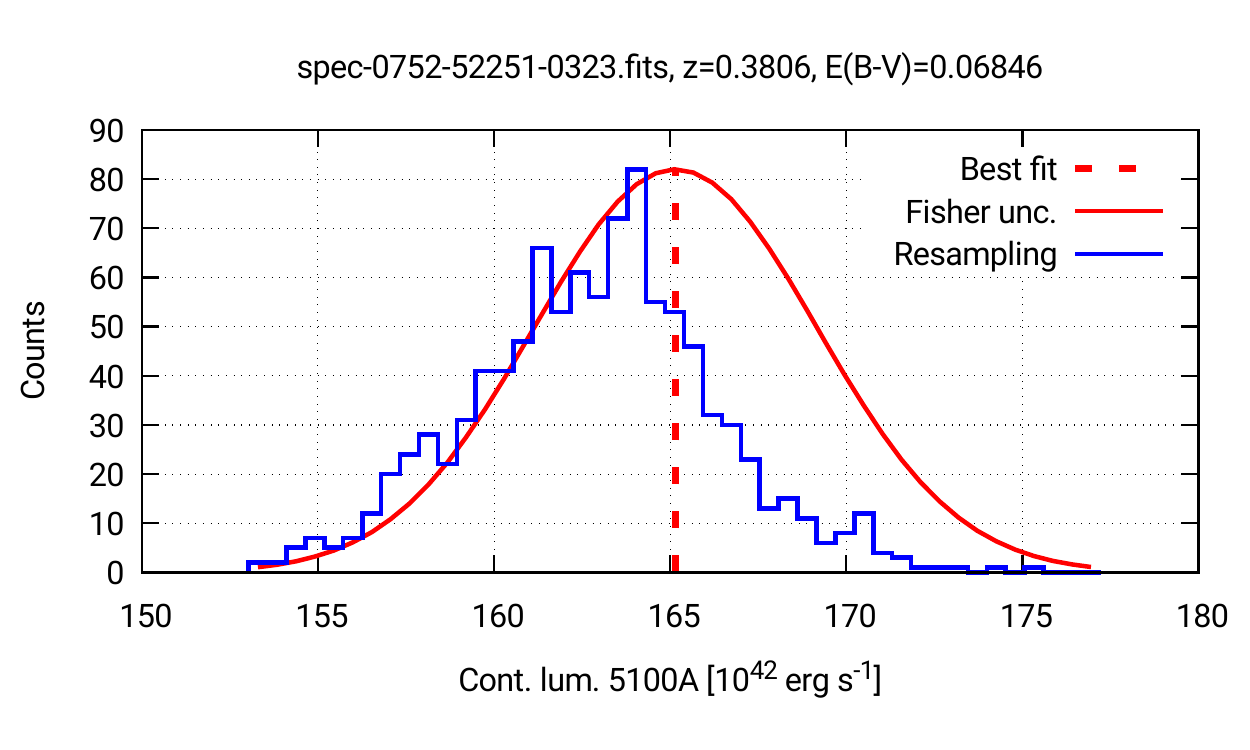}\\
  \includegraphics[width=.45\ww]{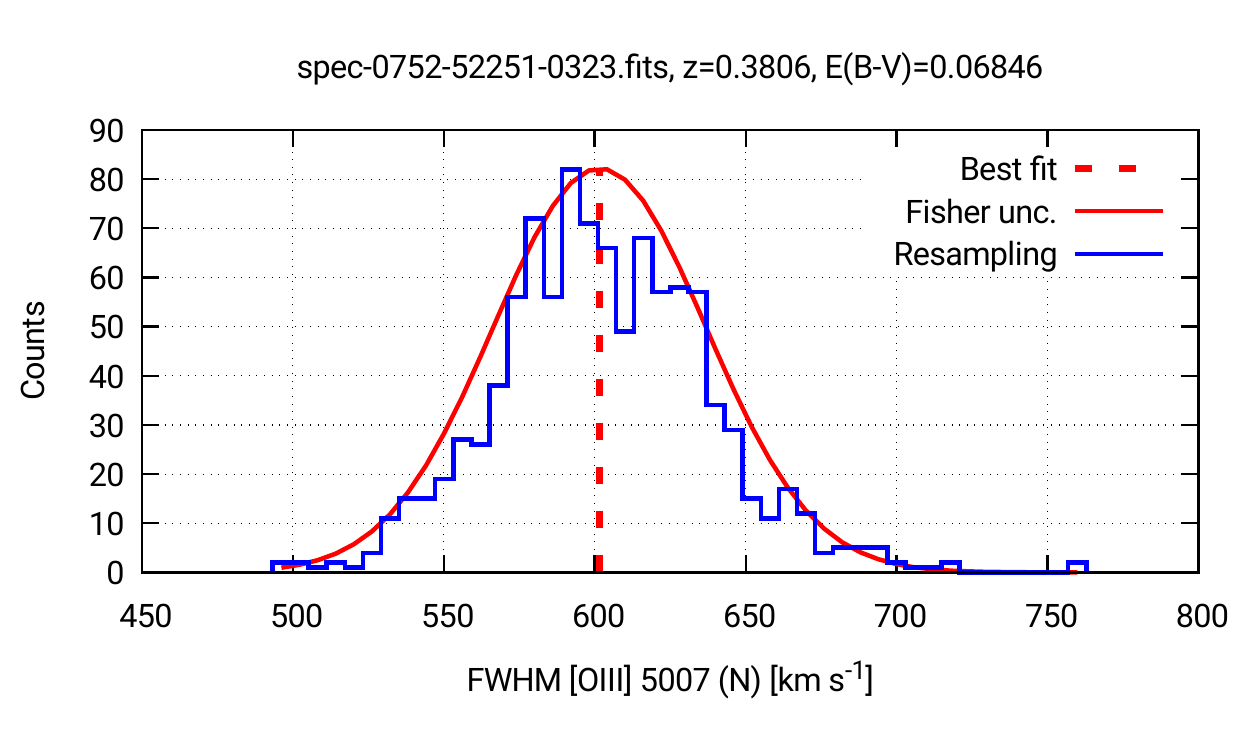}
  \includegraphics[width=.45\ww]{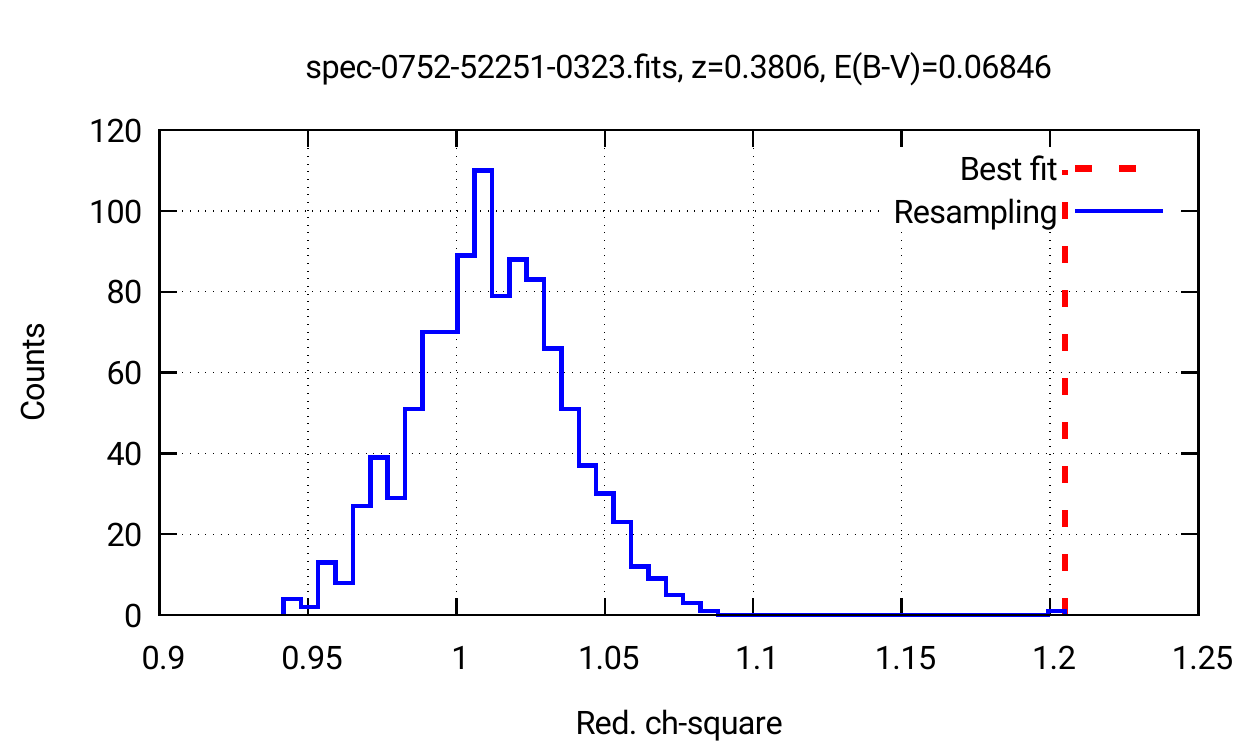}\\
  \includegraphics[width=.45\ww]{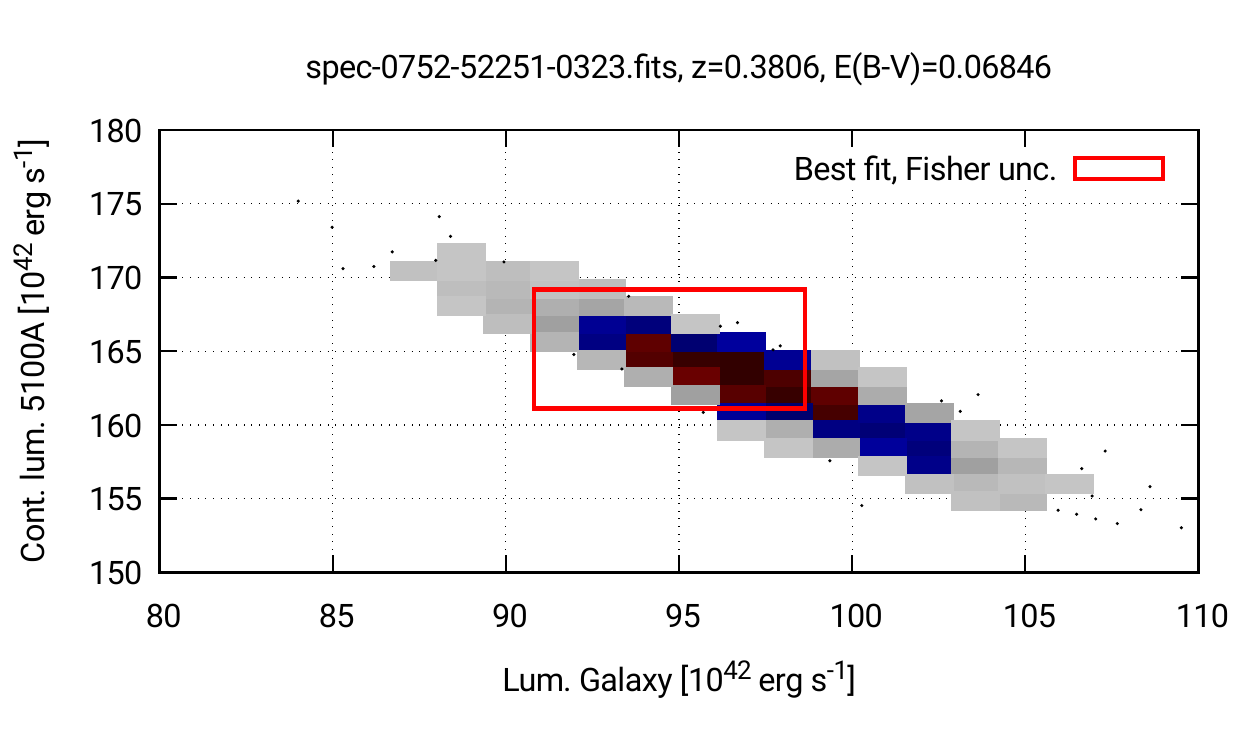}
  \includegraphics[width=.45\ww]{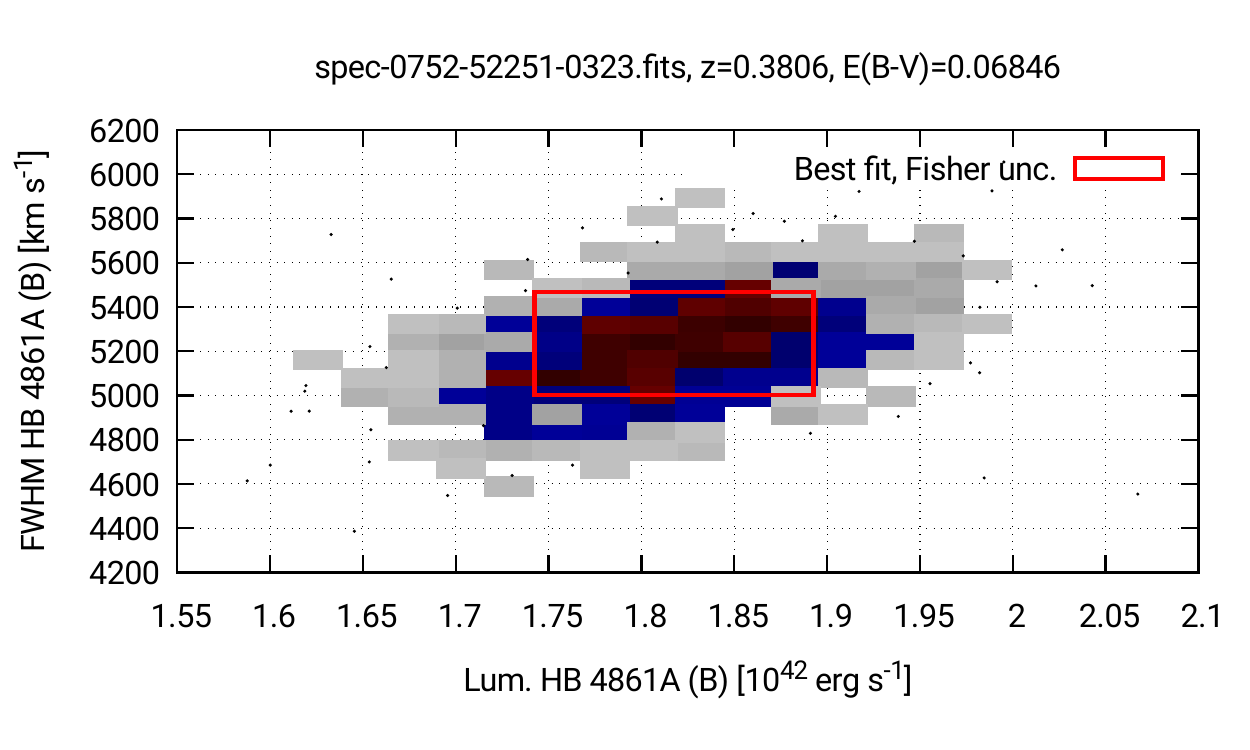}\\
  \caption{Comparison of expected distributions of
    \qsfit{} best fit parameters and associated Fisher uncertainties
    (red lines) and best fit parameters from resampled data (blue
    line).  The sixth panel shows the distribution of the reduced
    $\chi^2$ values (blue) of resampled data and the \qsfit{} value
    (red).  The last two panels show the 2--D distribution of two
    correlated parameters following the same rules described in
    \S\ref{sec:ex_plot}.  The red square identifies the region of $\pm
    1 \sigma$ Fisher uncertainties for both parameters.}
  \label{fig:MC-lowZ}
\end{figure*}

\begin{figure*}
  \includegraphics[width=.45\ww]{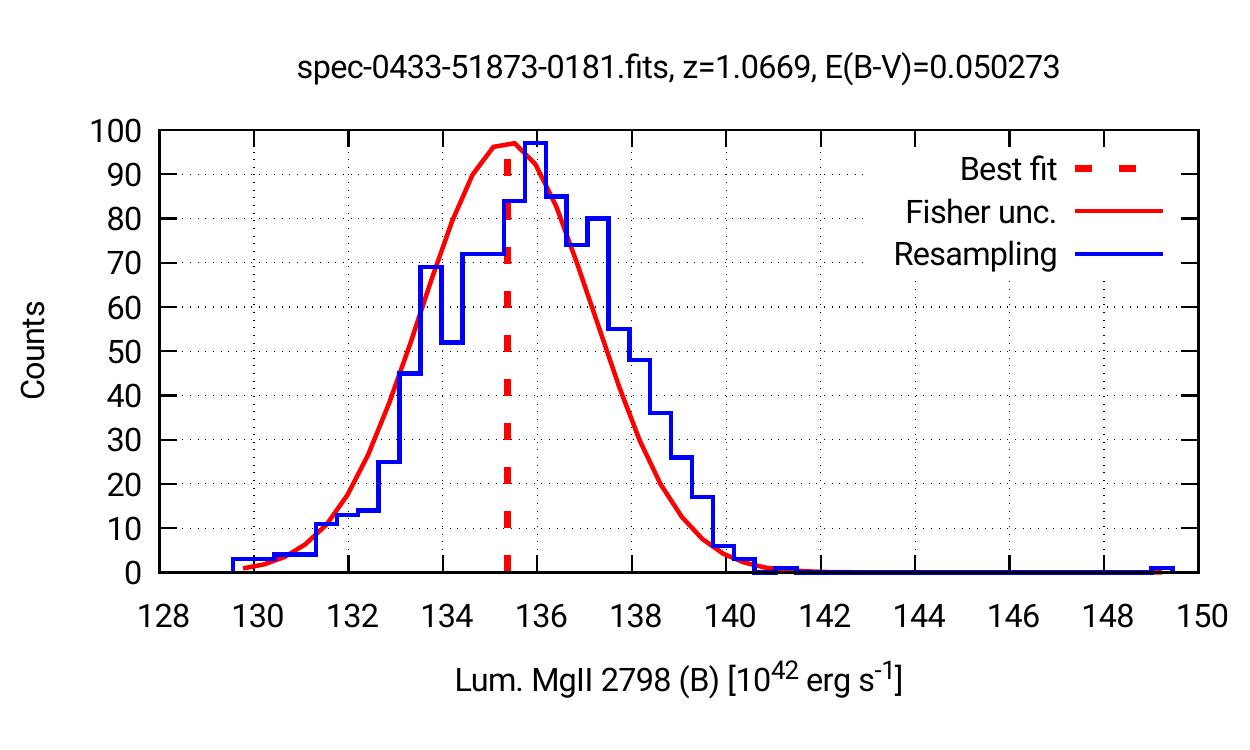}
  \includegraphics[width=.45\ww]{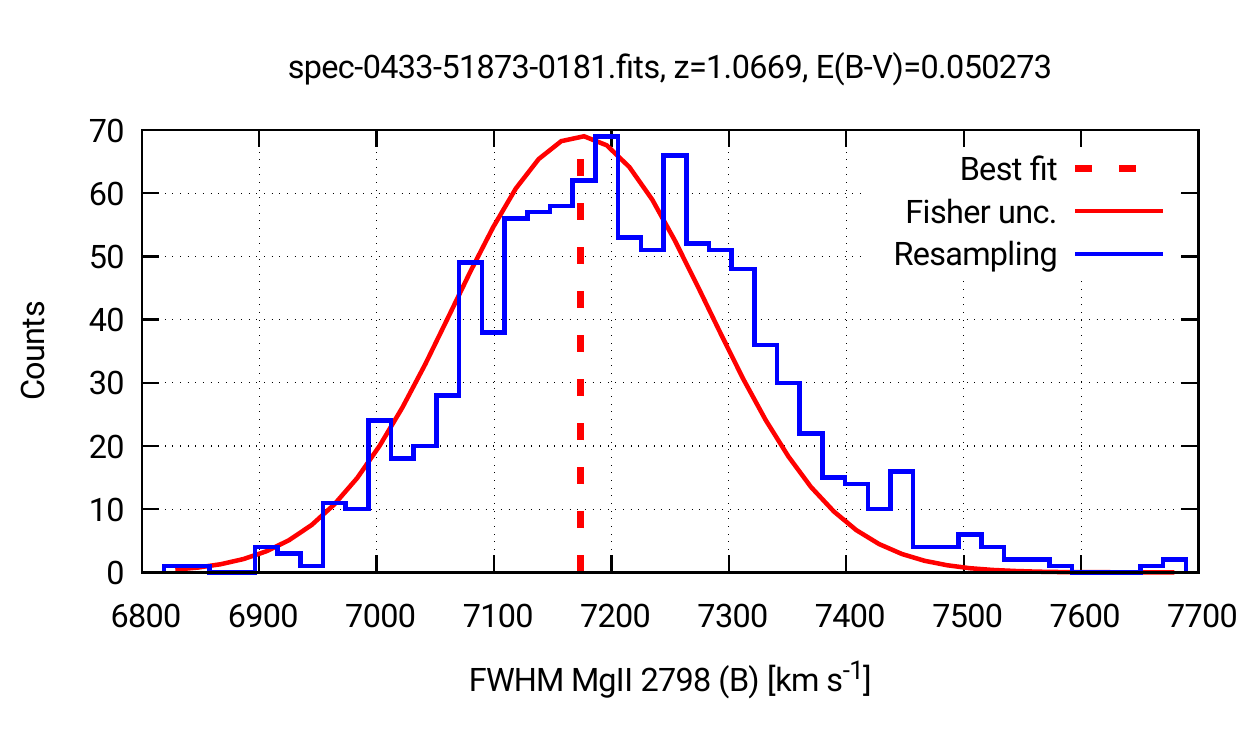}\\
  \includegraphics[width=.45\ww]{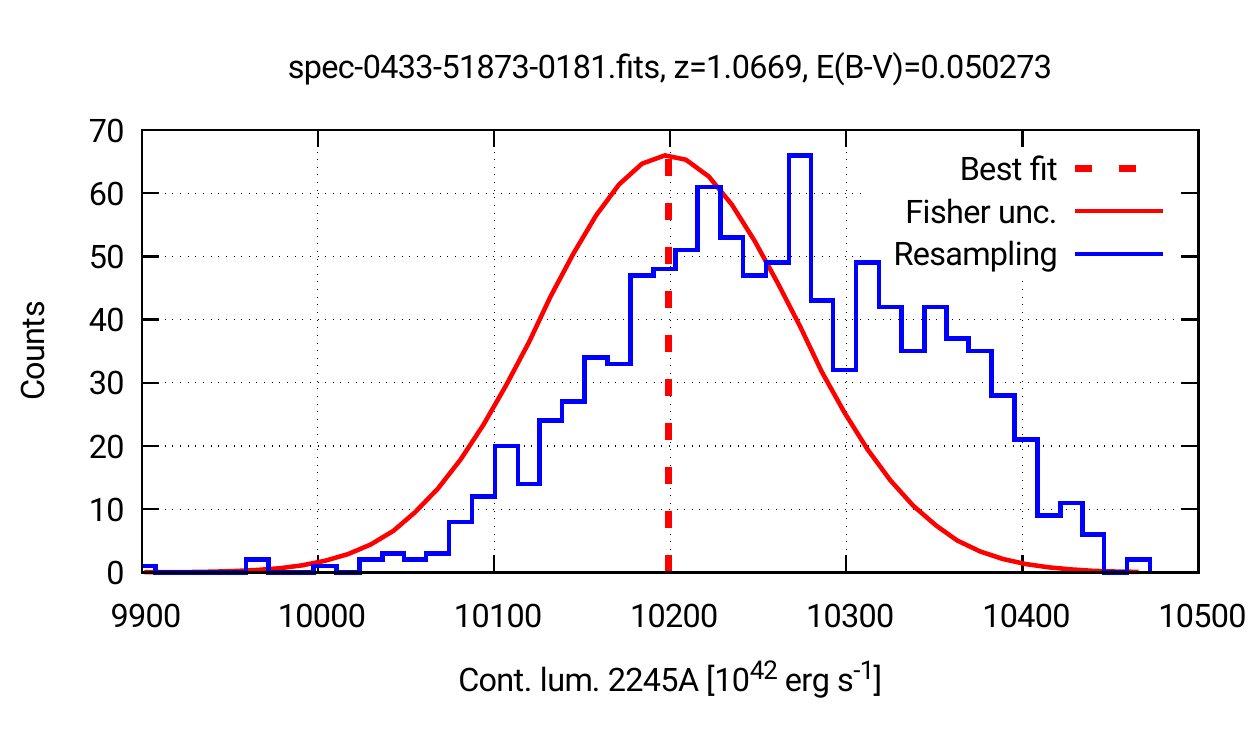}
  \includegraphics[width=.45\ww]{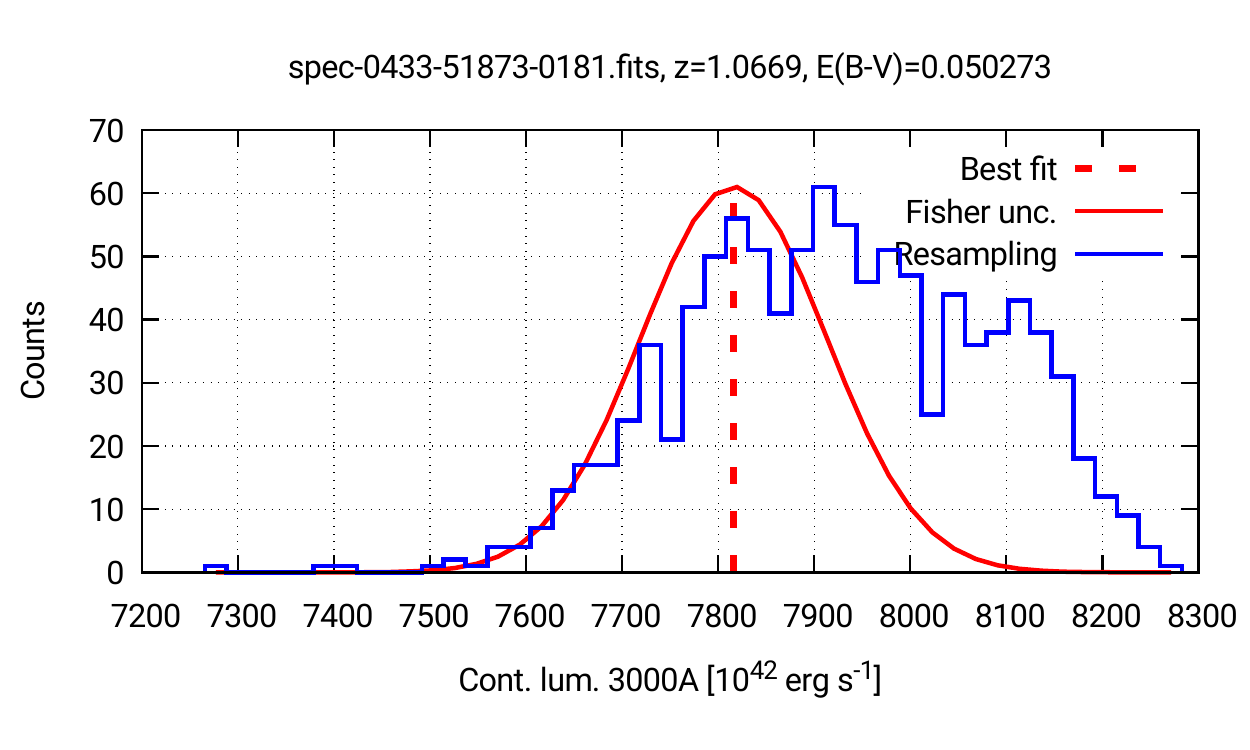}\\
  \includegraphics[width=.45\ww]{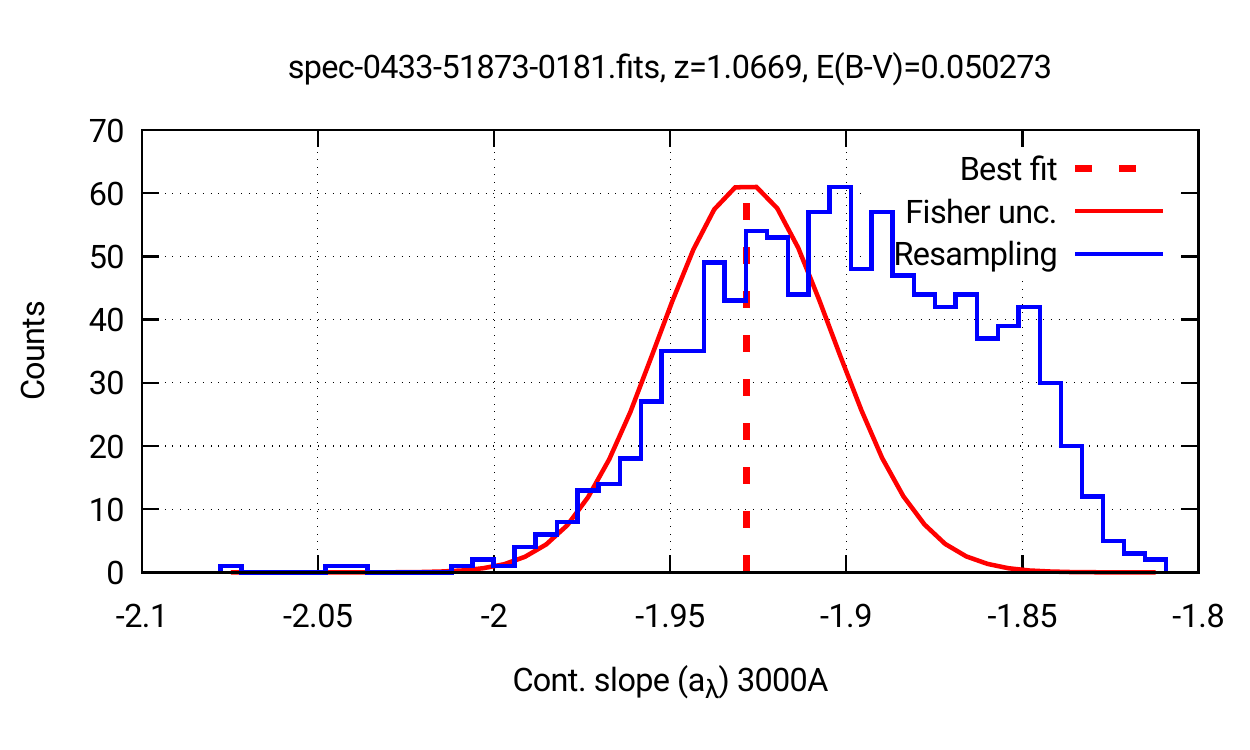}
  \includegraphics[width=.45\ww]{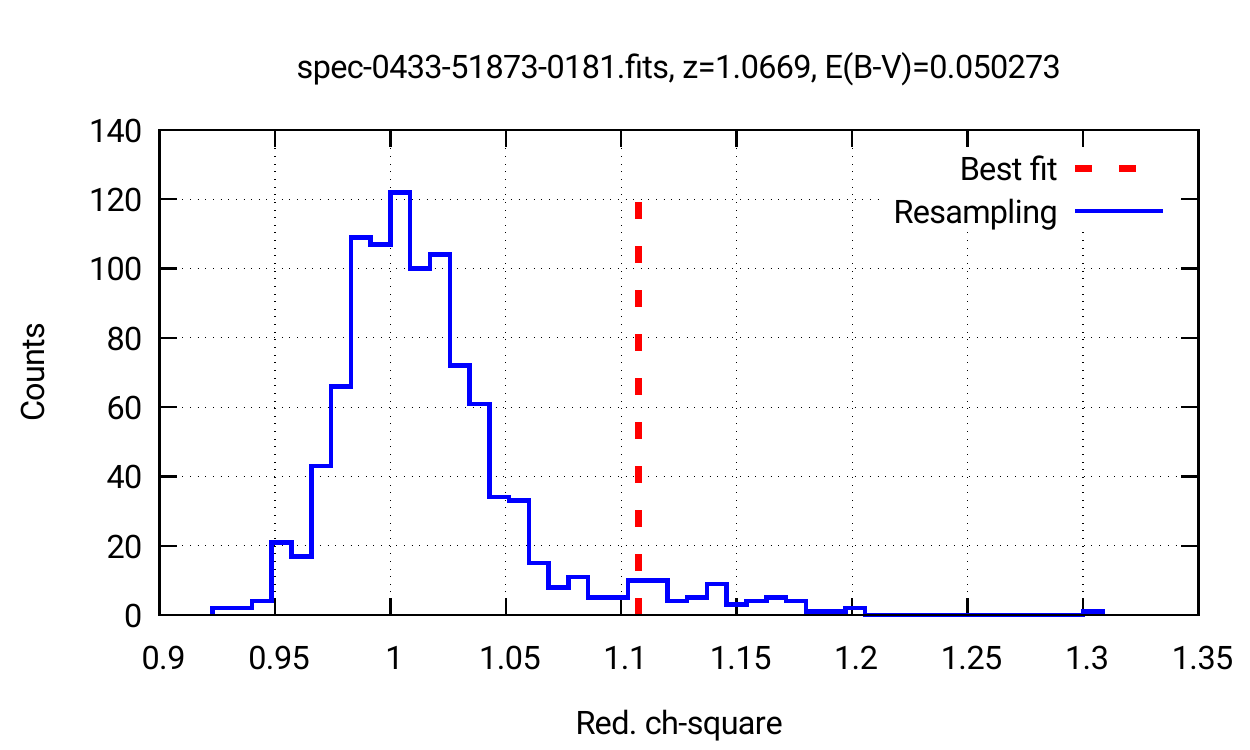}\\
  \includegraphics[width=.45\ww]{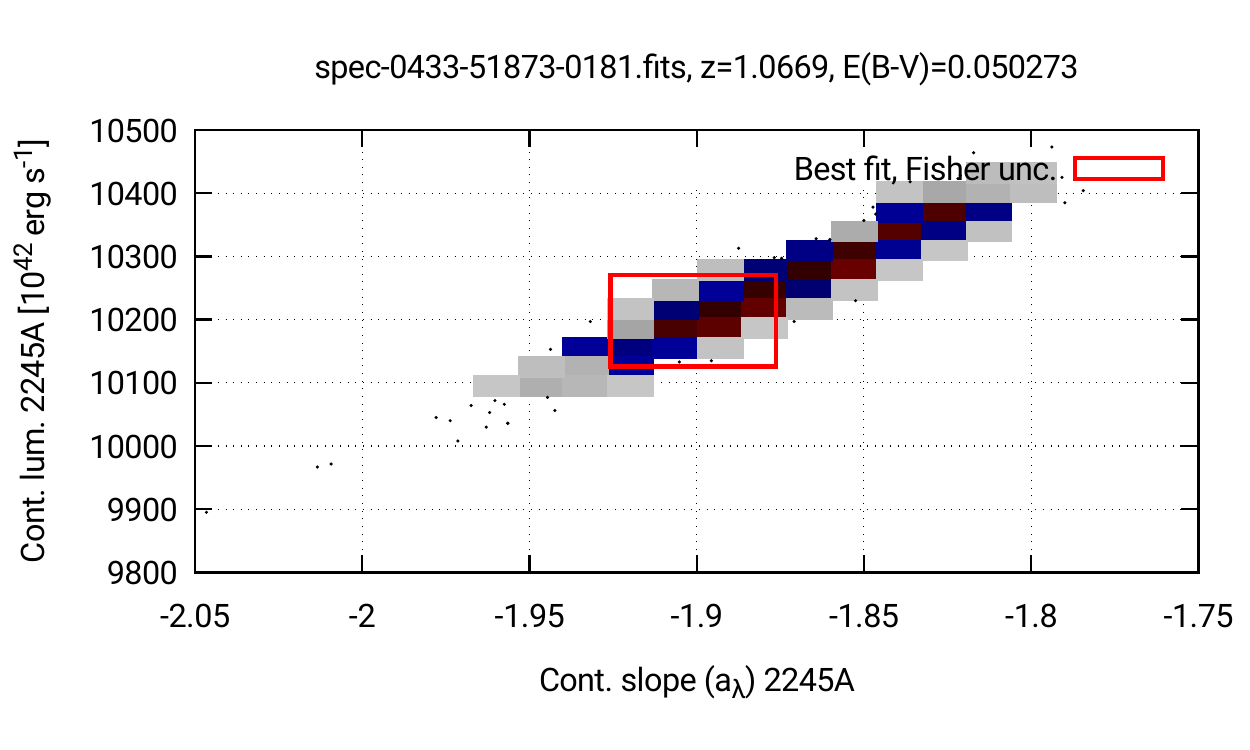}
  \includegraphics[width=.45\ww]{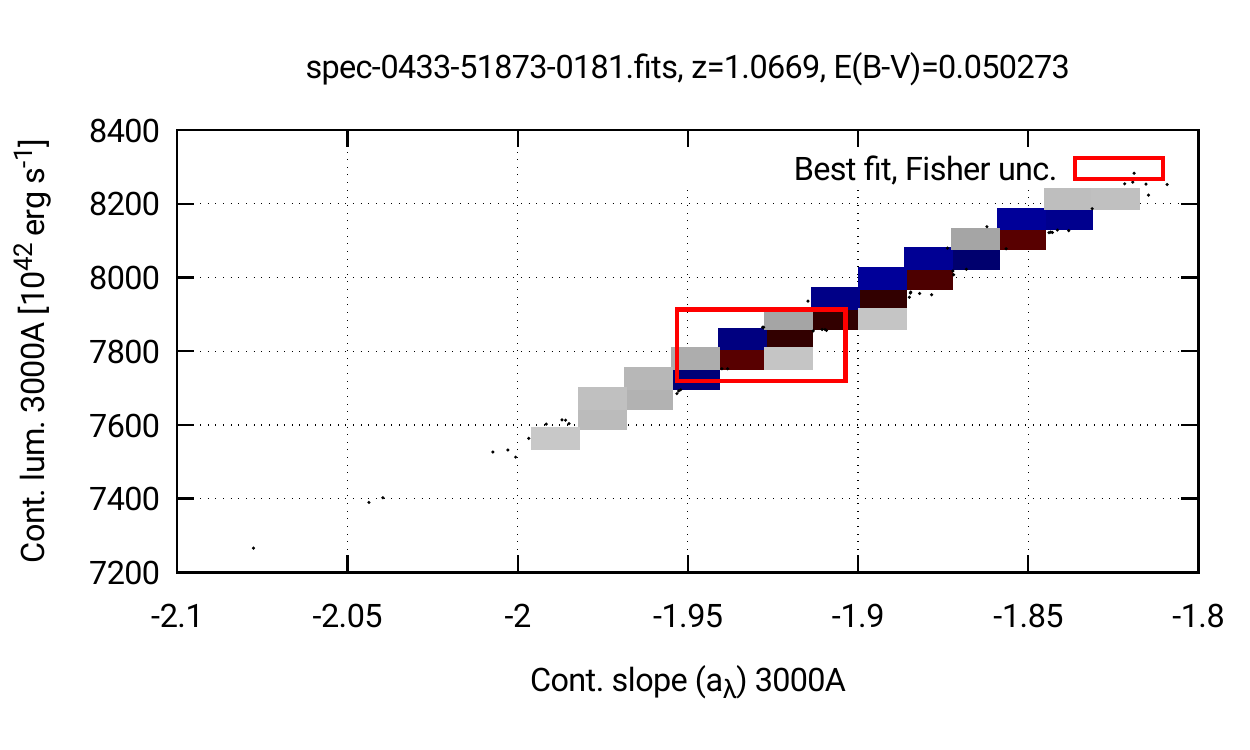}\\
  \caption{Same as Fig.~\ref{fig:MC-lowZ} but considering the analysis
    of another source.}
  \label{fig:MC-highZ}
\end{figure*}

Examples of such distributions are shown in Fig.~\ref{fig:MC-lowZ} and
\ref{fig:MC-highZ} for the same source analyzed in Fig.~\ref{fig:ex1}
and for {\texttt spec-0433-51873-0181} ($z$=1.07) respectively.  The
first five histograms show the comparison between the Gaussian
distribution expected from the best fit parameter value and its
associated Fisher uncertainty (shown in red) as returned by \qsfit{},
and the actual distribution of best fit parameters when the data are
resampled $10^3$ times (shown in blue).  The sixth panel shows the
distribution of the reduced $\chi^2$ values (in blue) which is
obviously centered on one, while the \qsfit{} value (in red) may be
slightly larger than one.  The last two panels shows the 2--D
distribution of two correlated parameters following the same rules
described in \S\ref{sec:catalog}.  The red square identifies the
region of $\pm 1 \sigma$ Fisher uncertainties for both parameters.  In
both Fig.~\ref{fig:MC-lowZ} and \ref{fig:MC-highZ} the Fisher values
provide reasonable estimates of parameter uncertainties, in a
significantly smaller fraction of time with respect to Monte Carlo
estimates.  The latter, however, provide uncertainties which are
significantly more reliable and should be used whenever the parameters
of interest are correlated.

To run a Monte Carlo resampling method a thousand times simply add the
\verb|resample=1000| keyword to the \verb|qsfit| call
(\S\ref{sec:qsfit-usage}).  The return value will be an array of
structures (1000 in this case), instead of a single one.

\subsection{The \qsfit{} continuum component}
\label{sec:sbpl}

The AGN continuum component used in the analysis of the catalog
(\S\ref{sec:catalog}) is a simple power law.  However, the \qsfit{}
continuum component can be customized to behave as a smoothly broken
power law model, to suit those cases where the spectral coverage is
sufficiently large to constrain the extra continuum parameters.

The smoothly broken power law is modeled as follows:
\begin{equation}
  \label{eq-sbpl}
  L_{\lambda} = A
  \cpar{\frac{\lambda}{\lambda_{\rm b}}}^{\alpha_{\lambda}}
  \spar{\frac{1 + (\lambda / \lambda_{\rm b})^{|\Delta \alpha|\gamma}}{2}}^{S/\gamma}
\end{equation}
where $\lambda_{\rm b}$ is the break wavelength, $A$ is the luminosity
density at $\lambda=\lambda_{\rm b}$, $\alpha_{\lambda}$ is the
spectral slope at $\lambda \ll \lambda_{\rm b}$, $\Delta
\alpha_{\lambda}$ is the slope change at $\lambda \gg \lambda_{\rm
  b}$, $S$ is either $+1$ or $-1$, depending on the sign of $\Delta
\alpha_{\lambda}$, and $\gamma$ is the curvature parameter. When
$\lambda \ll \lambda_{\rm b}$ or $\lambda \gg \lambda_{\rm b}$
Eq.\ref{eq-sbpl} reduces to:
\begin{equation}
  \label{eq:sbpl}
  L_{\lambda} = A \times \left\{
  \begin{array}{lr}
    \cpar{\frac{\lambda}{\lambda_{\rm b}}}^{\alpha_{\lambda}}                           & \lambda \ll \lambda_{\rm b}\\
    \cpar{\frac{\lambda}{\lambda_{\rm b}}}^{\alpha_{\lambda} + \Delta \alpha_{\lambda}} & \lambda \gg \lambda_{\rm b}
  \end{array}
  \right .
\end{equation}
The change of slopes occurs between the wavelengths $\lambda_1$ ($<
\lambda_{\rm b}$) and $\lambda_2$ ($> \lambda_{\rm b}$) such that:
\begin{equation}
  \log_{10} \frac{\lambda_2}{\lambda_1} \sim \frac{2}{|\Delta \alpha_{\lambda}| \times \gamma}
\end{equation}
hence $\gamma$ controls the ``sharpness'' of the change in slope.  In
the current \qsfit{} implementation the $A$ parameter is constrained
to be positive; $\lambda_{\rm b}$ is constrained in the range
$[\lmin{} + 0.2 \times(\lmax{}-\lmin{})$, $\lmin{} + 0.8
  \times(\lmax{}-\lmin{})]$; $\alpha_{\lambda}$ is constrained in the
range [$-3$, 1]; $\Delta \alpha_{\lambda}$ is constrained in the range
[$-0.2$, 0.2], although the appropriate limits depends on the
available spectral coverage.

\subsection{\qsfit{} design, internals and the \gfit{} framework}
\label{sec:gfit}

In this section we will provide a brief description of the principles
followed while developing \qsfit{}, its most relevant routines, and
the \gfit{} framework (used for model evaluation).  Also, we will
provide the very basic information required to understand the code in
the \verb|qsfit.pro| file.  This is not intended to be a complete
documentation of the code, but rather a quick and very short
introduction to \qsfit{} internals.

\medskip

Unlike the field of X--ray spectral analysis, there is not yet a well
established and widely accepted procedure to automatically analyze the
Type 1 AGN optical spectra.  This is due to the many possible ways the
problem can be approached, and to the different purposes pursued by
the researchers in the field.  \qsfit{} is just a first attempt to
fill this gap but, given the above premises, it is likely that the way
it operates will not be suitable to fit particular needs, and likely
the code will require some customization.  This implies that the best
way to design \qsfit{} is by providing a neat and short source code
with extensive documentation, easy to read and customize.  We went a
little further and carefully separated the procedures which are rather
well established and will unlikely be modified by the user, from those
that are actually related to the specific task of AGN spectral
fitting.  The former procedures were collected in a library in the
\verb|IDL/Glib| and \verb|IDL/qsfit/components| directories of the
\qsfit{} package, while the latter are all contained in a single file
named \verb|qsfit.pro| (located in the \verb|IDL/qsfit| directory).
In the vast majority of cases this is the only file one needs to
modify to customize the fitting process.  As a consequence, it is
actually pointless to develop a well defined and user--friendly
interface to customize the fitting process (as is the case for,
e.g. {\sc XSPEC}, {\sc Sherpa}, etc.), since the underlying code can
be modified in an unpredictable way.  The drawback of this approach is
that the users should become acquainted with the \qsfit{} source code.
On the other hand it provides the greatest flexibility to the user
since it allows to modify every detail of the analysis and implement
very complex recipes.  However, in order to facilitate the most common
customization tasks even for unexperienced users, we created a IDL
global variable named \verb|!QSFIT_OPT| containing the settings a user
may wish to change.  The default values for such settings, and their
meaning, can be found in the \verb|qsfit_prepare_options| routine in
the \verb|qsfit.pro| file.  Moreover, we marked the places in the
source file \verb|qsfit.pro| where an easy customization can be made,
the with the comment \verb|CUSTOMIZABLE|.  These lines can be easily
modified even without a complete understanding of the remaining code.

\medskip

The \qsfit{} code relies on the \gfit{} framework to handle the model
evaluation, parameter definitions and constraints, and the generations
of plots to compare the data and the model.  The \qsfit{} framework
relies on the MPFIT minimization routine \citep{2009-Markwardt-MPFIT},
and its use allowed us to write a very neat code and to always obtain
the best performances in model evaluation, regardless of the
complexity of the model.  The documentation for the \gfit{} framework
can be found in the \verb|IDL/Glib/gfit| directory.  Here we will
shortly describe its usage:
\begin{itemize}
  \item the \gfit{} software maintains its whole state in a single
    structure named \verb|gfit|.  The information related to the
    input data, the components to build the model, the parameters in
    each component, the parameters constraints, the plotting settings,
    etc. are all stored in this variable.  The \verb|gfit| structure
    is available in an IDL common block with the same name, hence
    each program unit willing to use it should state the following
    line at its beginning:
\begin{verbatim}
  COMMON GFIT
\end{verbatim}

\item the \verb|gfit| structure is modified by the calls to one of the
  \verb|gfit_*| routines, e.g. \verb|gfit_add_data| to add a data set,
  \verb|gfit_add_comp| to add a component to the model, etc.

\item the values in the \verb|gfit| structure can be directly modified
  by the user, e.g. the mathematical expression for the model will be
  set into \verb|gfit.expr.(0).model|, the limits for the host galaxy
  normalization parameter will be set into
  \verb|gfit.comp.galaxy.norm.limits|, etc.  An absorption line can
  easily be included in the model by setting the normalization limits
  to negative values, e.g.:
  \verb|gfit.comp.LINE_NAME.norm.limits = [-1, 0]|;

\item each time the model is modified (either by adding a component,
  by freezing or thawing a parameter) the \verb|gfit_compile|
  procedure should be called.  This procedure automatically generates
  and compile the IDL code needed to evaluate the model, ensuring the
  best possible performance even with very complex model definitions;

\item the \verb|gfit_report| and the \verb|gfit_plot| procedures are
  respectively used to generate a report of the \gfit{} state, and to
  generate the comparison plot between the model and the data;

\item the \verb|gfit_run| procedure is used to perform the model
  minimization through a call to \verb|mpfit|.  Once finished, the
  parameter values and uncertainties are updated.  The latter are
  available in the \verb|gfit.comp| substructure,
  e.g. \verb|gfit.comp.continuum.norm.val|,
  \verb|gfit.comp.continuum.norm.err|, etc.
\end{itemize}

\medskip

As discussed above, all the relevant code for \qsfit{} customization
is located in the \verb|qsfit.pro| file.  In the following we will
briefly mention the most relevant routines:
\begin{itemize}
\item \verb|qsfit_log|: this procedure print log messages.  All
  messages are also stored in an internal buffer to be dumped at the
  end of the analysis;

\item \verb|qsfit_compile|: this procedure (re--)compiles the \gfit{}
  model.  It should be called each time the model is modified;

\item \verb|qsfit_lineset|: this function returns the list of
  ``known'' emission lines to be considered in the analysis.  It can
  be customized to add/remove emission line components to the model.
  Disabling a few lines, e.g. the less luminous ones, is the simplest
  way to significantly speed up the \qsfit{} analysis;

\item \verb|qsfit_prepare|: this procedure reads a SDSS DR--10 FITS
  file and loads data into \gfit{} internal structure (see
  \S\ref{sec:prep}).  If you wish to read data from another source you
  should modify this procedure;

\item \verb|qsfit_run|: this procedure implements the actual \qsfit{}
  recipe for model fitting (\S\ref{sec:modelFitting}).  If you wish to
  modify the sequence by which each component is added you should
  modify this procedure;

\item \verb|qsfit_reduce|: this function reduces the data in the
  \gfit{} structure (\S\ref{sec:reduction}) and returns the final
  nested structure (\S\ref{sec:nested});

\item \verb|qsfit|: this is the main function to perform the \qsfit{}
  analysis, and actually the only one supposed to be directly called
  by the final user;

\item \verb|qsfit_report|: this procedure prints the final report of the
  \qsfit{} analysis;

\item \verb|qsfit_plot|: this procedure generates the plots of the
  results of the analysis (\S\ref{sec:qsfit-usage});

\item \verb|qsfit_version|: this function simply returns a string with
  the current \qsfit{} version, also stored in the nested structure
  returned by the \verb|qsfit| function (see \S\ref{sec:nested}).  If
  you modify the \verb|qsfit.pro| file we suggest to modify this
  function as well, in order to quickly identify which version
  generated a given result.
\end{itemize}

\section{Quality flags in the \qsfitcat{} catalog}
\label{sec:qualityflags}

The columns in the \qsfitcat{} catalog whose name has a
\verb|__QUALITY| suffix report the ``quality flags'' for the
associated quantity (see \S\ref{sec:flattened-struc}).  In this
section we report both the total number and the fraction of sources
(over the total number of sources in the catalog: \nn{}) which raised
a flag. Also shown are the total number and the fraction of sources
with no flag raised (i.e. whose \verb|__QUALITY| entry is equal to
zero), also called ``Good'' sources with respect to a specific
quantity.

The statistics for the continuum quantities are summarized in
Tab.~\ref{tab:QualityCont}, where each of the 5 wavelengths (equally
spaced in the logarithmic wavelength range) are shown in the upper
part of the table, in the columns labeled W1...W5.  The lower part of
the table reports the statistics for the 1450\AA, 2245\AA{}, 3000\AA,
4210\AA{} and 5100\AA{} fixed wavelengths.  The statistics for the
host galaxy quantities are summarized in
Tab.~\ref{tab:QualityGalaxy}. The statistics for the four UV iron
components and for the two optical iron quantities are shown in
Tab.~\ref{tab:QualityIron} (upper and lower table respectively).
Finally, the statistics for the emission line quantities are shown in
Tab.~\ref{tab:QualityLine} and Tab.~\ref{tab:QualityLine2}.
\begin{table*}
  \begin{center}
    \caption{Total number and fraction of sources (over the total
      number of sources in the catalog: \nn{}) which raised a flag in
      the \texttt{CONT*\_\_QUALITY} bitmask.  The upper table shows
      the statistics for the five equally spaced wavelengths in the
      available range (see \S\ref{sec:reduction}), while the lower
      table shows the results for the 1450\AA, 2245\AA{}, 3000\AA,
      4210\AA{} and 5100\AA{} fixed wavelengths.  The first row in
      each table shows the total number and fraction of sources which
      raised no flag in the bitmask (``Good'' sources with
      respect to the continuum quantities).}
    \label{tab:QualityCont}
    \begin{tabular}{l|rr|rr|rr|rr|rr}
      \hline\hline
      {\bf Quality}                     &
      \multicolumn{2}{c|}{\bf W1}       &
      \multicolumn{2}{c|}{\bf W2}       &
      \multicolumn{2}{c|}{\bf W3}       &
      \multicolumn{2}{c|}{\bf W4}       &
      \multicolumn{2}{c}{ \bf W5}       \\
      \hline
Good$^a$  &   70590 &  99.07\% &   70590 &  99.07\% &   70588 &  99.07\% &   70587 &  99.07\% &   70586 &  99.07\% \\
Bit 0$^b$ &       0 &   0.00\% &       0 &   0.00\% &       0 &   0.00\% &       0 &   0.00\% &       0 &   0.00\% \\
Bit 1$^c$ &      32 &   0.04\% &      32 &   0.04\% &      32 &   0.04\% &      32 &   0.04\% &      32 &   0.04\% \\
Bit 2$^d$ &       4 &   0.01\% &       4 &   0.01\% &       4 &   0.01\% &       4 &   0.01\% &       5 &   0.01\% \\
Bit 3$^e$ &     277 &   0.39\% &     277 &   0.39\% &     277 &   0.39\% &     277 &   0.39\% &     277 &   0.39\% \\
Bit 4$^f$ &      38 &   0.05\% &      38 &   0.05\% &      40 &   0.06\% &      41 &   0.06\% &      41 &   0.06\% \\
Bit 5$^g$ &     320 &   0.45\% &     320 &   0.45\% &     320 &   0.45\% &     320 &   0.45\% &     320 &   0.45\% \\
Bit 6$^h$ &       0 &   0.00\% &       0 &   0.00\% &       0 &   0.00\% &       0 &   0.00\% &       0 &   0.00\% \\
      \hline\hline
      \noalign{\vskip 5mm}
      \hline\hline
      {\bf Quality}                     &
      \multicolumn{2}{c|}{\bf 1450\AA}  &
      \multicolumn{2}{c|}{\bf 2245\AA}  &
      \multicolumn{2}{c|}{\bf 3000\AA}  &
      \multicolumn{2}{c|}{\bf 4210\AA}  &
      \multicolumn{2}{c}{\bf 5100\AA}  \\
      \hline
Good$^a$  &    6587 &   9.24\% &   52105 &  73.13\% &   60791 &  85.32\% &   25325 &  35.54\% &   14117 &  19.81\% \\
Bit 0$^b$ &   64647 &  90.73\% &   18962 &  26.61\% &    9839 &  13.81\% &   45403 &  63.72\% &   56806 &  79.73\% \\
Bit 1$^c$ &       0 &   0.00\% &       0 &   0.00\% &       6 &   0.01\% &      32 &   0.04\% &      31 &   0.04\% \\
Bit 2$^d$ &       0 &   0.00\% &       0 &   0.00\% &       0 &   0.00\% &       4 &   0.01\% &       4 &   0.01\% \\
Bit 3$^e$ &      11 &   0.02\% &     123 &   0.17\% &     272 &   0.38\% &     179 &   0.25\% &     102 &   0.14\% \\
Bit 4$^f$ &       4 &   0.01\% &      21 &   0.03\% &      34 &   0.05\% &      18 &   0.03\% &      13 &   0.02\% \\
Bit 5$^g$ &       2 &   0.00\% &      41 &   0.06\% &     318 &   0.45\% &     297 &   0.42\% &     186 &   0.26\% \\
Bit 6$^h$ &       0 &   0.00\% &       0 &   0.00\% &       0 &   0.00\% &       0 &   0.00\% &       0 &   0.00\% \\
      \hline\hline
    \end{tabular}
  \end{center}
  \begin{flushleft}
    $^a$: All bits set to 0;\\
    $^b$: Bit 0: wavelength is outside the observed range;\\
    $^c$: Bit 1: either the luminosity or its uncertainty are NaN or equal to zero;\\
    $^d$: Bit 2: luminosity relative uncertainty $>$ 1.5;\\
    $^e$: Bit 3: either the slope or its uncertainty are NaN or equal to zero;\\
    $^f$: Bit 4: slope hits a limit in the fit;\\
    $^g$: Bit 5: slope uncertainty $>$ 0.3;\\
  \end{flushleft}
\end{table*}
\begin{table*}
  \begin{center}
    \caption{Total number and fraction of sources (over the total
      number of sources in the catalog: \nn{}) which raised a flag in
      the \texttt{GALAXY\_\_QUALITY} bitmask.  The first row shows
      the total number and fraction of sources which raised no flag in
      the bitmask (``Good'' sources with respect to the host
      galaxy quantities).}
    \label{tab:QualityGalaxy}
    \begin{tabular}{l|rr}
      \hline\hline
      {\bf Quality} & \multicolumn{2}{c}{\bf \# sources} \\
      \hline
      Good$^a$  &   13015 &   18.3\% \\
      Bit 0$^b$ &   53887 &   75.6\% \\
      Bit 1$^c$ &    3727 &    5.2\% \\
      Bit 2$^d$ &     622 &    0.9\% \\
      \hline\hline
    \end{tabular}
  \end{center}
  \begin{flushleft}
    $^a$: All bits set to 0;\\
    $^b$: Bit 0: fit of galaxy template is not sensible at this redshift;\\
    $^c$: Bit 1: either the luminosity or its uncertainty are NaN or equal to zero;\\
    $^d$: Bit 2: luminosity relative uncertainty $>$ 1.5;\\
  \end{flushleft}
\end{table*}
\begin{table*}
  \begin{center}
    \caption{Total number and fraction of sources (over the total
      number of sources in the catalog: \nn{}) which raised a flag in
      the \texttt{IRONUV\_\_QUALITY} bitmasks (upper table), or in any
      of the \texttt{IRONOPT\_BR\_\_QUALITY} and
      \texttt{IRONOPT\_NA\_\_QUALITY} bitmasks (lower table).  The
      first row in each table shows the total number and fraction of
      sources that raised no flag in the bitmask (``Good'' sources
      with respect to the iron quantities).}
    \label{tab:QualityIron}
    \begin{tabular}{l|rr|rr}
      \hline\hline
          {\bf Quality}                   &
          \multicolumn{2}{c|}{\bf Iron UV}\\
          \hline
          Good$^a$  &   46540 &   65.3\% \\
          Bit 0$^b$ &       0 &    0.0\% \\
          Bit 1$^c$ &   24676 &   34.6\% \\
          Bit 2$^d$ &      35 &    0.0\% \\
          \hline\hline
          \noalign{\vskip 5mm}
          \hline\hline
          {\bf Quality}                   &
          \multicolumn{2}{c|}{\bf Broad}  &
          \multicolumn{2}{c|}{\bf Narrow}\\
          \hline
          Good$^a$  &   21449 &   30.1\% &   12070 &   16.9\% \\
          Bit 0$^b$ &   48416 &   68.0\% &   48416 &   68.0\% \\
          Bit 1$^c$ &     871 &    1.2\% &    7466 &   10.5\% \\
          Bit 2$^d$ &     515 &    0.7\% &    3299 &    4.6\% \\
          \hline\hline
    \end{tabular}
  \end{center}
  \begin{flushleft}
    $^a$: All bits set to 0;\\
    $^b$: Bit 0: fit of iron template is not sensible at this redshift;\\
    $^c$: Bit 1: either the luminosity or its uncertainty are NaN or equal to zero;\\
    $^d$: Bit 2: luminosity relative uncertainty $>$ 1.5;\\
  \end{flushleft}
\end{table*}
\begin{table*}
  \begin{center}
    \caption{Total number and fraction of sources (over the total
      number of sources reported in the row labeled ``Total'') whose
      emission lines were modeled with either 1 component (${\rm
        N}_{\rm comp} = 1$), 2 components (${\rm N}_{\rm comp} = 2$),
      etc. The line labeled ``Total'' reports the total number of
      sources for which the \texttt|*\_\_NCOMP| is not equal to zero,
      i.e. for the sources which are in the appropriate redshift range
      for the emission line to be detectable (see
      \S\ref{sec:comp-lines}).  In the second part of the tables we
      show the total number and fraction of sources which raised a
      flag in any of the emission line \texttt{\_\_QUALITY} bitmasks.
      The fractions refer always to the row labeled ``Total''.
      Finally, the lines labeled ``Good'' shows the total number and
      fraction of sources which raised no flag in the bitmask (``Good''
      sources with respect to the emission line quantities).}
    \label{tab:QualityLine}
    \begin{tabular}{l|rr|rr|rr|rr|rr}
      \hline\hline
{\bf Line}       &\multicolumn{2}{c|}{\bf SiIV (B)}           &\multicolumn{2}{c|}{\bf CIV (B)}           &\multicolumn{2}{c|}{\bf CIII (B)}           &\multicolumn{2}{c|}{\bf MgII (B)}           &\multicolumn{2}{c|}{\bf NeVI (N)}            \\
Wavelen. [\AA]   &\multicolumn{2}{c|}{    1400}               &\multicolumn{2}{c|}{    1549}              &\multicolumn{2}{c|}{    1909}               &\multicolumn{2}{c|}{    2798}               &\multicolumn{2}{c|}{    3426}                \\
                 &                                  &         &                                 &         &                                  &         &                                  &         &                                  &          \\
N$_{\rm comp}$=1 &    7375                          &(97.1\%) &   12001                         &(58.7\%) &   40138                          &(93.1\%) &   48826                          &(74.1\%) &   49970                          &(100.0\%) \\
N$_{\rm comp}$=2 &     220                          &(2.9\%)  &    7709                         &(37.7\%) &    2916                          &(6.8\%)  &   16397                          &(24.9\%) &                                  &          \\
N$_{\rm comp}$=3 &       1                          &(0.0\%)  &     685                         &(3.4\%)  &      49                          &(0.1\%)  &     661                          &(1.0\%)  &                                  &          \\
N$_{\rm comp}$=4 &                                  &         &      39                         &(0.2\%)  &       1                          &(0.0\%)  &      14                          &(0.0\%)  &                                  &          \\
N$_{\rm comp}$=5 &                                  &         &       2                         &(0.0\%)  &                                  &         &                                  &         &                                  &          \\
N$_{\rm comp}$=6 &                                  &         &       1                         &(0.0\%)  &                                  &         &                                  &         &                                  &          \\
N$_{\rm comp}$=7 &                                  &         &       2                         &(0.0\%)  &                                  &         &                                  &         &                                  &          \\
Total            &    7596                          &         &   20439                         &         &   43104                          &         &   65898                          &         &   49970                          &          \\
                 &                                  &         &                                 &         &                                  &         &                                  &         &                                  &          \\
{\bf Quality}    &                                  &         &                                 &         &                                  &         &                                  &         &                                  &          \\
Good$^a$         &    5151                          &(67.8\%) &   15193                         &(74.3\%) &   25208                          &(58.5\%) &   52399                          &(79.5\%) &   11134                          &(22.3\%)  \\
Bit 0$^b$        &      99                          &(1.3\%)  &      67                         &(0.3\%)  &      94                          &(0.2\%)  &      49                          &(0.1\%)  &    4073                          &(8.2\%)   \\
Bit 1$^c$        &      62                          &(0.8\%)  &      53                         &(0.3\%)  &      87                          &(0.2\%)  &      86                          &(0.1\%)  &    4373                          &(8.8\%)   \\
Bit 2$^d$        &     565                          &(7.4\%)  &     930                         &(4.6\%)  &    6493                          &(15.1\%) &     654                          &(1.0\%)  &   22811                          &(45.6\%)  \\
Bit 3$^e$        &     661                          &(8.7\%)  &     861                         &(4.2\%)  &    4357                          &(10.1\%) &    2005                          &(3.0\%)  &   13585                          &(27.2\%)  \\
Bit 4$^f$        &      37                          &(0.5\%)  &      33                         &(0.2\%)  &      45                          &(0.1\%)  &      77                          &(0.1\%)  &    1718                          &(3.4\%)   \\
Bit 5$^g$        &     699                          &(9.2\%)  &    1922                         &(9.4\%)  &    9337                          &(21.7\%) &    8495                          &(12.9\%) &   15206                          &(30.4\%)  \\
Bit 6$^h$        &     108                          &(1.4\%)  &      74                         &(0.4\%)  &     913                          &(2.1\%)  &    1329                          &(2.0\%)  &    2161                          &(4.3\%)   \\
Bit 7$^i$        &    1642                          &(21.6\%) &    2609                         &(12.8\%) &    4239                          &(9.8\%)  &    3426                          &(5.2\%)  &    5105                          &(10.2\%)  \\

      \hline\hline
      \noalign{\vskip 5mm}
      \hline\hline
{\bf Line}       &\multicolumn{2}{c|}{\bf OII (N)}            &\multicolumn{2}{c|}{\bf NeIII (N)}            &\multicolumn{2}{c|}{\bf Hd (B)}           &\multicolumn{2}{c|}{\bf Hg (B)}           &\multicolumn{2}{c|}{\bf Hb (N)}            \\
Wavelen. [\AA]   &\multicolumn{2}{c|}{    3727}               &\multicolumn{2}{c|}{    3869}                 &\multicolumn{2}{c|}{    4103}             &\multicolumn{2}{c|}{    4342}             &\multicolumn{2}{c|}{    4863}              \\
                 &                                 &          &                                   &          &                                &         &                                &         &                                &          \\
N$_{\rm comp}$=1 &   40712                         &(100.0\%) &   36424                           &(100.0\%) &   29215                        &(95.2\%) &   23450                        &(91.5\%) &   17780                        &(100.0\%) \\
N$_{\rm comp}$=2 &                                 &          &                                   &          &    1465                        &(4.8\%)  &    2159                        &(8.4\%)  &                                &          \\
N$_{\rm comp}$=3 &                                 &          &                                   &          &       4                        &(0.0\%)  &      18                        &(0.1\%)  &                                &          \\
N$_{\rm comp}$=4 &                                 &          &                                   &          &                                &         &                                &         &                                &          \\
N$_{\rm comp}$=5 &                                 &          &                                   &          &                                &         &                                &         &                                &          \\
N$_{\rm comp}$=6 &                                 &          &                                   &          &                                &         &                                &         &                                &          \\
N$_{\rm comp}$=7 &                                 &          &                                   &          &                                &         &                                &         &                                &          \\
Total            &   40712                         &          &   36424                           &          &   30684                        &         &   25627                        &         &   17780                        &          \\
                 &                                 &          &                                   &          &                                &         &                                &         &                                &          \\
{\bf Quality}    &                                 &          &                                   &          &                                &         &                                &         &                                &          \\
Good$^a$         &   21029                         &(51.7\%)  &   11498                           &(31.6\%)  &    8790                        &(28.6\%) &   20393                        &(79.6\%) &    5123                        &(28.8\%)  \\
Bit 0$^b$        &    2156                         &(5.3\%)   &    1028                           &(2.8\%)   &    3375                        &(11.0\%) &      32                        &(0.1\%)  &     376                        &(2.1\%)   \\
Bit 1$^c$        &    2184                         &(5.4\%)   &    1525                           &(4.2\%)   &    2049                        &(6.7\%)  &      89                        &(0.3\%)  &     451                        &(2.5\%)   \\
Bit 2$^d$        &    9292                         &(22.8\%)  &   13495                           &(37.0\%)  &    8711                        &(28.4\%) &    2011                        &(7.8\%)  &    8355                        &(47.0\%)  \\
Bit 3$^e$        &    9393                         &(23.1\%)  &   10442                           &(28.7\%)  &    7994                        &(26.1\%) &     374                        &(1.5\%)  &    4197                        &(23.6\%)  \\
Bit 4$^f$        &     819                         &(2.0\%)   &     574                           &(1.6\%)   &     892                        &(2.9\%)  &      42                        &(0.2\%)  &     162                        &(0.9\%)   \\
Bit 5$^g$        &    6039                         &(14.8\%)  &    5044                           &(13.8\%)  &    9380                        &(30.6\%) &    1298                        &(5.1\%)  &     546                        &(3.1\%)   \\
Bit 6$^h$        &     953                         &(2.3\%)   &    1064                           &(2.9\%)   &    1967                        &(6.4\%)  &     167                        &(0.7\%)  &     104                        &(0.6\%)   \\
Bit 7$^i$        &    3176                         &(7.8\%)   &    2670                           &(7.3\%)   &   11407                        &(37.2\%) &    3307                        &(12.9\%) &     199                        &(1.1\%)   \\

      \hline\hline
    \end{tabular}
  \end{center}
  \begin{flushleft}
    $^a$: All bits set to 0;\\
    $^b$: Bit 0: either the luminosity or its uncertainty are NaN or equal to zero;\\
    $^c$: Bit 1: luminosity relative uncertainty $>$ 1.5;\\
    $^d$: Bit 2: either the FWHM or its uncertainty are NaN or equal to zero;\\
    $^e$: Bit 3: FWHM value hits a limit in the fit;\\
    $^f$: Bit 4: FWHM relative uncertainty $>$ 2;\\
    $^g$: Bit 5: either the V$_{\rm off}$ or its uncertainty are NaN or equal to zero;\\
    $^h$: Bit 6: V$_{\rm off}$ value hits a limit in the fit;\\
    $^i$: Bit 7: V$_{\rm off}$ uncertainty $>$ 500 km s$^{-1}$;\\
  \end{flushleft}
\end{table*}
\begin{table*}
  \begin{center}
    \caption{...continue from Tab.~\ref{tab:QualityLine}.}
    \label{tab:QualityLine2}
    \begin{tabular}{l|rr|rr|rr|rr|rr}
      \hline\hline
{\bf Line}       &\multicolumn{2}{c|}{\bf Hb (B)}           &\multicolumn{2}{c|}{\bf OIII (N)}            &\multicolumn{2}{c|}{\bf OIII (N)}            &\multicolumn{2}{c|}{\bf HeI (B)}           &\multicolumn{2}{c|}{\bf NII (N)}            \\
Wavelen. [\AA]   &\multicolumn{2}{c|}{    4863}             &\multicolumn{2}{c|}{    4959}                &\multicolumn{2}{c|}{    5007}                &\multicolumn{2}{c|}{    5876}              &\multicolumn{2}{c|}{    6549}               \\
                 &                                &         &                                  &          &                                  &          &                                 &         &                                 &          \\
N$_{\rm comp}$=1 &   15405                        &(86.7\%) &   16372                          &(100.0\%) &   16233                          &(100.0\%) &    7209                         &(96.2\%) &    2841                         &(100.0\%) \\
N$_{\rm comp}$=2 &    2323                        &(13.1\%) &                                  &          &                                  &          &     280                         &(3.7\%)  &                                 &          \\
N$_{\rm comp}$=3 &      50                        &(0.3\%)  &                                  &          &                                  &          &       2                         &(0.0\%)  &                                 &          \\
N$_{\rm comp}$=4 &                                &         &                                  &          &                                  &          &                                 &         &                                 &          \\
N$_{\rm comp}$=5 &                                &         &                                  &          &                                  &          &                                 &         &                                 &          \\
N$_{\rm comp}$=6 &                                &         &                                  &          &                                  &          &                                 &         &                                 &          \\
N$_{\rm comp}$=7 &                                &         &                                  &          &                                  &          &                                 &         &                                 &          \\
Total            &   17778                        &         &   16372                          &          &   16233                          &          &    7491                         &         &    2841                         &          \\
                 &                                &         &                                  &          &                                  &          &                                 &         &                                 &          \\
{\bf Quality}    &                                &         &                                  &          &                                  &          &                                 &         &                                 &          \\
Good$^a$         &   14714                        &(82.8\%) &       0                          &(0.0\%)   &   14377                          &(88.6\%)  &    1351                         &(18.0\%) &    1602                         &(56.4\%)  \\
Bit 0$^b$        &      37                        &(0.2\%)  &     351                          &(2.1\%)   &      33                          &(0.2\%)   &    1164                         &(15.5\%) &      50                         &(1.8\%)   \\
Bit 1$^c$        &      42                        &(0.2\%)  &     322                          &(2.0\%)   &      30                          &(0.2\%)   &     407                         &(5.4\%)  &      64                         &(2.3\%)   \\
Bit 2$^d$        &    1194                        &(6.7\%)  &    3384                          &(20.7\%)  &     500                          &(3.1\%)   &    3127                         &(41.7\%) &     690                         &(24.3\%)  \\
Bit 3$^e$        &     280                        &(1.6\%)  &    1838                          &(11.2\%)  &    1321                          &(8.1\%)   &    1631                         &(21.8\%) &     476                         &(16.8\%)  \\
Bit 4$^f$        &      16                        &(0.1\%)  &     111                          &(0.7\%)   &      11                          &(0.1\%)   &     236                         &(3.2\%)  &       6                         &(0.2\%)   \\
Bit 5$^g$        &    1072                        &(6.0\%)  &   16372                          &(100.0\%) &      87                          &(0.5\%)   &    2838                         &(37.9\%) &      84                         &(3.0\%)   \\
Bit 6$^h$        &      76                        &(0.4\%)  &       0                          &(0.0\%)   &      32                          &(0.2\%)   &     858                         &(11.5\%) &      20                         &(0.7\%)   \\
Bit 7$^i$        &    1533                        &(8.6\%)  &       0                          &(0.0\%)   &      17                          &(0.1\%)   &    2956                         &(39.5\%) &      64                         &(2.3\%)   \\
      \hline\hline
      \noalign{\vskip 5mm}
      \hline\hline
{\bf Line}       &\multicolumn{2}{c|}{\bf Ha (N)}            &\multicolumn{2}{c|}{\bf Ha (B)}           &\multicolumn{2}{c|}{\bf NII (N)}            &\multicolumn{2}{c|}{\bf SII (N)}            &\multicolumn{2}{c|}{\bf SII (N)}            \\
Wavelen. [\AA]   &\multicolumn{2}{c|}{    6565}              &\multicolumn{2}{c|}{    6565}             &\multicolumn{2}{c|}{    6583}               &\multicolumn{2}{c|}{    6716}               &\multicolumn{2}{c|}{    6731}               \\
                 &                                &          &                                &         &                                 &          &                                 &          &                                 &          \\
N$_{\rm comp}$=1 &    2924                        &(100.0\%) &    2780                        &(95.1\%) &    2843                         &(100.0\%) &    2483                         &(100.0\%) &    2417                         &(100.0\%) \\
N$_{\rm comp}$=2 &                                &          &     140                        &(4.8\%)  &                                 &          &                                 &          &                                 &          \\
N$_{\rm comp}$=3 &                                &          &       3                        &(0.1\%)  &                                 &          &                                 &          &                                 &          \\
N$_{\rm comp}$=4 &                                &          &                                &         &                                 &          &                                 &          &                                 &          \\
N$_{\rm comp}$=5 &                                &          &                                &         &                                 &          &                                 &          &                                 &          \\
N$_{\rm comp}$=6 &                                &          &                                &         &                                 &          &                                 &          &                                 &          \\
N$_{\rm comp}$=7 &                                &          &                                &         &                                 &          &                                 &          &                                 &          \\
Total            &    2924                        &          &    2923                        &         &    2843                         &          &    2483                         &          &    2417                         &          \\
                 &                                &          &                                &         &                                 &          &                                 &          &                                 &          \\
{\bf Quality}    &                                &          &                                &         &                                 &          &                                 &          &                                 &          \\
Good$^a$         &    1818                        &(62.2\%)  &    2818                        &(96.4\%) &    2014                         &(70.8\%)  &    1136                         &(45.8\%)  &    1090                         &(45.1\%)  \\
Bit 0$^b$        &      14                        &(0.5\%)   &       7                        &(0.2\%)  &      12                         &(0.4\%)   &      42                         &(1.7\%)   &      91                         &(3.8\%)   \\
Bit 1$^c$        &      42                        &(1.4\%)   &       5                        &(0.2\%)  &      38                         &(1.3\%)   &      64                         &(2.6\%)   &      80                         &(3.3\%)   \\
Bit 2$^d$        &     196                        &(6.7\%)   &      39                        &(1.3\%)  &     161                         &(5.7\%)   &     703                         &(28.3\%)  &     470                         &(19.4\%)  \\
Bit 3$^e$        &     889                        &(30.4\%)  &       9                        &(0.3\%)  &     629                         &(22.1\%)  &     608                         &(24.5\%)  &     829                         &(34.3\%)  \\
Bit 4$^f$        &      13                        &(0.4\%)   &       1                        &(0.0\%)  &       5                         &(0.2\%)   &      23                         &(0.9\%)   &      32                         &(1.3\%)   \\
Bit 5$^g$        &      17                        &(0.6\%)   &      30                        &(1.0\%)  &      19                         &(0.7\%)   &     356                         &(14.3\%)  &     205                         &(8.5\%)   \\
Bit 6$^h$        &       2                        &(0.1\%)   &       1                        &(0.0\%)  &      21                         &(0.7\%)   &      19                         &(0.8\%)   &      14                         &(0.6\%)   \\
Bit 7$^i$        &      14                        &(0.5\%)   &      44                        &(1.5\%)  &      24                         &(0.8\%)   &      55                         &(2.2\%)   &      51                         &(2.1\%)   \\
      \hline\hline
    \end{tabular}
  \end{center}
  \begin{flushleft}
    $^a$: All bits set to 0;\\
    $^b$: Bit 0: either the luminosity or its uncertainty are NaN or equal to zero;\\
    $^c$: Bit 1: luminosity relative uncertainty $>$ 1.5;\\
    $^d$: Bit 2: either the FWHM or its uncertainty are NaN or equal to zero;\\
    $^e$: Bit 3: FWHM value hits a limit in the fit;\\
    $^f$: Bit 4: FWHM relative uncertainty $>$ 2;\\
    $^g$: Bit 5: either the V$_{\rm off}$ or its uncertainty are NaN or equal to zero;\\
    $^h$: Bit 6: V$_{\rm off}$ value hits a limit in the fit;\\
    $^i$: Bit 7: V$_{\rm off}$ uncertainty $>$ 500 km s$^{-1}$;\\
  \end{flushleft}
\end{table*}

\bsp

\label{lastpage}

\begin{thebibliography}{}

\bibitem[\protect\citeauthoryear{{Ahn}, {Alexandroff}, {Allende Prieto},
  {Anders}, {Anderson}, {Anderton}, {Andrews}, {Aubourg}, {Bailey}, {Bastien}
  \& et al.}{{Ahn} et~al.}{2014}]{2014-Ahn-SDSS-DR10}
{Ahn} C.~P.,  {Alexandroff} R.,  {Allende Prieto} C.,  {Anders} F.,  {Anderson}
  S.~F.,  {Anderton} T.,  {Andrews} B.~H.,  {Aubourg} {\'E}.,  {Bailey} S.,
  {Bastien} F.~A.,    et al. 2014, \apjs, 211, 17

\bibitem[\protect\citeauthoryear{{Andrae}}{{Andrae}}{2010}]{2010-Andrae-ErrorEstimation}
{Andrae} R.,  2010, \texttt{(arXiv:1009.2755)}

\bibitem[\protect\citeauthoryear{{Baldwin}}{{Baldwin}}{1977}]{1977-BaldwinEffect}
{Baldwin} J.~A.,  1977, \apj, 214, 679

\bibitem[\protect\citeauthoryear{{Baldwin}, {Wampler} \& {Gaskell}}{{Baldwin}
  et~al.}{1989}]{1989-Baldwin-EmissionLineProperties}
{Baldwin} J.~A.,  {Wampler} E.~J.,    {Gaskell} C.~M.,  1989, \apj, 338, 630

\bibitem[\protect\citeauthoryear{{Barth} et~al.,}{{Barth}
  et~al.}{2015}]{2015-Barth-LickMonitoring}
{Barth} A.~J.,  et~al., 2015, \apjs, 217, 26

\bibitem[\protect\citeauthoryear{{Boroson}}{{Boroson}}{2002}]{2002-Boroson}
{Boroson} T.~A.,  2002, \apj, 565, 78

\bibitem[\protect\citeauthoryear{{Boroson} \& {Green}}{{Boroson} \&
  {Green}}{1992}]{1992-boroson-emlineprop-irontempl}
{Boroson} T.~A.,  {Green} R.~F.,  1992, \apjs, 80, 109

\bibitem[\protect\citeauthoryear{{Burtscher}, {Jaffe}, {Raban}, {Meisenheimer},
  {Tristram} \& {R{\"o}ttgering}}{{Burtscher}
  et~al.}{2009}]{2009-Burtscher-MC-Resampling}
{Burtscher} L.,  {Jaffe} W.,  {Raban} D.,  {Meisenheimer} K.,  {Tristram}
  K.~R.~W.,    {R{\"o}ttgering} H.,  2009, \apjl, 705, L53

\bibitem[\protect\citeauthoryear{{Calderone}, {Sbarrato} \&
  {Ghisellini}}{{Calderone} et~al.}{2012}]{2012-calderone-torus}
{Calderone} G.,  {Sbarrato} T.,    {Ghisellini} G.,  2012, \mnras, 425, L41

\bibitem[\protect\citeauthoryear{{Calistro Rivera}, {Lusso}, {Hennawi} \&
  {Hogg}}{{Calistro Rivera} et~al.}{2016}]{2016-CalistroRivera-MCMC-SED}
{Calistro Rivera} G.,  {Lusso} E.,  {Hennawi} J.~F.,    {Hogg} D.~W.,  2016,
  arXiv:1606.05648

\bibitem[\protect\citeauthoryear{{Cardelli}, {Clayton} \& {Mathis}}{{Cardelli}
  et~al.}{1989}]{1989-cardelli-extinction}
{Cardelli} J.~A.,  {Clayton} G.~C.,    {Mathis} J.~S.,  1989, \apj, 345, 245

\bibitem[\protect\citeauthoryear{{Cristiani} \& {Vio}}{{Cristiani} \&
  {Vio}}{1990}]{1990-Cristiani-CompositeQuasar}
{Cristiani} S.,  {Vio} R.,  1990, \aap, 227, 385

\bibitem[\protect\citeauthoryear{{Davis}, {Woo} \& {Blaes}}{{Davis}
  et~al.}{2007}]{2007-Davis-UVContinuum-ModelsSlopes}
{Davis} S.~W.,  {Woo} J.-H.,    {Blaes} O.~M.,  2007, \apj, 668, 682

\bibitem[\protect\citeauthoryear{{Dietrich}, {Appenzeller}, {Vestergaard} \&
  {Wagner}}{{Dietrich} et~al.}{2002}]{2002-Dietrich}
{Dietrich} M.,  {Appenzeller} I.,  {Vestergaard} M.,    {Wagner} S.~J.,  2002,
  \apj, 564, 581

\bibitem[\protect\citeauthoryear{{Elvis} et~al.,}{{Elvis}
  et~al.}{1994}]{1994-Elvis-atlasQuasar}
{Elvis} M.,  et~al., 1994, \apjs, 95, 1

\bibitem[\protect\citeauthoryear{{Floyd}, {Kukula}, {Dunlop}, {McLure},
  {Miller}, {Percival}, {Baum} \& {O'Dea}}{{Floyd}
  et~al.}{2004}]{2004-Floyd-HostOfLumQSO}
{Floyd} D.~J.~E.,  {Kukula} M.~J.,  {Dunlop} J.~S.,  {McLure} R.~J.,  {Miller}
  L.,  {Percival} W.~J.,  {Baum} S.~A.,    {O'Dea} C.~P.,  2004, \mnras, 355,
  196

\bibitem[\protect\citeauthoryear{{Francis}, {Hewett}, {Foltz}, {Chaffee},
  {Weymann} \& {Morris}}{{Francis} et~al.}{1991}]{1991-francis-composite}
{Francis} P.~J.,  {Hewett} P.~C.,  {Foltz} C.~B.,  {Chaffee} F.~H.,  {Weymann}
  R.~J.,    {Morris} S.~L.,  1991, \apj, 373, 465

\bibitem[\protect\citeauthoryear{{Grandi}}{{Grandi}}{1982}]{1982-Grandi-3000Bump}
{Grandi} S.~A.,  1982, \apj, 255, 25

\bibitem[\protect\citeauthoryear{{Green}, {Forster} \& {Kuraszkiewicz}}{{Green}
  et~al.}{2001}]{2001-Green-BaldwinEffect}
{Green} P.~J.,  {Forster} K.,    {Kuraszkiewicz} J.,  2001, \apj, 556, 727

\bibitem[\protect\citeauthoryear{{Grupe}, {Komossa}, {Leighly} \&
  {Page}}{{Grupe} et~al.}{2010}]{2010-Grupe-SEDOfXRaySelectedAGN}
{Grupe} D.,  {Komossa} S.,  {Leighly} K.~M.,    {Page} K.~L.,  2010, \apjs,
  187, 64

\bibitem[\protect\citeauthoryear{{Heavens}}{{Heavens}}{2009}]{2009-Heavens-StatisticalTechniques}
{Heavens} A.,  2009, \texttt{(arXiv:0906.0664)}

\bibitem[\protect\citeauthoryear{{Kellermann}, {Sramek}, {Schmidt}, {Shaffer}
  \& {Green}}{{Kellermann} et~al.}{1989}]{1989-Kellermann-def_radio_loudness}
{Kellermann} K.,  {Sramek} R.,  {Schmidt} M.,  {Shaffer} D.~B.,    {Green} R.,
  1989, \aj, 98, 1195

\bibitem[\protect\citeauthoryear{{Khachikian} \& {Weedman}}{{Khachikian} \&
  {Weedman}}{1974}]{1974-Khachikian-type1vs2}
{Khachikian} E.~Y.,  {Weedman} D.~W.,  1974, \apj, 192, 581

\bibitem[\protect\citeauthoryear{{Kinney}, {Rivolo} \& {Koratkar}}{{Kinney}
  et~al.}{1990}]{1990-Kinney-Baldwin}
{Kinney} A.~L.,  {Rivolo} A.~R.,    {Koratkar} A.~P.,  1990, \apj, 357, 338

\bibitem[\protect\citeauthoryear{{Kriss}}{{Kriss}}{1994}]{1994-Kriss-specfit}
{Kriss} G.,  1994, in {Crabtree} D.~R.,  {Hanisch} R.~J.,   {Barnes} J.,  eds,
  Astronomical Data Analysis Software and Systems III Vol.~61 of Astronomical
  Society of the Pacific Conference Series, {Fitting Models to UV and Optical
  Spectral Data}.
p.~437

\bibitem[\protect\citeauthoryear{{Markwardt}}{{Markwardt}}{2009}]{2009-Markwardt-MPFIT}
{Markwardt} C.~B.,  2009, in {Bohlender} D.~A.,  {Durand} D.,   {Dowler} P.,
  eds, Astronomical Data Analysis Software and Systems XVIII Vol.~411 of
  Astronomical Society of the Pacific Conference Series, {Non-linear
  Least-squares Fitting in IDL with MPFIT}.
p.~251

\bibitem[\protect\citeauthoryear{{Matsuoka}, {Strauss}, {Shen}, {Brandt},
  {Greene}, {Ho}, {Schneider}, {Sun} \& {Trump}}{{Matsuoka}
  et~al.}{2015}]{2015-Matsuoka-SDSS-RM}
{Matsuoka} Y.,  {Strauss} M.~A.,  {Shen} Y.,  {Brandt} W.~N.,  {Greene} J.~E.,
  {Ho} L.~C.,  {Schneider} D.~P.,  {Sun} M.,    {Trump} J.~R.,  2015, \apj,
  811, 91

\bibitem[\protect\citeauthoryear{{O'Donnell}}{{O'Donnell}}{1994}]{1994-odonnell-updateCCM}
{O'Donnell} J.~E.,  1994, \apj, 422, 158

\bibitem[\protect\citeauthoryear{{Pogge} \& {Peterson}}{{Pogge} \&
  {Peterson}}{1992}]{1992-Pogge-IntrinsicBaldwin}
{Pogge} R.~W.,  {Peterson} B.~M.,  1992, \aj, 103, 1084

\bibitem[\protect\citeauthoryear{{Polletta} et~al.,}{{Polletta}
  et~al.}{2007}]{2007-Polletta-SWIRETemplates}
{Polletta} M.,  et~al., 2007, \apj, 663, 81

\bibitem[\protect\citeauthoryear{{Richards} et~al.,}{{Richards}
  et~al.}{2006}]{2006-richards-meanSED}
{Richards} G.~T.,  et~al., 2006, \apjs, 166, 470

\bibitem[\protect\citeauthoryear{{Sanders}, {Phinney}, {Neugebauer}, {Soifer}
  \& {Matthews}}{{Sanders} et~al.}{1989}]{1989-sanders-torusReproBBB}
{Sanders} D.~B.,  {Phinney} E.~S.,  {Neugebauer} G.,  {Soifer} B.~T.,
  {Matthews} K.,  1989, \apj, 347, 29

\bibitem[\protect\citeauthoryear{{Schneider} et~al.,}{{Schneider}
  et~al.}{2010}]{2010-Schneider-catdr7}
{Schneider} D.~P.,  et~al., 2010, \aj, 139, 2360

\bibitem[\protect\citeauthoryear{{Shankar} et~al.,}{{Shankar}
  et~al.}{2016}]{2016-Shankar-OptUVEmissivityQSO}
{Shankar} F.,  et~al., 2016, \apjl, 818, L1

\bibitem[\protect\citeauthoryear{{Shen} et~al.,}{{Shen}
  et~al.}{2011}]{2011-shen-catdr7}
{Shen} Y.,  et~al., 2011, \apjs, 194, 45

\bibitem[\protect\citeauthoryear{{Shen} \& {Liu}}{{Shen} \&
  {Liu}}{2012}]{2012-Shen-ComparingSEVEstimators}
{Shen} Y.,  {Liu} X.,  2012, \apj, 753, 125

\bibitem[\protect\citeauthoryear{{Silva}, {Granato}, {Bressan} \&
  {Danese}}{{Silva} et~al.}{1998}]{1998-Silva-GRASIL}
{Silva} L.,  {Granato} G.~L.,  {Bressan} A.,    {Danese} L.,  1998, \apj, 509,
  103

\bibitem[\protect\citeauthoryear{{Storey} \& {Hummer}}{{Storey} \&
  {Hummer}}{1995}]{1995-Storey-BalmerLines}
{Storey} P.~J.,  {Hummer} D.~G.,  1995, \mnras, 272, 41

\bibitem[\protect\citeauthoryear{{Sulentic}, {Marziani} \&
  {Dultzin-Hacyan}}{{Sulentic} et~al.}{2000}]{2000-Sulentic-PhenomenologyBLR}
{Sulentic} J.~W.,  {Marziani} P.,    {Dultzin-Hacyan} D.,  2000, \araa, 38, 521

\bibitem[\protect\citeauthoryear{{Sulentic}, {Zamfir}, {Marziani}, {Bachev},
  {Calvani} \& {Dultzin-Hacyan}}{{Sulentic}
  et~al.}{2003}]{2003-Sulentic-EV1-RadioLoud}
{Sulentic} J.~W.,  {Zamfir} S.,  {Marziani} P.,  {Bachev} R.,  {Calvani} M.,
  {Dultzin-Hacyan} D.,  2003, \apjl, 597, L17

\bibitem[\protect\citeauthoryear{{Telfer}, {Zheng}, {Kriss} \&
  {Davidsen}}{{Telfer} et~al.}{2002}]{2002-telfer-UVPropOfQSO}
{Telfer} R.~C.,  {Zheng} W.,  {Kriss} G.~A.,    {Davidsen} A.~F.,  2002, \apj,
  565, 773

\bibitem[\protect\citeauthoryear{{Urry} \& {Padovani}}{{Urry} \&
  {Padovani}}{1995}]{1995-Urry-unifiedscheme}
{Urry} C.~M.,  {Padovani} P.,  1995, \pasp, 107, 803

\bibitem[\protect\citeauthoryear{{Vanden Berk} et~al.,}{{Vanden Berk}
  et~al.}{2001}]{2001-vanden-composite}
{Vanden Berk} D.~E.,  et~al., 2001, \aj, 122, 549

\bibitem[\protect\citeauthoryear{{V{\'e}ron-Cetty}, {Joly} \&
  {V{\'e}ron}}{{V{\'e}ron-Cetty} et~al.}{2004}]{2004-veron-spectra-izw1}
{V{\'e}ron-Cetty} M.-P.,  {Joly} M.,    {V{\'e}ron} P.,  2004, \aap, 417, 515

\bibitem[\protect\citeauthoryear{{Vestergaard}, {Fan}, {Tremonti}, {Osmer} \&
  {Richards}}{{Vestergaard} et~al.}{2008}]{2008-Vestergaard-BHMassFunction}
{Vestergaard} M.,  {Fan} X.,  {Tremonti} C.~A.,  {Osmer} P.~S.,    {Richards}
  G.~T.,  2008, \apjl, 674, L1

\bibitem[\protect\citeauthoryear{{Vestergaard} \& {Wilkes}}{{Vestergaard} \&
  {Wilkes}}{2001}]{2001-vestergaard-UV-iron}
{Vestergaard} M.,  {Wilkes} B.~J.,  2001, \apjs, 134, 1

\bibitem[\protect\citeauthoryear{{Wang}, {Brinkmann} \& {Bergeron}}{{Wang}
  et~al.}{1996}]{1996-Wang-AGNwStrongFeII}
{Wang} T.,  {Brinkmann} W.,    {Bergeron} J.,  1996, \aap, 309, 81

\bibitem[\protect\citeauthoryear{{Wills}, {Netzer} \& {Wills}}{{Wills}
  et~al.}{1985}]{1985-Wills-SBB}
{Wills} B.~J.,  {Netzer} H.,    {Wills} D.,  1985, \apj, 288, 94

\bibitem[\protect\citeauthoryear{{York} et~al.,}{{York}
  et~al.}{2000}]{2000-york-sdss-summary}
{York} D.~G.,  et~al., 2000, \aj, 120, 1579

\end{thebibliography}
\end{document}